\documentclass[twocolumn]{autart}

\journal{Physics Reports}

\usepackage{graphicx}
\usepackage{amsmath}
\usepackage{amssymb}
\usepackage[centerlast]{subfigure}

\numberwithin{equation}{section}

\newcommand{\bv}{{\vec{v}}}
\newcommand{\bx}{{\vec{x}}}

\newcommand{\rset}{{\mathbb{R}}}
\newcommand{\tset}{{\mathbb{T}}}
\newcommand{\zset}{{\mathbb{Z}}}
\newcommand{\la}{\left\langle}
\newcommand{\ra}{\right\rangle}

\begin{document}

\begin{frontmatter}

\title{Burgers Turbulence}

\author{J\'er\'emie Bec} \address{Laboratoire Cassiop\'ee UMR6202,
CNRS, OCA; BP4229, 06304 Nice Cedex 4, France.}
\ead{jeremie.bec@obs-nice.fr}

\author{Konstantin Khanin} \address{Department of Mathematics,
University of Toronto, M5S 3G3 Toronto, Ontario, Canada.}
\ead{khanin@math.toronto.edu}


\begin{abstract}
  The last decades witnessed a renewal of interest in the Burgers
  equation.  Much activities focused on extensions of the original
  one-dimensional pressureless model introduced in the thirties by the
  Dutch scientist J.M.\ Burgers, and more precisely on the problem of
  \emph{Burgers turbulence}, that is the study of the solutions to the
  one- or multi-dimensional Burgers equation with random initial
  conditions or random forcing.  Such work was frequently motivated by
  new emerging applications of Burgers model to statistical physics,
  cosmology, and fluid dynamics.  Also Burgers turbulence appeared as
  one of the simplest instances of a nonlinear system out of
  equilibrium.  The study of random Lagrangian systems, of stochastic
  partial differential equations and their invariant measures, the
  theory of dynamical systems, the applications of field theory to the
  understanding of dissipative anomalies and of multiscaling in
  hydrodynamic turbulence have benefited significantly from progress
  in Burgers turbulence. The aim of this review is to give a unified
  view of selected work stemming from these rather diverse
  disciplines.
\end{abstract}

\end{frontmatter}

\tableofcontents

%%%%%%%%%%%%%%%%%%%%%%%%%%%%%%%%%%%%%%%%%%%%%%

\section{From interface dynamics to cosmology}

At the end of the thirties, the Dutch scientist J.M.~Burgers
\cite{b39} introduced a one-dimensional model for pressure-less gas
dynamics.  He was hoping that the use of a simple model having much in
common with the Navier--Stokes equation would significantly contribute
to the study of fluid turbulence.  This model now known as the Burgers
equation
\begin{equation}
  \partial_t v + v\,\partial_x v = \nu \,\partial_x^2 v,
  \label{eq:burg1d}
\end{equation}
has not only the same type of hydrodynamical (or advective) quadratic
nonlinearity as the Navier--Stokes equation that is balanced by a
diffusive term, but it also has similar invariances and conservation
laws (invariance under translations in space and time, parity
invariance, conservation of energy and momentum in one dimension for
$\nu=0$).

Such hopes appeared to be shattered when in the fifties, Hopf
\cite{h50} and Cole \cite{c51} showed that the Burgers equation can be
integrated explicitly.  This model thus lacks one of the essential
properties of Navier--Stokes turbulence: sensitivity to small
perturbations in the initial data and thus the spontaneous arise of
randomness by chaotic dynamics.  Unable to cope with such a
fundamental aspect, the Burgers equation then lost its interest in
``explaining'' fluid turbulence.

In spite of this, the Burgers equation reappeared in the eighties as
the asymptotic form of various nonlinear dissipative systems.
Physicists and astrophysicists then devoted important effort to the
understanding of its multi-dimensio\-nal form and to the study of its
random solutions arising from random initial conditions or a random
forcing.  The goal of this paper is to review selected works that
exemplify this strong renewal of interest in Burgers turbulence.

The rest of this section is dedicated to the description of several
physical situations where the Burgers equation arises. We will then
see in section~\ref{s:basictools} that in any dimension and in the
limit of vanishing viscosity, the solutions to the Burgers equation
can be expressed in an explicit manner in the decaying case or in an
implicit manner in the forced case, in terms of a variational
principle that permits a systematic classification of its various
singularities (shocks and others) and of their local structure (normal
form).  Section~\ref{s:decay} is dedicated to the study of the decay
of the solutions to the one-dimensional unforced Burgers equation with
random initial data. The multi-dimensional decaying problem is
discussed in section~\ref{s:mass}. The motivation comes from cosmology
where large-scale structures can be described in terms of mass
transport by solutions to the Burgers equation. The basic principles
of the forced Burgers turbulence are discussed in
section~\ref{s:force} where the notions of global minimizer and
topological shocks are introduced. Section~\ref{s:timeperiodic} is
dedicated to the study of the solutions to the periodically kicked
Burgers equation and their relation with Aubry--Mather theory for
commensurate-incommensurate phase transitions.
Section~\ref{s:1Dstatistics} reviews various studies of the
stochastically forced Burgers equation in one dimension with a
particular emphasize on questions that are arising from the
statistical study of turbulent flows. Finally,
section~\ref{s:conclusion} encompasses concluding remarks and a
non-exhaustive list of open questions on the problem of the Burgers
turbulence.

\subsection{The Burgers equation in statistical mechanics}

The Burgers equation appears in condensed matter, in statistical
physics, and also beyond physics in vehicle traffic models (see
\cite{css00}, for a review on this topic). When a random forcing term
is added - usually a white noise in time - it is used to describe
various problems of interface deposition and growth (see, for
instance, \cite{bs95}). An instance frequently studied is the
Kardar--Parisi--Zhang (KPZ) model \cite{kpz86}.  This continuous
version of ballistic deposition models accounts for the lateral growth
of the interface.  Let us indeed consider an interface where particles
deposit with a random flux $F$ that depends both on time $t$ and on
the horizontal position $\bx$.  The growth of the local height $h$
happens in the direction normal to the interface and its time
evolution is given by
\begin{equation}
  \partial_t h - \frac{1}{2} \left| \nabla h \right|^2 = \nu \nabla^2
  h + F,
  \label{eq:kpz}
\end{equation}
where the first term of the right-hand side represents the relaxation
due to a surface tension $\nu$.  The gradient of (\ref{eq:kpz}) gives
the multidimensional Burgers equation
\begin{equation}
  \partial_t \bv + \bv\cdot\nabla\bv = \nu\,\nabla^2 \bv - \nabla F,
  \qquad \bv = -\nabla h,
  \label{eq:burgdd}
\end{equation}
forced by the random potential $F$.  As we will see later, shocks
generically appear in the solution to the Burgers equation in the
inviscid limit $\nu\to0$. They correspond to discontinuities of the
derivative of the height $h$.  The KPZ model is hence frequently used
to understand the appearence of roughness in various interface
problems, as for instance front propagation in randomly distributed
forests (see, e.g., \cite{pagep95}).

The Hopf--Cole transformation ${\mathcal Z} = \exp(h/2\nu)$ allows
rewriting (\ref{eq:kpz}) as a linear problem with random coefficients.
\begin{equation}
  \partial_t {\mathcal Z} = \nu\,\nabla^2{\mathcal Z} + \frac{1}{2\nu}
  F\, {\mathcal Z},
  \label{eq:directed}
\end{equation}
This equation appears in many complex systems, as for instance
directed polymers in random media \cite{kz87,bmp95}.  Indeed the
solution ${\mathcal Z}(\bx,t)$ is exactly the partition function of an
elastic string in the random potential $(1/2\nu)\,F(\bx,t)$, subject
to the constraint that its boundary is fixed at $(\bx,t)$. Note that
here, the time variable $t$ is actually a space variable in the main
direction of the polymer.

\subsection{The adhesion model in cosmology}
\label{ssec:adhesion}

The multidimensional Burgers equation has important applications in
cosmology where it is closely linked to what is usually referred to as
the Zel'dovich approximation \cite{z70}. In the limit of vanishing
viscosity $\nu\to0$ the Burgers equation is known as the
\emph{adhesion model} \cite{gs84}.  Right after the decoupling between
baryons and photons, the primitive Universe is a rarefied medium
without pressure composed mainly of non-collisional dust interacting
through Newtonian gravity. The initial density of this \emph{dark
matter} fluctuates around a mean value $\bar{\rho}$.  These
fluctuations are responsible for the formation of the large-scale
structures in which both the dark non-baryonic matter and the luminous
baryonic matter concentrate.  A hydrodynamical formulation of the
cosmological problem leads to a description where matter evolves with
a velocity $\bv$, solution of the Euler--Poisson equation (see, e.g.,
\cite{p93}, for further details).

\begin{figure}[ht]
  \centerline{\subfigure[\label{f:nbody}]{
    \includegraphics[width=0.3\textwidth]{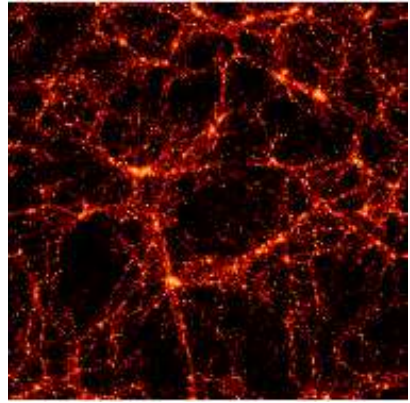}}}
    \centerline{\subfigure[\label{f:figkofman}]{
    \includegraphics[width=0.3\textwidth]{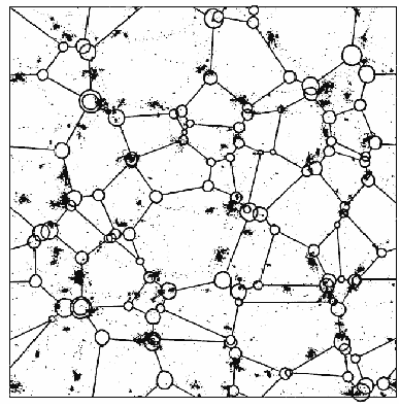}}}
  \caption{(a) Projection of the matter distribution in a slice
    obtained from an N-body simulation by the Virgo
    consortium~\cite{jetal98}. (b) Composite picture showing the
    superposition of the results of an N-body simulation with the
    skeleton of the results obtained from the adhesion model
    (from~\cite{kpsm92}).}
\end{figure}
In the linear theory of the gravitational instability, that is for
infinitesimally small initial fluctuations of the density field, an
instability is obtained with potential dominant modes (i.e.\ $\bv =
-\nabla\Psi$) and, in the suitable coordinates, the gravitational
interactions can be neglected.  In 1970, Zel'dovich proposed to extend
these two properties to the nonlinear r\'{e}gimes where density
fluctuations become important.  In this approximation, he also
postulates that the acceleration is a Lagrangian invariant, leading to
the formation of caustics. N-body simulations however show that the
large-scale structures of the Universe are much simpler than caustics:
they resemble sort of thin layers in which the particles tend to be
trapped (see figure~\ref{f:nbody}).

It was shown by Gurbatov and Saichev \cite{gs84} that these structures
are very well approximated by those obtained when constraining the
particles not to cross each other but to stick together.  Even if this
mechanism is not collisional but rather gravitational (probably due to
instabilities at small spatial scales), its effect can be modeled by a
small viscous diffusive term in the Euler--Poisson equation and thus
amounts to considering the Burgers equation in the limit of vanishing
viscosity.

\subsection{A benchmark for hydrodynamical turbulence}

As a nonlinear conservation law, and since its solution can be easily
known explicitly, the one-dimensional Burgers equation frequently
serves as a testing ground for numerical schemes, and especially for
those dedicated to compressible hydrodynamics. For instance, it is a
central example for the validation of finite-volumes schemes.

The Burgers equation was also used for testing statistical theories of
turbulence. For instance, field theoretical methods have frequently
been applied to turbulence (see \cite{o77,rs78}). These approaches had
very little impact until recently when they led to significant
advances in the understanding of intermittency in passive scalar
advection (see, e.g., \cite{fgv01} for a review).  In the past such
attempts were mostly based on a formal expansion of the nonlinearity
using, for instance, Feynman graphs. Since the Burgers equation has
the same type of quadratic nonlinearity as the Navier--Stokes
equation, such methods are applicable in both instances.  From this
point of view, it is important to know answers for Burgers turbulence
to questions that are generally asked for Navier--Stokes turbulence.
For instance, Burgers turbulence with a random forcing is the
counterpart of the hydrodynamical turbulence model where a steady
state is maintained by an external forcing.  The Burgers equation has
frequently been used as a model where the dissipation of kinetic
energy remains finite in the limit of vanishing viscosity (dissipative
anomaly). This allows singling out artifacts arising from manipulation
that ignore shock waves (see, for instance, \cite{ff83,eve99b}).

Beyond statistical theory, Burgers turbulence gives a simple
hydrodynamical training ground for developing mathematical tools to
study not only Navier--Stokes turbulence but also various
hydrodynamical or Lagrangian problems.  The forced Burgers equation
has recently been at the center of studies that allowed unifying
different branches of mathematics.  Mainly used in the past as a
simple illustration of the notion of entropy (or viscosity) solution
for conservation laws \cite{l57,o57,l82}, the Burgers equation was
related in the eighties to the theory of Hamiltonian systems developed
by Kolmogorov \cite{k57}, Arnold \cite{a63} and Moser \cite{m62}
(KAM), through the introduction of the weak KAM theory
\cite{e89,f97,fbook}. More recently, the study of the solutions to the
Burgers equation with a random forcing was at the center of a
``random'' Aubry--Mather theory related to random Lagrangian systems
\cite{ekms00,ik03}.  A particular emphasis on these aspects of Burgers
turbulence is given throughout the present review. For the application
of the Burgers equation to the propagation of random nonlinear waves
in nondispersive media, we refer the reader to the book written by
Gurbatov, Malakhov, and Saichev~\cite{gms91}. For a complete state of
the art on most mathematical apsects of Burgers turbulence, we refer
the reader to the lecture notes by Woyczy\'{n}ski~\cite{w98}.

%%%%%%%%%%%%%%%%%%%%%%%%%%%%%%%%%%%%%%%%%%%%%%
\section{Basic tools}
\label{s:basictools}

In this section we introduce various analytical, geometrical and
numerical tools that are useful for constructing solutions to the
Burgers equation, with and without forcing, in the limit of vanishing
viscosity.  All these tools are derived from a variational principle
that allows writing in an implicit way the solution at any time.  This
variational principle leads to a straightforward classification of the
various singularities that are generically present in the solution to
the Burgers equation.
 
\subsection{Inviscid limit and variational principle}
\label{subs:varprinc}

We consider here the multidimensional viscous Burgers equation with
forcing
\begin{equation}
  \partial_t \bv + (\bv \cdot \nabla) \bv = \nu \nabla^2 \bv - \nabla
  F(\bx,t),
  \label{eq:burgddf}
\end{equation}
where $\bx$ lives on a prescribed configuration space $\Omega$ of
dimension $d$.  For a potential initial condition, $\bv(\bx,t_0) =
-\nabla\Psi_0(\bx)$, the velocity field remains potential by
construction at any later time, $\bv = -\nabla\Psi$, where the
potential $\Psi$ satisfies the equation
\begin{equation}
  \partial_t \Psi - \frac{1}{2} \left| \nabla\Psi \right|^2 = \nu
  \nabla^2 \Psi + F.
  \label{eq:visckpz}
\end{equation}
Note that if one sets abruptly $\nu=0$ in (\ref{eq:visckpz}), then
$-\Psi$ solves the Hamilton--Jacobi equation associated to the
Hamiltonian $\mathcal{H}(\vec{q}, \vec{p}) = |\vec{p}|^2 +
F(\vec{q},t)$. In the unforced case, $-\Psi$ is a solution of the
Hamilton--Jacobi equation associated to the dynamics of free
particles.  The Hopf--Cole transformation \cite{h50,c51} uses a change
of unknown $\Psi(\vec{x},t) = 2\nu\,\ln \Theta(\bx,t)$. The new
unknown scalar field $\Theta$ is solution of the (imaginary-time)
Schr\"odinger equation
\begin{equation}
  \partial_t \Theta = \nu \nabla^2 \Theta + \frac{1}{2\nu}\,F\,\Theta,
  \label{eq:schrodinger}
\end{equation}
with the initial condition $\Theta(\bx,t_0) =
\exp(\Psi_0(\bx)/(2\nu))$. The solution can be expressed through the
Feynman-Kac formula
\begin{eqnarray}
  && \Theta(\bx,t) \!=\! \left\langle\! \exp\!\left[\!\frac{1}{2\nu}
  \Psi_0(\vec{W}_{t_0}) \!-\! \frac{1}{2\nu}\! \int_{t_0}^t\!\!\!\!
  F(\vec{W}_{s},s)\,\mathrm{d}s \right] \!\right\rangle\!,
  \label{eq:feynman-kac}
\end{eqnarray}
where the brackets $\langle\cdot\rangle$ denote the ensemble average
with respect to the realizations of the $d$-dimensional Brownian
motion $\vec{W}_s$ with variance $2\nu$ defined on the configuration
space $\Omega$ and which starts at $\vec{x}$ at time $t$. The limit
$\nu\to0$ is obtained by a classical saddle-point argument. The main
contribution will come from the trajectories $\vec{W}$ minimizing the
argument of the exponential; the velocity potential can then be
expressed as a solution of the \emph{variational principle}
\begin{equation}
  \Psi(\bx,t) = -\inf_{\vec{\gamma}(\cdot)}\left[
  \mathcal{A}(\vec{\gamma}, t_0, t) -\Psi_0(\vec{\gamma}(t_0)) \right],
  \label{eq:varprinc}
\end{equation}
where the infimum is taken over all trajectories $\vec{\gamma}$ that
are absolutely continuous (e.g.\ piece-wise differentiable) with
respect to the time variable and that satisfy $\vec{\gamma}(t) =
\bx$. The action $\mathcal{A}(\vec{\gamma},t_0,t)$ associated to the
trajectory $\vec{\gamma}$ is defined by
\begin{equation}
  \mathcal{A}(\vec{\gamma},t_0,t) = \int_{t_0}^t \left[ \frac{1}{2}
  |\dot{\vec{\gamma}}(s)|^2 -
  F(\vec{\gamma}(s),s)\right]\,\mathrm{d}s,
  \label{eq:defaction}
\end{equation}
where the dot stands for time derivative. The kinetic energy term
$|\dot{\vec{\gamma}}|^2/2$ comes from the propagator of the
$d$-dimensional Brownian motion $\vec{W}$.  This variational
formulation of the solution to the Burgers equation was obtained first
by Hopf \cite{h50}, Lax \cite{l57} and Oleinik \cite{o57} for scalar
conservation laws.  Its generalization to multidimensional
Hamilton--Jacobi equations was done by Kruzhkov~\cite{k75} (see also
\cite{l82}). In the case of a random forcing potential $F$, it was
shown by E, Khanin, Mazel and Sinai \cite{ekms00} that this
formulation is still valid after replacing the action by a stochastic
integral.  It is also important to notice that the variational
formulation (\ref{eq:varprinc}) in the limit of vanishing viscosity is
valid irrespective of the configuration space $\Omega$ on which the
solution is defined.

The \emph{minimizing trajectories} $\vec{\gamma}(\cdot)$ necessarily
satisfy at times $s<t$ the Newton (or Euler--Lagrange) equation
\begin{equation}
  \ddot{\vec{\gamma}} = -\nabla F(\vec{\gamma}(s), s),
  \label{eq:eullag}
\end{equation}
with the boundary conditions (at the final time $t$)
\begin{equation}
  \vec{\gamma}(t) = \bx \quad\mbox{and}\quad \dot{\vec{\gamma}}(t) =
  \bv(\bx,t).
  \label{eq:eullaginit}
\end{equation}
Note that these equations are only valid backward in time. Extending
them to times larger than $t$ requires knowing that the Lagrangian
particle will neither cross the trajectory of another particle, nor be
absorbed by a mature shock.  This requires global knowledge of the
solution that satisfies the variational principle (\ref{eq:varprinc}).

When the forcing term is absent from (\ref{eq:burgddf}), it is easily
checked that the variational principle reduces to
\begin{equation}
  \Psi(\bx,t) = \max_{\bx_0} \left( \Psi_0(\bx_0) -
  \frac{|\bx-\bx_0|^2}{2t} \right)\, ,
  \label{eq:varprincdecay}
\end{equation}
where the maximum is taken over all initial positions $\bx_0$ in the
configuration space $\Omega$. The Euler--Lagrange equation takes then
the particularly simple form
\begin{equation}
  \ddot{\vec{\gamma}} = 0,\quad\mbox{i.e.} \quad\bx = \bx_0 + t
  \,\bv_0(\bx_0),
  \label{eq:eullagdecay}
\end{equation}
which simply means that the initial velocity is conserved along
characteristics.

Typically there exist Eulerian locations $\bx$ where the minimum in
(\ref{eq:varprinc}) -- or the maximum in (\ref{eq:varprincdecay}) in
the unforced case -- is reached for several different trajectories
$\vec{\gamma}$. Such locations correspond to singularities in the
solution to the Burgers equation. After their appearance, the velocity
potential $\Psi$ contains angular points corresponding to
discontinuities of the velocity field $\bv = -\nabla \Psi$.

\subsection{Variational principle for the viscous case}
\label{subs:varprincvisc}

The derivation of the variational principle (\ref{eq:varprinc}) makes
use of the Hopf--Cole transformation and of the Feynman--Kac formula.
There is in fact another approach which also yields a variational
formulation of the solution to the viscous Hamilton--Jacobi equation
(\ref{eq:visckpz}).  Indeed it turns out that the solution to
(\ref{eq:visckpz}) can be obtained in the following way. Consider
solutions to the stochastic differential equation
\begin{equation}
  \mathrm{d} \vec{\gamma}_{\vec{u}} = \vec{u}(\vec{\gamma}_{\vec{u}},
  s)\, \mathrm{d}s + \sqrt{2\nu}\, \mathrm{d}\vec{W}_s\,,
\end{equation}
where $\vec{u}$ is a stochastic control, that is an arbitrary
time-dependent velocity field which depends (progressively measurably)
on the noise $\vec{W}$. Limiting ourselves to solutions satisfying the final
condition $\vec{\gamma}_{\vec{u}}(t) = \vec{x}$, we can write
\begin{equation}
\Psi(\bx,t) = -\inf_{\vec{u}} \la
\mathcal{A}_{\vec{u}}(\vec{\gamma}_{\vec{u}}, t_0, t)
-\Psi_0(\vec{\gamma}_{\vec{u}}(t_0)) \ra,
  \label{eq:varprincvisc}
\end{equation}
where the brackets $\la \cdot\ra$ now denote average with respect to
$\vec{W}_s$ and the action is given by
\begin{equation}
  \mathcal{A}_{\vec{u}}(\vec{\gamma}_{\vec{u}},t_0,t) = \int_{t_0}^t
  \left[ \frac{1}{2} |\vec{u}(s)|^2 -
  F(\vec{\gamma}_{\vec{u}}(s),s)\right]\,\mathrm{d}s.
  \label{eq:defactionvisc}
\end{equation}
It is obvious that this variational principle gives
(\ref{eq:varprinc}) in the inviscid limit $\nu\to0$. Note that this
approach has the advantage to be applicable not only to Burgers
dynamics but to any convex Lagrangian (see~\cite{fs93,gikp05}).

\subsection{Singularities of Burgers turbulence}
\label{subs:singularities}

The singularities appearing in the course of time play an essential
role in understanding various aspects of the statistical properties in
the inviscid limit. The shocks -- discontinuities of the velocity
field -- and other singularities, such as preshocks, generally not
associated to discontinuities, are often responsible for non-trivial
universal behaviors. In order to understand the contribution of each
kind of singularities, it is first important to know in a detailed
manner their genericity and their type.

As we have seen in the previous section, the potential solutions to
the multidimensional Burgers equation can be expressed in the inviscid
limit in terms of the variational principle (\ref{eq:varprinc}) (that
reduces to (\ref{eq:varprincdecay}) in the unforced case). There
typically exist Eulerian locations $\bx$ where the minimum is either
degenerate or attained for several trajectories. A co-dimension can be
associated to such points by counting the number of relations that are
necessary to determine them. The singular locations of co-dimension
$c$ form manifolds of the Eulerian space-time with dimension $(d-c)$.
The singularities with the lower co-dimension are the \emph{shocks}
corresponding to the Eulerian positions where two different
trajectories minimize (\ref{eq:varprinc}); they form Eulerian
manifolds of dimension $(d-1)$: in one dimension the shocks are
isolated points, in two dimensions they are lines, in three dimensions
surfaces, etc. There also exist Eulerian manifolds with three
different minimizing trajectories. In one dimension, they are isolated
space-time events corresponding to the merger of two shocks. In two
dimensions, they are \emph{triple points} where three shock lines
meet.  In three dimensions they are filaments corresponding to the
intersection of three shock surfaces. There also exist Eulerian
locations where the minimum in (\ref{eq:varprinc}) is reached for four
different trajectories, etc.

\begin{figure}[ht]
  \centerline{\subfigure[\label{f:class2d}]{
      \includegraphics[width=0.25\textwidth]{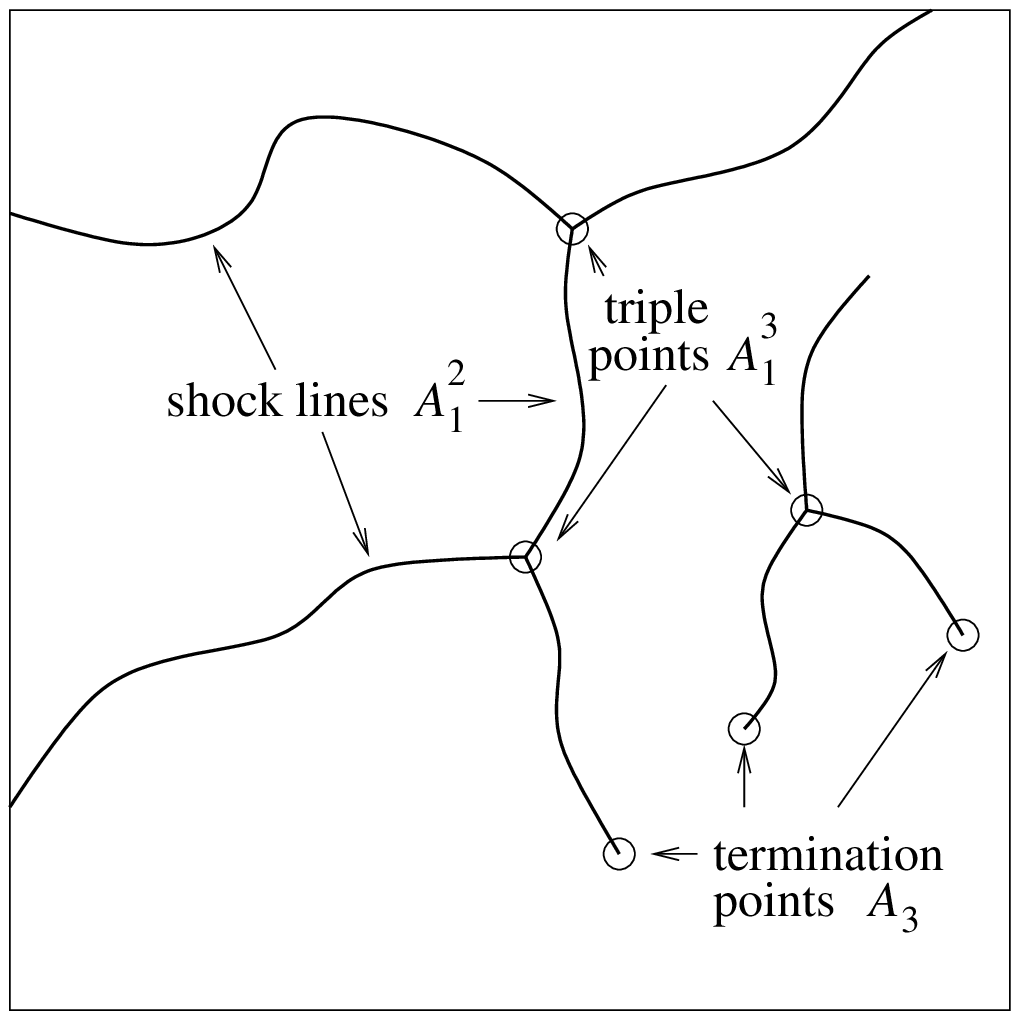}}}
  \centerline{\subfigure[\label{f:class3d}]{
      \includegraphics[width=0.3\textwidth]{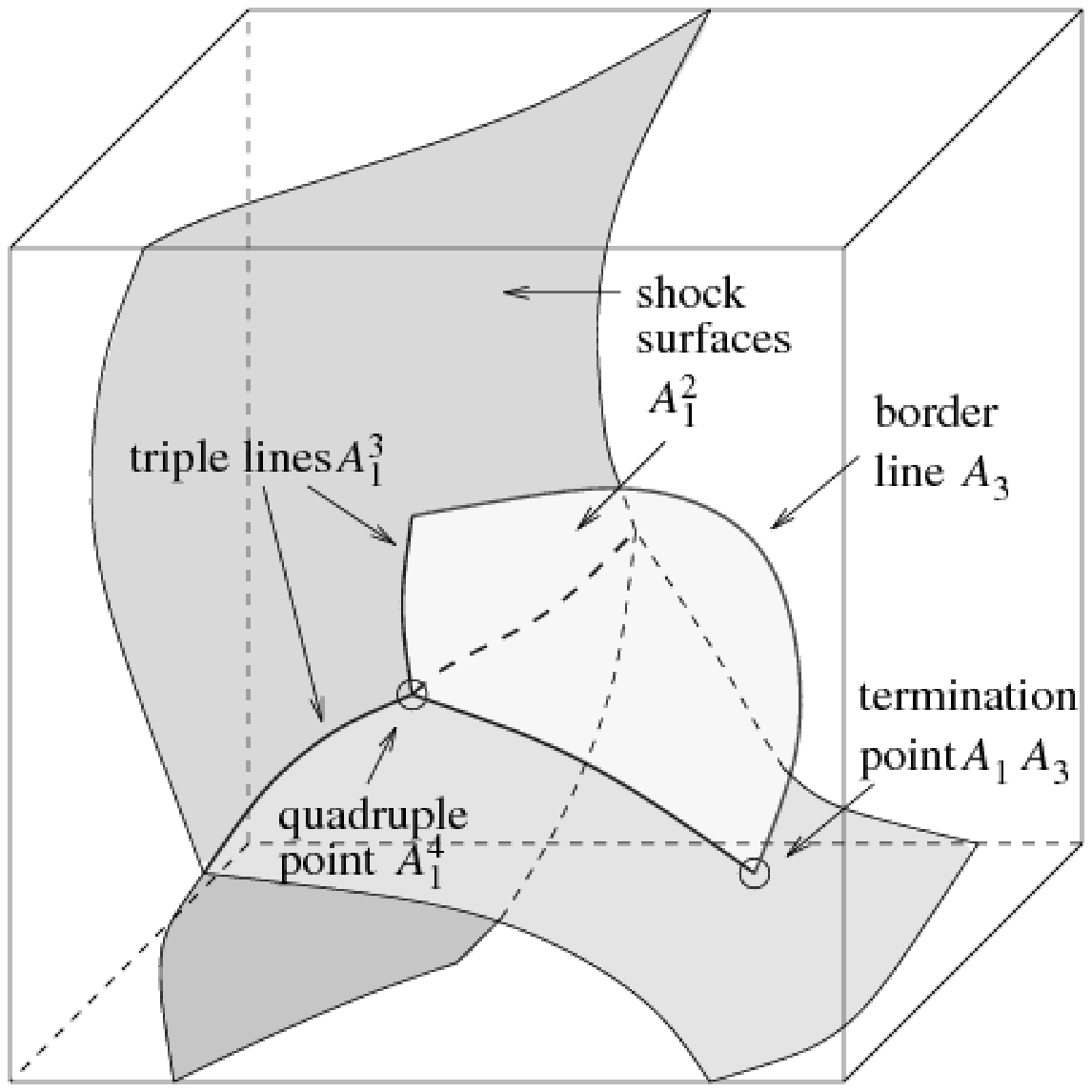}}}
  \caption{\label{f:class} Typical aspect of the singularities present
  at a fixed time in the solution for (a) $d=2$ and (b) $d=3$.}
\end{figure}
The generic form of such singularities and their typical metamorphoses
occurring in the course of time were studied in details and classified
for $d=2$ and $d=3$ by Arnold, Baryshnikov and Bogaevsky in the
Appendix of \cite{gs84} and in a more detailed paper by Bogaevsky
\cite{boga-physicad}.  This classification is based on two criteria:
(i) the number of trajectories minimizing (\ref{eq:varprinc}) and (ii)
the multiplicity of each of these minima. The shocks corresponding to
locations with two distinct minimizers are hence denoted by
$A_1^2$. At a fixed time, the $A_1^2$ singularities are discrete
points in one dimension. In two dimensions (see
figure~\ref{f:class2d}) they form curve segments with extremities that
can be either triple points $A_1^3$ or isolated termination points of
the type $A_3$ corresponding to a degenerate minimum.  In three
dimensions (see figure~\ref{f:class3d}) the singular manifold is
formed by shock surfaces of $A_1^2$ points. The boundaries of these
surfaces are either made of degenerate $A_3$ points or of triple
lines made of $A_1^3$ points.  The triple lines intersect at isolated
$A_1^4$ points or intersect shock boundaries at particular
singularities called $A_1 A_3$ where the minimum is attained in two
points, one of which is degenerate.

It is important to remark here that degenerate singularities (of the
type $A_3$ or of higher orders $A_5$, $A_7$, etc.) introduce in the
solution points where the velocity gradients becomes arbitrarily
large. This is not the case of the $A_1^n$ singularities which
correspond to discontinuities of the velocity but are associated to
bounded values of its gradients. As we will see in
sections~\ref{s:mass} and \ref{s:1Dstatistics}, these degenerate
singularities are responsible for an algebraic behavior of the
probability density function of velocity gradients, velocity
increments and of the mass density.

\begin{figure}[ht]
  \centerline{\subfigure[\label{f:comp1d}]{
      \includegraphics[width=0.25\textwidth]{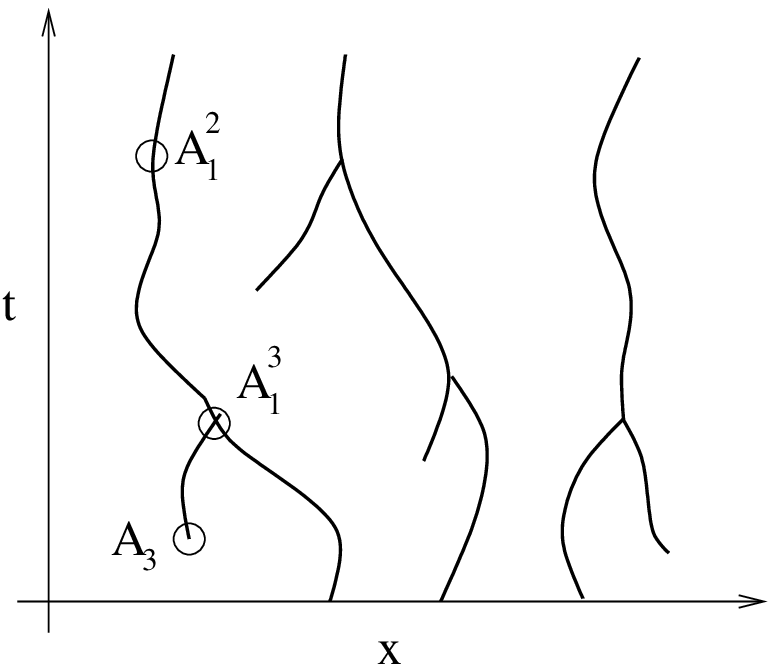}}}
    \centerline{\subfigure[\label{f:comp2d}]{
      \includegraphics[width=0.25\textwidth]{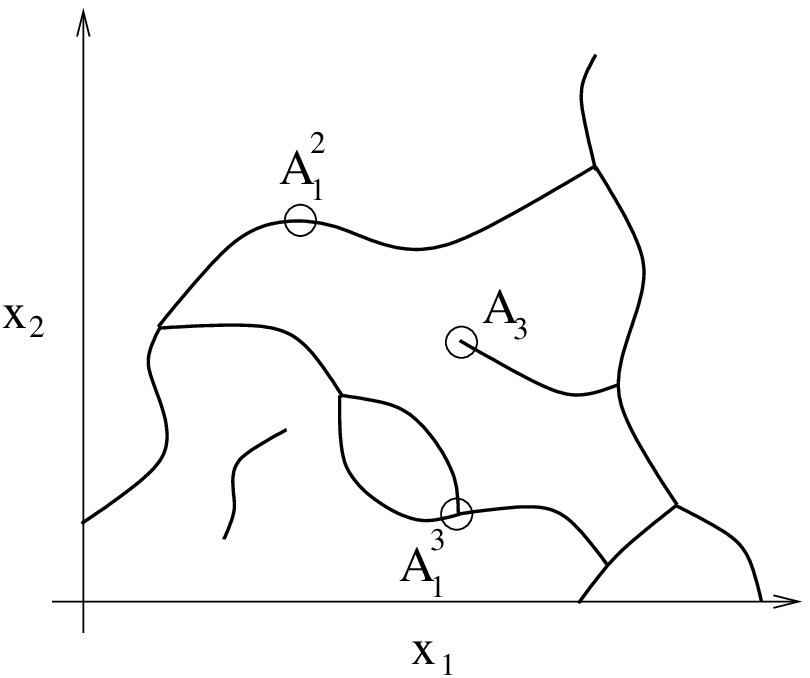}}}
  \caption{\label{f:comp} Illustration of the similarities between the
  singular manifold in space time for $d=1$ and at fixed time for
  $d=2$ (b).  The two manifolds contain the same type of singularities
  with the same co-dimensions. The restrictions on the possible
  metamorphoses in dimension $d=1$ are the following: a point of the
  type $A_3$ can only exist at the bottom extremity of a shock
  trajectory; the $A_1^3$ points necessarily correspond to the merger
  of two shocks; shock trajectories cannot have a horizontal tangent.}
\end{figure}
The singularities with co-dimensions $(d+1)$ generically appear in the
solution at isolated times.  They correspond to instantaneous changes
in the topological structure of the singular manifold, called
\emph{metamorphoses} and can be also classified (see
\cite{boga-physicad}). In one dimension, there are two generic
metamorphoses: shock formations (the \emph{preshocks}) corresponding
to a specific space-time location where the minimum is degenerate
($A_3$ singularities) and shock mergers associated to space-time
positions where the minimum is attained for three different
trajectories ($A_1^3$ singularities).  We see that some of the
singularities generically present in two dimensions appear at isolated
times in three dimensions.  Actually, all the singularities
generically present in dimension $(d+1)$ appear in dimension $d$ on a
discrete set of space time, that is at isolated positions and instants
of time.  However, it has been shown in \cite{boga-physicad} that the
irreversible dynamics of the Burgers equation restricts the set of
possible metamorphoses. The admissible metamorphoses are characterized
by the following property: after the bifurcation, the singular
manifold must remain locally contractible (homotopic to a point in the
neighborhood of the Eulerian location of the metamorphosis).  This
topological restriction is illustrated for the one-dimensional case in
figure~\ref{f:comp}. Note that this constraint actually holds for all
solutions to the Hamilton--Jacobi equation in the limit of vanishing
viscosity, as long as the Hamiltonian is a convex function.

In order to determine precisely how all these singularities contribute
to the statistical properties of the solution, it is important to know
the local structure of the velocity (or potential) field in their
vicinity. Various \emph{normal forms} can be obtained from the
multiplicity of the minimum in the variational formulation of the
solution (\ref{eq:varprinc}).  In the case without forcing, they can
be obtained from a Taylor expansion of the initial velocity
potential. This will be used in next section to determine the tail of
the probability distribution of a mass density field advected by a
velocity solution to the Burgers equation.

\subsection{Remarks on numerical methods}
\label{subs:numerics}

All the traditional methods used to solve equations of fluid dynamics,
or more generally any partial differential equations, can be used to
obtain the solutions to the Burgers equation. However, as we have seen
above, the solution typically has singularities (discontinuities of
the velocity) in the limit of vanishing viscosity.  Hence methods
which rely on the smoothness of the solution require a non-vanishing
viscosity, which is introduced either in an explicit way to ensure
stability (as, e.g., for pseudo-spectral methods) or in an implicit
way through the discretization procedure (as for finite-differences
methods). In both cases the value of the viscosity is determined from
the mesh size and, even in one dimension, their uses might be very
disadvantageous. We will now demonstrate various numerical methods
that allow approximating the solutions to the Burgers equation
directly in the limit of vanishing viscosity $\nu\to0$.

\subsubsection{Finite volumes}

The one-dimensional Burgers equation with no forcing is a scalar
conservation law. Its entropic solutions (or viscosity solutions) can
thus be approximated numerically by finite-volume methods. Instead of
constructing a discrete approximation of the solution on a grid, such
methods consist in considering an approximation of its mean value on a
discrete partitioning of the system into finite volumes.  One then
needs to evaluate for each of these volumes the fluxes exchanged with
each of its neighbors.  Various approximations of these fluxes were
introduced by Godunov, Roe, and Lax and Wendroff (see, e.g.,
\cite{dl85}, Vol.~3, for a review). These methods require to dicretize
both space and time. The time step being then related to the spatial
mesh size by a Courant--Friedrichs--Lewy type condition.  Thus to
integrate the equation during times comparable to one eddy turnover
time, they require a computational time $O(N^2)$ where $N$ is the
resolution.  As we now show there actually exist numerical schemes
that allow constructing the solution to the decaying Burgers equation
for arbitrary times without any need to compute the solution at
intermediate times.

\subsubsection{Fast Legendre transform}
\label{sssec:flt}

As we have seen in section~\ref{subs:varprinc}, the solution to the
unforced Burgers equation is given explicitly by the variational
principle (\ref{eq:varprincdecay}). A method based on the idea of
using this formulation together with a monotonicity property of the
Lagrangian map $\vec{x}_0\to\vec{x} = \vec{X}(\vec{x}_0,t)$ was given
in \cite{nv94}. It is called the \emph{fast Legendre transform} whose
principles were already sketched in \cite{b89}. Both Eulerian and
Lagrangian positions are discretized on regular grids. Then, for a
fixed Eulerian location $\vec{x}^{(i)}$ on the grid, one has to find
the corresponding Lagrangian coordinate $\vec{x}_0^{(j)}$ maximizing
(\ref{eq:varprincdecay}). A naive implementation would require
$O(N_{\rm E}^d\,N_{\rm L}^d)$ operations if the Eulerian and the
Lagrangian grids contain $N_{\rm E}^d$ and $N_{\rm L}^d$ points
respectively. Actually the number of operations can be reduced to
$O((N_{\rm E}\,\ln N_{\rm L})^d)$ by using the monotonicity of the
Lagrangian map, that is the fact that for any pair of Lagrangian
positions $\vec{x}_0^{(1)}$ and $\vec{x}_0^{(2)}$, one has at any time
$[\vec{X}(\vec{x}_0^{(1)},t)- \vec{X}(\vec{x}_0^{(2)},t)]
\cdot(\vec{x}_0^{(1)}-\vec{x}_0^{(2)})\ge0$.  In the case of
orthogonal grids, this property allows performing the maximization by
exploring along a binary tree the various possibilities; thus the
number of operations is reduced to $\ln N_{\rm L}$ for each of the
$N_{\rm E}$ positions on the Eulerian grid. Such algorithms give
access to the solution not only directly in the limit of vanishing
viscosity but also by jumping directly from the initial time to an
arbitrary time.

This method can also be used for the forced Burgers equation,
approximating the forcing by a sum of impulses at discrete times and
letting the solution decay between two such kicks.  This gives an
efficient algorithm for the forced Burgers equation directly
applicable in the limit of vanishing viscosity.

\subsubsection{Particle tracking methods}

In one dimension, Lagrangian methods can be implemented in a
straightforward manner after noticing that particles cannot cross each
other and that it is advisable to track not only fluid particles but
also shocks (see, e.g., \cite{b01}). Lagrangian methods can in
principle be used to solve the Burgers equation in any dimension.
However the shock dynamics is meaningful only for potential solutions.
Outside the potential framework almost nothing is known about the
construction of the solution beyond the first crossing of
trajectories.  In the potential case, a particle method can be
formulated by choosing to represent the solution in the
position-potential $(\vec{x},\Psi)$ space instead of the
position-velocity $(\vec{x},\vec{v})$ space. An idea in two
dimensions, which was not yet implemented, consists in considering a
meshing of the hyper-surface defined by the velocity potential. If
such a mesh contains only triple points, such singularities are
preserved by the dynamics and can be tracked using the results
discussed below in section~\ref{subs:mass} and by checking at all time
steps in an exhaustive manner at all the metamorphoses encountered by
triple points.

%%%%%%%%%%%%%%%%%%%%%%%%%%%%%%%%%%%%%%%%%%%%%%
\section{Decaying Burgers turbulence}
\label{s:decay}

We focus in this section on the solutions to the $d$-dimensional
unforced potential Burgers equation
\begin{eqnarray}
  &&\partial_t \vec{v} + \vec{v}\!\cdot\!\nabla \vec{v} = \nu \nabla^2
  \vec{v},\ \ \vec{v}(\vec{x},0) \!=\! \vec{v}_0(\vec{x})\! =\!
  -\nabla\Psi_0(\vec{x}).
  \label{eq:eqburgdecay}
\end{eqnarray}
As showed in section~\ref{subs:varprinc}, the solution can be
expressed in the limit of vanishing viscosity $\nu\to 0$ in terms of a
variational principle that relates the velocity potential at time $t$
to its initial value:
\begin{equation}
  \Psi(\bx,t) = \max_{\bx_0} \left( \Psi_0(\bx_0) -
  \frac{|\bx-\bx_0|^2}{2t} \right).
  \label{eq:varprincdecay2}
\end{equation}
The next subsection describes several geometrical constructions of the
solution that are helpful to determine various statistical properties
of the decaying problem~(\ref{eq:eqburgdecay}). This is illustrated in
subsections \ref{subs:kida} and \ref{subs:sinai} which are devoted to
the study of the decay of smooth homogeneous and of Brownian initial
data, respectively.

The study of the solutions to the Burgers equation transporting a
density field is of particular interest in the application of the
Burgers equation in cosmology within the framework of the adhesion
model. This question will be discussed in section \ref{s:mass}.

\subsection{Geometrical constructions of the solution}
\label{subs:geometrical}

\subsubsection{The potential Lagrangian manifold}
\begin{figure}[ht]
  \centerline{\subfigure[\label{f:figpot1d}]{
      \includegraphics[width=0.25\textwidth]{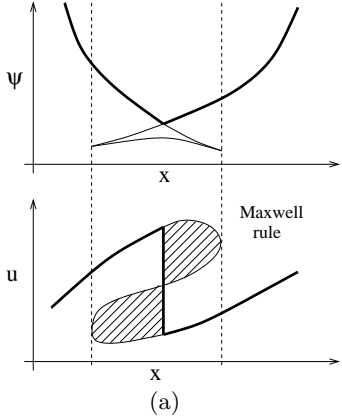}}}
   \centerline{\subfigure[\label{f:figpot2d}]{
      \includegraphics[width=0.45\textwidth]{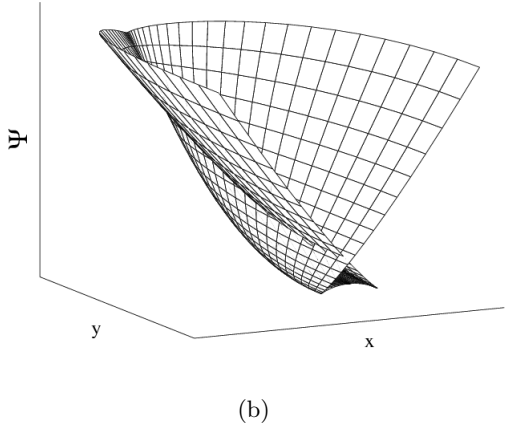}}}
  \caption{(a) Lagrangian manifold for $d=1$ in the $(x,\Psi)$ plane
  (upper) and in the $(x,v)$ plane (lower); the heavy lines correspond
  to the correct Eulerian solutions. (b) Lagrangian manifold in the
  $(\bx,\Psi)$ space for $d=2$.}
\end{figure}

The variational formulation of the solution (\ref{eq:varprincdecay2})
has a simple geometrical interpretation in the position-potential
space $(\bx,\Psi)$ of dimension $d+1$.  Indeed, consider the
$d$-dimensional manifold parameterized by the Lagrangian coordinate
$\bx_0$ and defined by
\begin{equation}
  \left\{ \begin{array}{lll} \bx & = & \bx_0 - t\,\nabla\Psi_0(\bx_0)\\
    \Psi & = & \displaystyle \Psi_0(\bx_0) -\frac{t}{2}
    |\nabla\Psi_0(\bx_0)|^2.
  \end{array} \right. 
  \label{eq:defxpsimanifold}
\end{equation}
The first line corresponds to the position where the gradient of the
argument of the maximum function in (\ref{eq:varprincdecay2}) vanishes
while the second line is just its argument evaluated at the maximum.
For a sufficiently regular initial potential $\Psi_0$ (at least twice
differentiable) and for sufficiently small times, equation
(\ref{eq:defxpsimanifold}) unambiguously defines a single-valued
function $\Psi(\bx,t)$. However, there exists generically a time
$t_\star$ at which the manifold is folding. Figure \ref{f:figpot1d}
(upper) shows in one space dimension the typical shape of the
Lagrangian manifold defined by (\ref{eq:defxpsimanifold}) after the
critical time $t_\star$.  For some Eulerian positions $\bx$, there is
more than one branch and cusps are present at Eulerian locations where
the number of branches change.

The situation is very similar in higher dimensions as illustrated for
$d=2$ in figure~\ref{f:figpot2d}. Clearly from the variational principle
(\ref{eq:varprincdecay}), the correct solution to the inviscid Burgers
equation is obtained by taking the maximum, that is the highest
branch.  The velocity potential is by construction always continuous
but it contains angular points corresponding to discontinuities of the
velocity $\bv = -\nabla\Psi$. Such singularities are located at
Eulerian locations where the maximum in (\ref{eq:varprincdecay}) is
degenerate and attained for different $\bx_0$. As already discussed
in section~\ref{subs:singularities} the different singularities
appearing in the solutions can be classified in any dimension.

Below we describe other geometrical constructions of the solutions to
the decaying Burgers equation in the limit of vanishing viscosity that
are based on the variational principle (\ref{eq:varprincdecay}).

\subsubsection{The velocity Lagrangian manifold}
In one dimension, when the velocity field is always potential, the
method based on the study of the potential manifold in the $(x,\Psi)$
space described above can be straightforwardly extended to the
position-velocity phase space. Consider the Lagrangian manifold
defined by
\begin{equation}
  \left\{ \begin{array}{lll} x & = & x_0 - t\,v_0(x_0)\\ v & = &
  v_0(x_0).
  \end{array} \right. 
  \label{eq:defxumanifold}
\end{equation}
The regular parts of the graph of the solution are necessarily
contained in this manifold.  However, for times larger than $t_\star$,
folding appears and the naive solution would be multi-valued. To
construct the true solution one should find a way to choose among the
different branches.  In one dimension, there is a simple relation
between the potential Lagrangian manifold in the $(x,\Psi)$ plane and
those of the $(x,v)$ plane defined by (\ref{eq:defxumanifold}): the
potential manifold is obtained by taking the ``multi-valued integral''
that can be defined by transforming the spatial integral into an
integral with respect to the arc length.  The maximum representation
(\ref{eq:varprincdecay}) implies that the velocity potential is
continuous. Hence a shock corresponds to an Eulerian position $x$
where two points belonging to different branches define equal areas in
the $(x,v)$ plane.  In the case of a single loop of the manifold, this
is equivalent to applying the \emph{Maxwell rule} to determine the
shock position (see figure~\ref{f:figpot1d} - lower).  This
construction of the solution can become rather involved as soon as the
number of shocks becomes large or that several mergers have taken
place.  For the moment there is no generalization to dimension higher
than one of this Maxwell rule construction of the solution. For such
an extension, one needs to develop a geometrical framework to describe
the Lagrangian manifold in the $(\bx,\bv)$ space. Such approaches
would certainly shed some light on the problem of constructing
non-potential solutions to the Burgers equation in the limit of
vanishing viscosity.

\begin{figure}[t]
  \centerline{\subfigure[\label{f:convexhull1d}]{
      \includegraphics[width=0.3\textwidth]{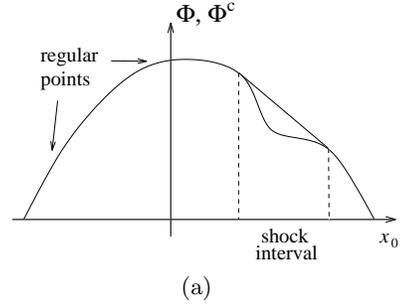}}}
  \centerline{\subfigure[\label{f:convexhull2d}]{
      \includegraphics[width=0.45\textwidth]{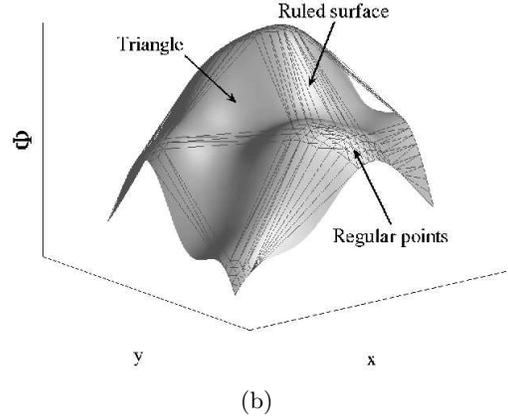}}}
  \caption{Convex hull construction in terms of the Lagrangian
  potential (a) for $d=1$ and (b) for $d=2$.}
\end{figure}
\subsubsection{The convex hull of the Lagrangian potential}
Another geometrical construction of the solution, which is valid in
any dimension makes use of the \emph{Lagrangian potential}
\begin{equation}
  \Phi(\bx_0,t) = t \Psi_0(\bx_0) - \frac{|\bx_0|^2}{2}.
  \label{eq:deflagpot}
\end{equation}
Clearly, the negative gradient of the Lagrangian potential gives the
naive Lagrangian map
\begin{equation}
  \vec{X}(\bx_0,t) = -\nabla_{\bx_0} \Phi(\bx_0,t) = \bx_0+
  t\bv_0(\bx_0),
  \label{eq:defnaivelagmap}
\end{equation}
that is satisfied by Lagrangian trajectories as long as they do not
enter shocks. The maximum formulation of the solution
(\ref{eq:varprincdecay}) can be rewritten as
\begin{equation}
  t\Psi(\bx,t) + \frac{|\bx|^2}{2} = \max_{\bx_0} (\Phi(\bx_0) +
  \bx_0\cdot\bx),
  \label{eq:varprinclagpot}
\end{equation}
which represents the potential as, basically, a Legendre transform of
the Lagrangian potential.  An important property of the Legendre
transform is that the right-hand side.\ of (\ref{eq:varprinclagpot})
is unchanged if the Lagrangian potential $\Phi$ is replaced by its
convex hull, that is the intersection of all the half planes
containing its graph.  In other terms, the convex hull $\Phi^{\rm c}$
of the Lagrangian potential $\Phi$ is defined as $\Phi^{\rm
c}(\bx_0,t) = \inf g(\bx_0)$, where the infimum is taken over all
convex functions $g$ satisfying $g(\cdot)\ge\Phi(\cdot,t)$.  This is
illustrated in one dimension in figure~\ref{f:convexhull1d} which
shows both regular points (Lagrangian points which have not fallen
into a shock) and one shock interval, situated below the segment which
is a part of the convex hull.  In two dimensions, as illustrated in
figure~\ref{f:convexhull2d}, the convex hull is typically formed by
regular points, by ruled surfaces, and by triangles which correspond,
to the regular part of the velocity field, the shock lines, and the
shock nodes, respectively.

Note that in one dimension, there exists an equivalent construction
which is directly based on the Lagrangian map $x_0\mapsto X(x_0,t)$
defined by (\ref{eq:defnaivelagmap}). Working with the convex hull is
equivalent to the Maxwell rule applied to the non-invertible regions of
the Lagrangian map. A shock corresponds to a whole Lagrangian interval
having a single point as an Eulerian image. One then talks about a
Lagrangian \emph{shock interval}.

\subsubsection{The paraboloid construction}
\begin{figure}[ht]
  \centerline{\subfigure[\label{f:parabola1d}]{
      \includegraphics[width=0.45\textwidth]{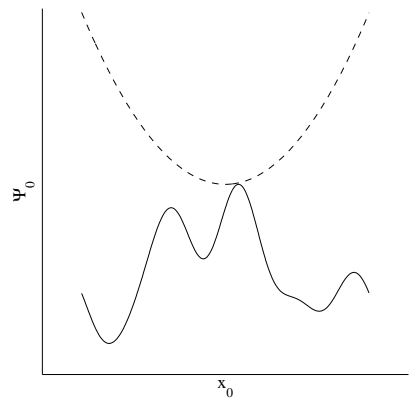}}}
      \centerline{\subfigure[\label{f:parabola2d}]{
      \includegraphics[width=0.45\textwidth]{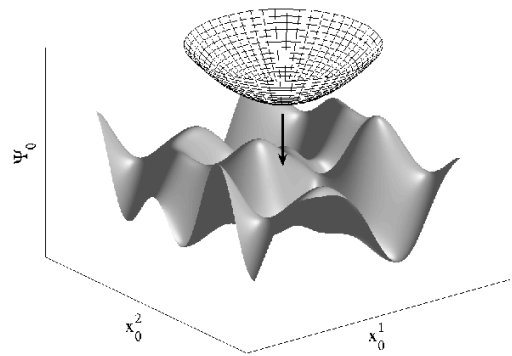}}}
  \caption{\label{f:parabola} Paraboloid construction of solution for
  (a) $d=1$ and (b) $d=2$.}
\end{figure}

Finally, the maximum representation (\ref{eq:varprinclagpot}) leads in
a straightforward way to another geometrical construction of the
solution. As illustrated in figure~\ref{f:parabola} in both one and
two dimensions, a paraboloid with apex at $\bx$ and radius of
curvature proportional to $t$ is moved down in the $(\bx_0,\Psi_0)$
space until it touches the surface defined by the initial velocity
potential $\Psi_0$ at the Lagrangian location associated to $\bx$. The
location $\bx_0$ where the paraboloid touches the graph of the
potential is exactly the pre-image of $\bx$. If it touches
simultaneously at several locations, a shock is located at the
Eulerian position $\bx$. One constructs in this way the inverse
Lagrangian map.

\subsection{Kida's law for energy decay}
\label{subs:kida}

An important issue in turbulence is that of the law of decay at long
times when the viscosity is very small. Before turning to the Burgers
equation it is useful to recall some of the features of decay for the
incompressible Navier--Stokes case.  It is generally believed that
high-Reynolds number turbulence has universal and non-trivial
small-scale properties. In contrast, large scales, important for
practical applications such as transport of heat or pollutants, are
believed to be non-universal. This is however so only for the toy
model of turbulence maintained by prescribed large-scale random
forces.  Very high-Reynolds number turbulence, decaying away from its
production source, and far from boundaries can relax under its
internal nonlinear dynamics to a (self-similarly evolving) state with
universal and non-trivial statistical properties \emph{at all scales}.
Von K\'arm\'an and Howarth \cite{kh38}, investigating the decay for
the case of high-Reynolds number homogeneous isotropic
three-dimensional turbulence, proposed a self-preservation
(self-similarity) ansatz for the spatial correlation function of the
velocity: the functional shape of the correlation function remains
fixed, while the integral scale $L(t)$ grows in time and the mean
kinetic energy $E(t)=V^2(t)$ decays, both following power laws; there
are two exponents which can be related by the condition that the
energy dissipation per unit mass $|\dot E(t)|$ should be proportional
to $V^3/L$.  But \emph{an additional relation} is needed to actually
determine the exponents.  The invariance in time of the energy
spectrum at low wavenumbers, known as the ``permanence of large
eddies'' \cite{f95,l97,gsaft97} can be used to derive the law of
self-similar decay when the initial spectrum $E_0(k)\propto k^n$ at
small wavenumbers $k$ with $n$ below a critical value equal to 3 or 4,
the actual value being disputed because of the ``Gurbatov phenomenon''
(see the end of this section). One then obtains a law of decay $E(t)
\propto t^{-2(n+1)/(3+n)}$. (Kolmogorov \cite{k41decay} proposed a law
of energy decay $V^2(t) \propto t^{-10/7}$, which corresponds to $n=4$
and used in its derivation the so-called ``Loitsyansky invariant'', a
quantity actually not conserved, as shown by Proudman and Reid
\cite{pr54}.) When the initial energy spectrum at low wavenumbers goes
to zero too quickly, the permanence of large eddies cannot be used,
because the energy gets backscattered to low wavenumbers by nonlinear
interactions. For Navier--Stokes turbulence the true law of decay is
then known only within the framework of closure theories (see, e.g.,
\cite{l97}).

\begin{figure}[t]
  \centerline{\includegraphics[width=0.22\textwidth]
    {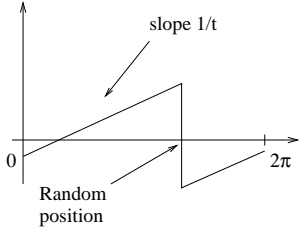}}
  \caption{Snapshot of solution of decaying Burgers turbulence at long
    times when spatial periodicity is assumed.}
  \label{f:declinper}
\end{figure}

For one-dimensional Burgers turbulence, many of the above issues are
completely settled. First, we observe that the problem of decay is
quite simple if spatial periodicity is assumed. Indeed, all the shocks
appearing in the solution will eventually merge into a single shock
per period, as shown in figure~\ref{f:declinper}.  The position of
this shock is random and the two ramps have slope $1/t$, as is easily
shown using the parabola construction of
subsection~\ref{subs:geometrical}. Hence, the law of decay is simply
$E(t) \propto t^{-2}$.
Nontrivial laws of decay are obtained if the Burgers turbulence is
homogeneous in an unbounded domain and has the ``mixing'' property
(which means, roughly, that correlations are vanishing when the
separation goes to infinity). The number of shocks is then typically
infinite but their density per unit length decreases in time because
shocks are constantly merging. The $E(t) \propto t^{-2(n+1)/(3+n)}$
law mentioned above can be derived for Burgers turbulence from the
permanence of large eddies when $n\le 1$ \cite{gsaft97}. For $n=0$,
this $t^{-2/3}$ law was actually derived by Burgers himself
\cite{b74}.

The hardest problem is again when permanence of large eddies does not
determine the outcome, namely for $n>1$. This problem was solved by
Kida \cite{k79} (see also~\cite{ff83,gms91,gsaft97}).

We now give some key ideas regarding the derivation of Kida's law of
energy decay. We assume Gaussian, homogeneous smooth initial
conditions, such that the potential is homogeneous. Note that a
homogeneous function is not, in general, the derivative of another
homogeneous function. Here this is guaranteed by assuming that the
initial spectrum of the kinetic energy is of the form
\begin{equation}
  E_0(k) \propto k^n \mbox{ for } k \to 0 \mbox{ with }n>1\,.
\label{ezeronpg1}
\end{equation}
This condition implies that the mean square initial potential $\int
k^{-2} E_0(k)\,\mathrm{d}k$ has no infrared (small-$k$) divergence
(the absence of an ultraviolet divergence is guaranteed by the assumed
smoothness).

\begin{figure}[t]
  \centerline{\includegraphics[width=0.45\textwidth]{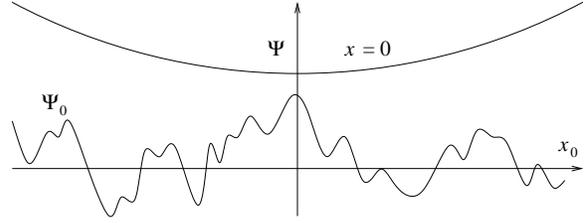}}
  \caption{An initial potential which is everywhere below the parabola
    $x_0^2/(2t) + \Psi$. The probability of such events gives the
    cumulative probability to have a potential at time $t$ less than
    $\Psi$.}
\label{f:parabkida}
\end{figure}

A very useful property of decaying Burgers turbulence, which has no
counterpart for Navier--Stokes turbulence, is the relation
\begin{equation}
  E(t) = \frac{\partial}{\partial t} \left\langle \Psi \right\rangle,
  \label{epsider}
\end{equation}
which follows by taking the mean of the Hamilton--Jacobi equation for
the potential (\ref{eq:visckpz}) in the absence of viscosity and of a
driving force.  Hence, the law of energy decay can be obtained from
the law for the mean potential. The latter can be derived from the
cumulative probability of the potential which, by homogeneity, does
not depend on the position. By (\ref{eq:varprincdecay}), its
expression at $x=0$ is
\begin{eqnarray}
  && \mbox{Prob}\!\left\{\mbox{Pot.}\!<\!\Psi\right\} \!=\!
    \mbox{Prob}\!\left\{\!\forall x_0,\,
    \Psi_0(x_0)\!<\!\frac{x_0^2}{2t} \!+\! \Psi \right\}\!.
\label{cumulpot}
\end{eqnarray}
Expressed in words, we want to find the probability that the initial
potential does not cross the parabola $x_0^2/(2t) + \Psi$ (see
figure~\ref{f:parabkida}).
Since, at large times $t$, the relevant $\Psi$ is going to be large,
the problem becomes that of not crossing a parabola with small
curvature and very high apex.  The crossings, more precisely the
up-crossings, are spatially quite rare and, for large $t$, form a
Poisson process \cite{msw95} for which
\begin{equation}
  \hbox{Prob. no crossing} \simeq \mathrm{e}^{-\left\langle
  N(t)\right\rangle},
  \label{poisson}
\end{equation}
where $\left\langle N(t)\right\rangle$ is the mean number of
up-crossings. By the Rice formula (a consequence of the identity
$\delta(\lambda x) = (1/|\lambda|) \delta(x)$),
\begin{eqnarray}
  && \left\langle N(t)\right\rangle \!=\! \left\langle
  \int_{-\infty}^{+\infty} \!\!\! \mathrm{d}x_0\, \delta\!\left(
  m(x_0) \!-\!  \Psi \right ) \frac{\mathrm{d}m}{\mathrm{d}x_0}
  \mathrm{H}\!\left(\!\frac{\mathrm{d}m}{\mathrm{d}x_0}\!\right)\!
  \right\rangle\!,
  \label{rice}
\end{eqnarray}
where $\mathrm{H}$ is the Heaviside function and
\begin{equation}
  m(x_0)\equiv \Psi_0(x_0) - \frac{x_0^2}{2t}.
  \label{defma}
\end{equation}
Since $\Psi_0(x_0)$ is Gaussian, the right-hand side of (\ref{rice}) can be
easily expressed in terms of integrals over the probability densities
of $\Psi_0(x_0)$ and of $\mathrm{d}\Psi_0(x_0)/\mathrm{d}x_0$ (as a
consequence of homogeneity these variables are uncorrelated and,
hence, independent). The resulting integral can then be expanded by
Laplace's method for large $t$, yielding
\begin{equation}
  \left\langle N(t)\right\rangle \sim t^{1/2} \Psi^{-1/2} e^{-\Psi^2},
  \quad t \to \infty.
  \label{asymptN}
\end{equation}
When this expression is used in (\ref{poisson}) and the result is
differentiated with respect to $\Psi$ to obtain the probability
density function (PDF) of $p(\Psi)$, the latter is found to be
concentrated around $\Psi_\star = (\ln t)^{1/2}$.  It then follows
that, at large times, we obtain Kida's log-corrected $1/t$ law for the
energy decay
\begin{eqnarray}
  && \left\langle \Psi \right\rangle \sim (\ln t)^{1/2}\!, \ E(t)\sim
    \frac{1}{t(\ln t)^{1/2}}\!, \ L(t)\sim\left[\frac{t}{\ln
    t}\right]^{1/4}\!\!\!\!\!\!\!\!\!.
  \label{psienLkida}
\end{eqnarray}
\begin{figure}[ht]
  \centerline{\includegraphics[width=0.4\textwidth]{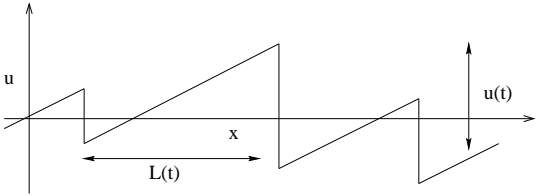}}
  \caption{The Eulerian solution at large times $t$. The ramps have
    slope $1/t$.  In time-independent scales, the figure would be
    stretched horizontally and squeezed vertically by a factor
    proportional to $t$.}
  \label{f:ramps}
\end{figure}
The Eulerian solution, at large times, has the ramp structure shown in
figure~\ref{f:ramps} with shocks of typical strength $V(t)=E^{1/2}(t)$,
separated by a distance $L(t)$.

The fact that Kida's law is valid for any $n>1$, and not just for
$n\ge 2$ as thought originally, gives rise to an interesting
phenomenon now known as the ``Gurbatov effect'': if $1<n<2$ the
large-time evolution of the energy spectrum cannot be globally
self-similar. Indeed, the permanence of large eddies, which is valid
for any $n<2$ dictates that the spectrum should preserve exactly its
initial $C_nk^n$ behavior at small wavenumbers $k$, with a
constant-in-time $C_n$. Global self-similarity would then imply a
$t^{-2(n+1)/(3+n)}$ law for the energy decay, which would contradict
Kida's law. Actually, as shown in~\cite{gsaft97}, there are two
characteristic wavenumbers with different time dependences, the
integral wavenumber $k_L(t) \sim (L(t))^{-1}$ and a switching
wavenumber $k_s(t)$ below which holds the permanence of large
eddies. It was shown that the same phenomenon is present also in the
decay of a passive scalar \cite{ex00}. Whether or not a similar
phenomenon is present in three-dimensional Navier--Stokes
incompressible turbulence, or even in closure models, is a
controversial matter \cite{et00,ol00}.

For decaying Burgers turbulence, if we leave aside the Gurbatov
phenomenon which does not affect energy-carrying scales, the following
may be shown. If we rescale distances by a factor $L(t)$ and velocity
amplitudes by a factor $V(t)=E^{1/2}(t)$ and then let $t\to \infty$,
the spatial (single-time) statistical properties of the whole random
velocity field become time-independent. In other words, there is a
self-similar evolution at large times. Hence, dimensionless ratios
such as the velocity flatness
\begin{equation}
  F(t)\equiv \frac{\left\langle v ^4(t)\right\rangle }{\left[
    \left\langle v ^2(t)\right\rangle \right]^2}
\label{defuflatness}
\end{equation}
have a finite limit as $t\to \infty$.  A similar property holds for
the decay of passive scalars \cite{cefv00}.  We do not know if this
property holds also for Navier--Stokes incompressible turbulence or
if, say, the velocity flatness grows without bounds at large times.

%%%%%%%%%%%%%%%%%%%%%%%%%%%%%%%%%%%%%%%%%%%%%%%%%%%%%%%%%%%%%%%%%%%%%%%
\subsection{Brownian initial velocities}
\label{subs:sinai}

\begin{figure}[ht]
  \centerline{\subfigure[\label{f:snapshot-brownian}]{
      \includegraphics[width=0.4\textwidth]{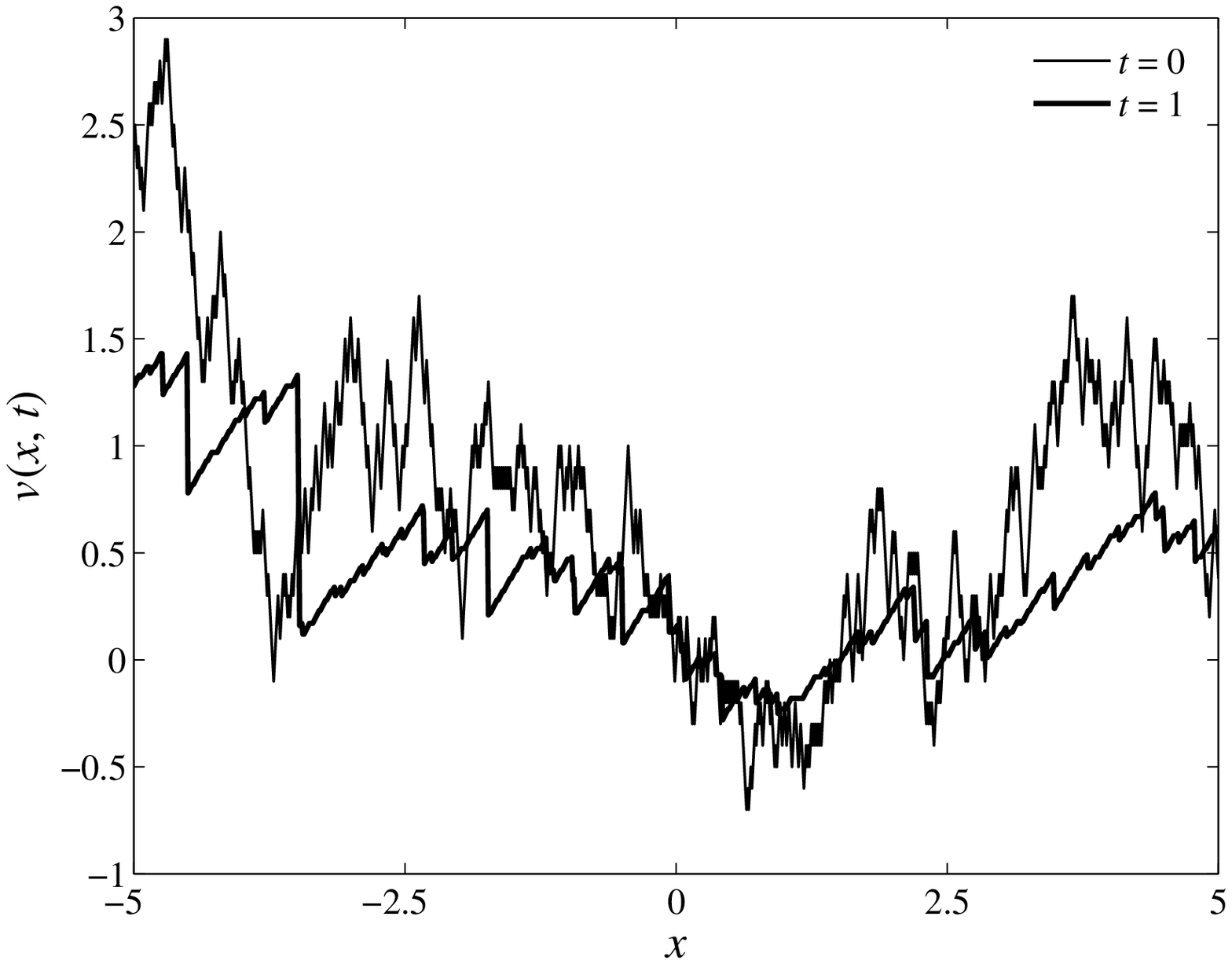}}}
  \vspace{-10pt}\centerline{\subfigure[\label{f:lagpot-brownian}]{
      \includegraphics[width=0.4\textwidth]{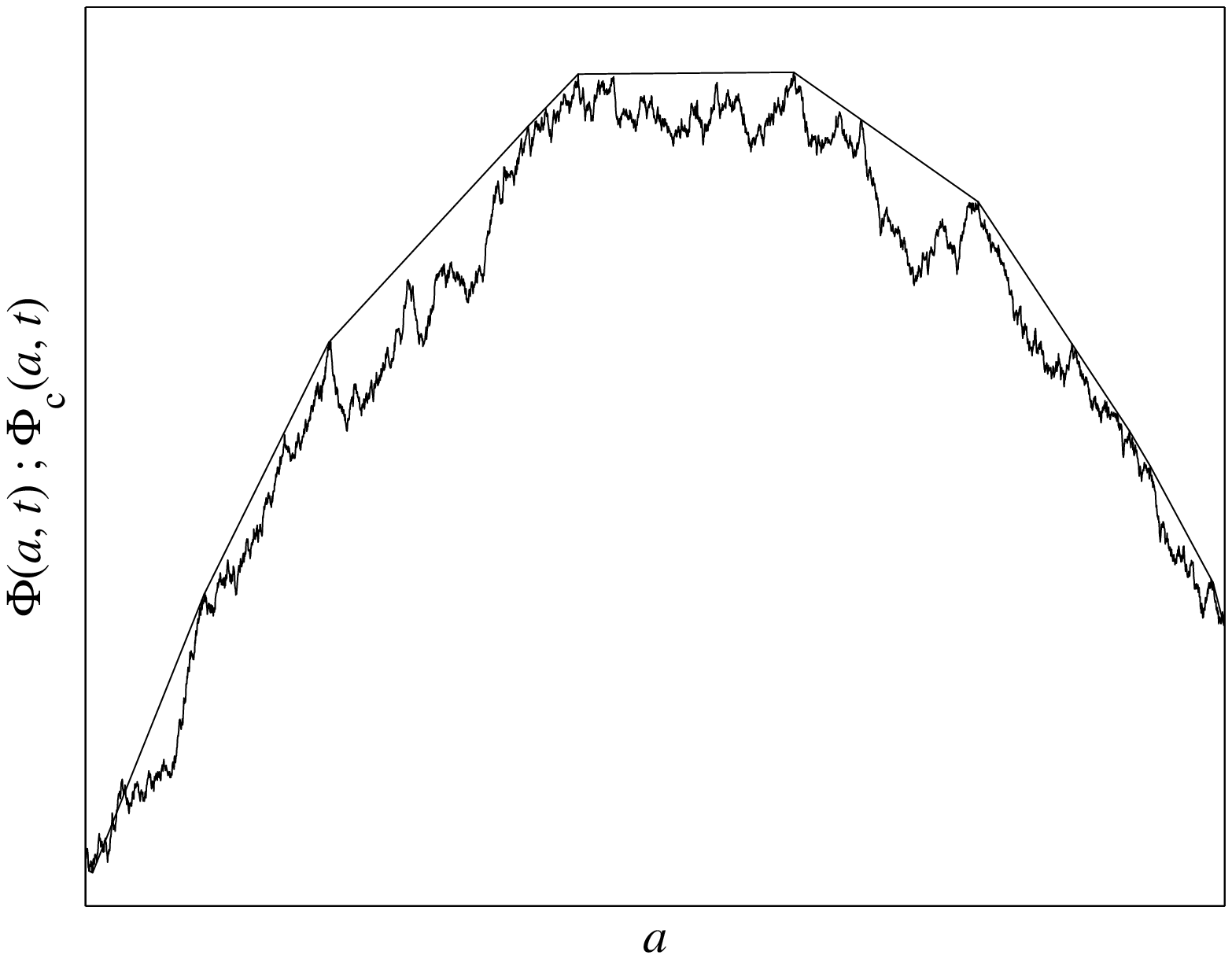}}}
  \caption{\label{f:browniansol} Snapshot of the solution resulting
  from Brownian initial data in one dimension. (a) Velocity profile at
  initial time $t=0$ and at time $t=1$; notice the dense proliferation
  of shocks. (b) Lagrangian potential together with its convex hull.}
\end{figure}
Initial conditions in the Burgers equation that are Gaussian with a
power-law spectrum $\propto k^{-\alpha}$ have been frequently studied
because they belong in cosmology to the class of \emph{scale-free}
initial conditions (see \cite{p93,cl95}). We consider here the
one-dimensional case with Brownian motion as initial velocity,
corresponding to $\alpha=2$.

\begin{figure}[ht]
  \begin{center}
    \begin{minipage}{0.3\textwidth}
	\includegraphics[width=\textwidth]{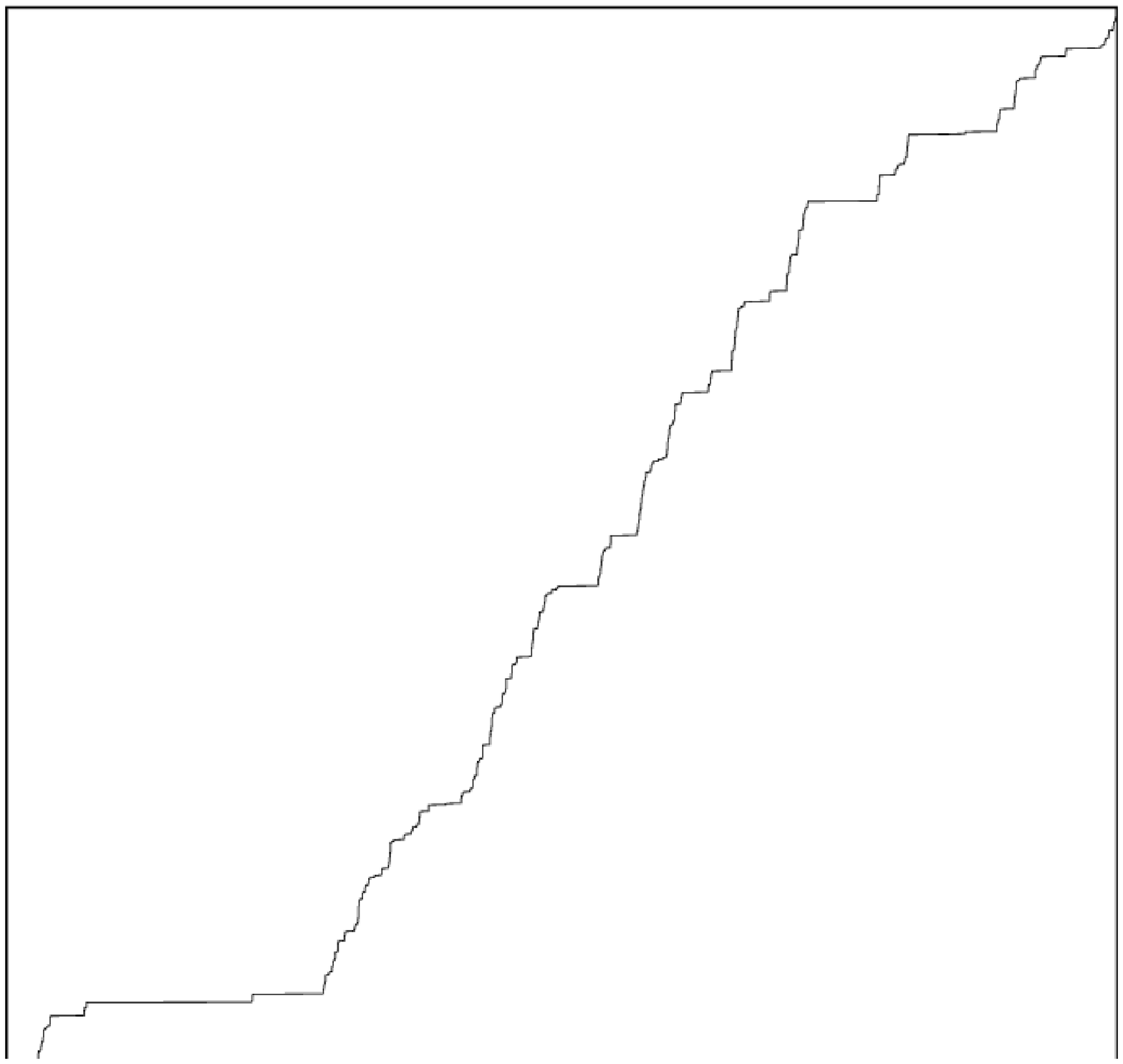}\\[5pt]
	\strut \end{minipage}
  \end{center}
\caption{\label{f:devil-staircases} The Lagrangian map looks like a
  Devil's staircase: it is constant almost everywhere, except on a
  fractal Cantor-like set (from~\cite{vdfn94}).}
\label{f:escaliers}
\end{figure}

Brownian motion is continuous but not differentiable; thus, shocks
appear after arbitrarily short times and are actually dense (see
figure~\ref{f:snapshot-brownian}). Numerically supported conjectures
made in \cite{saf92} have led to a proof by Sinai \cite{s92} of the
following result: in Lagrangian coordinates, the regular points, that
is fluid particles which have not yet fallen into shocks, form a
fractal set of Hausdorff dimension $1/2$. This implies that the
Lagrangian map forms a Devil's staircase of dimension $1/2$ (see
figure~\ref{f:escaliers}). Note that when the initial velocity is
Brownian, the Lagrangian potential has a second space derivative
delta-correlated in space; this can be approximately pictured as a
situation where the Lagrangian potential has very wild oscillations in
curvature. Hence, it is not surprising that very few points of its
graph can belong to its convex hull (see
figure~\ref{f:lagpot-brownian}).

\begin{figure}[ht]
\centerline{\includegraphics[width=0.4\textwidth]{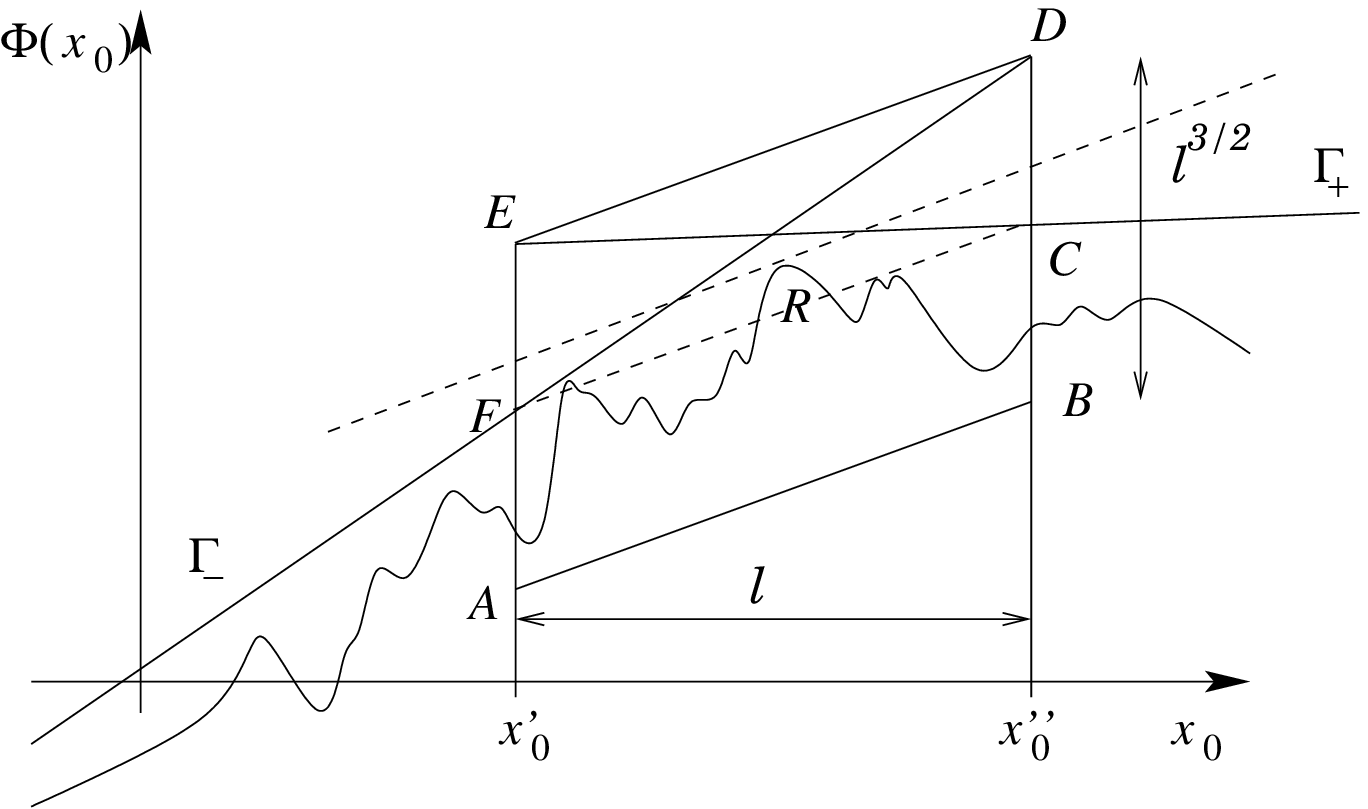}}
\caption{The box construction used to find a regular point $R$ within a
  Lagrangian interval of length $\ell$ (from~\cite{s92,vdfn94}).}
\label{f:leftboxright}
\end{figure}

We will now highlight some aspects of Sinai's proof of this
result. The idea is to use the construction of the solution in terms
of the convex hull of the Lagrangian potential (see
section~\ref{subs:geometrical}), so that regular points are exactly
points where the graph of the Lagrangian potential coincides with its
convex hull.  For this problem, the Hausdorff dimension of the regular
points is also equal to its box-counting dimension, which is easier to
determine. One obtains it by finding the probability that a small
Lagrangian interval of length $\ell$ contains at least one regular
point which belongs simultaneously to the graph of the Lagrangian
potential $\Phi$ and to its convex hull. In other words, one looks for
points, such as $R$, with the property that the graph of $\Phi$ lies
below its tangent at~$R$ (see figure~\ref{f:leftboxright}).  Following
Sinai, this can be equivalently formulated by the box construction
with the following constraints on the graph:\\ {\sl Left}: \ graph of
the potential below the half line $\Gamma_-$,\\ {\sl Right}: \ graph
of the potential below the half line $\Gamma_+$,\\ {\sl Box}: \
$\!\!\left\{\!\begin{array}{ll} 1:& \mbox{enter } (AF) \mbox{ with a
slope larger}\\[-5pt] & \mbox{than that of } \Gamma_- \mbox{ by }
O(\ell^{1/2})\\[-2pt] 2:& \mbox{exit } (CB) \mbox{ with a slope less
than} \\[-5pt] & \mbox{that of } \Gamma_+ \mbox{ by
}O(\ell^{1/2})\\[-2pt] 3: & \mbox{cross } (FC) \mbox{ and stay below }
(ED). \end{array}\right.$\\ It is obvious that such conditions ensure
the existence of at least one regular point, as seen by moving $(ED)$
down parallel to itself until it touches the graph.  Note that $A$ and
the slope of $(AB)$ are prescribed. Hence, one is calculating
conditional probabilities; but it may be shown that the conditioning
is not affecting the scaling dependence on $\ell$.

As the Brownian motion $v_0(x_0)$ is a {\em Markov process}, the
constraints {\sl Left}, {\sl Box} and {\sl Right} are independent and
hence,
\begin{eqnarray}
   P^{\rm reg.}&&\! (\ell) \equiv {\rm Prob}\!\left \{ \mbox{regular
  point in interval of length } \ell \right \}\nonumber \\ &&= {\rm
  Prob}\! \left \{ \mbox{\sl Left} \right \} \!\times\! {\rm Prob}\!
  \left \{ \mbox{\sl Box} \right \} \!\times \!{\rm Prob} \!\left \{
  \mbox{\sl Right} \right \}.
\label{leftboxright}
\end{eqnarray}
The sizes of the box were chosen so that ${\rm Prob} \left \{
  \mbox{\sl Box} \right \}$ is independent of $\ell$:
\begin{equation}
{\rm Prob} \left \{
  \mbox{\sl Box} \right \} \sim \ell^0.
\label{box0}
\end{equation}
Indeed, Brownian motion and its integral have scaling exponent $1/2$
and $3/2$, respectively, and the problem with $\ell \ll 1$ can be
rescaled into that with $\ell = 1$ without changing probabilities.

\begin{figure}[ht]
  \centerline{\includegraphics[width=0.47\textwidth]{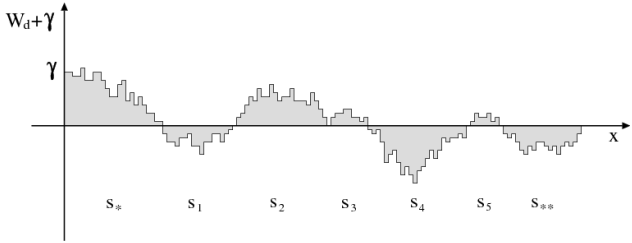}}
  \caption{The arches construction which uses the zero-passings of a
  random walk to estimate the integral of Brownian motion
  (from~\cite{s92,vdfn94}).}  \label{f:arches}
\end{figure}

It is clear by symmetry that ${\rm Prob}\! \left \{ \mbox{\sl Left}
\right \}$ and $ {\rm Prob}\! \left \{ \mbox{\sl Right} \right \}$
have the same scaling in $\ell$.  Let us concentrate on ${\rm Prob}\!
\left \{ \mbox{\sl Right} \right \}$. We can write the equation for
the half line $\Gamma_+$ in the form
\begin{eqnarray}
& \Gamma_+\!\!: x_0 \mapsto & \Phi(x_0^{\prime\prime}) \!+\! \delta
\ell^{3/2} \nonumber \\ &&\!+\! \left[ \frac{\mathrm{d}\Phi}{\mathrm{d}
x_0} (x_0^{\prime\prime})\!+\!\gamma \ell^{1/2} \right]\!
(x_0\!-\!x_0^{\prime\prime}),
\label{eqgammaplus}
\end{eqnarray}
where $\gamma$ and $\delta$ are positive $O(1)$ quantities. Hence,
introducing $\alpha \equiv x_0-x_0''$, the condition {\sl Right} can
be written to the leading order as
\begin{equation}
\int_0^\alpha \!\!\! \left [ v_0(x_0) + \gamma
\ell^{1/2}\right]\!\mathrm{d}x_0 +\delta \ell^{3/2} +
\frac{\alpha^2}{2} > 0,
\label{condright1}
\end{equation} 
for all $\alpha>0$. By the change of variable $\alpha = \beta\ell$
and use of the fact that the Brownian motion has scaling exponent
$1/2$, one can write the condition {\sl Right} as
\begin{equation}
  \int_0^\beta \left ( v_0(x_0) + \gamma \right) \mathrm{d}x_0 >
    -\delta, \mbox{ for all } \alpha  \in [0, \ell^{-1}]. 
\label{condright2}
\end{equation}
Without affecting the leading order, one can replace the Brownian
motion by a stepwise constant random walk with jumps of $\pm1$ at
integer $x_0$'s.  The integral in (\ref{condright2}) has a simple
geometric interpretation, as highlighted in figure~\ref{f:arches},
which shows a random walk starting from the ordinate $\gamma$ and the
arches determined by successive zero-passings. The areas of these
arches are denoted $S_\star, S_1, ... S_n, S_{\star\star}$.

It is easily seen that
\begin{eqnarray} {\rm Prob} \!\left\{ \mbox{\sl
Right} \right \} \sim {\rm Prob} \{ S_1>0,\,S_1+S_2>0,\,\ldots &&
\nonumber\\ S_1+\cdots+S_n>0 &&\!\},
\label{condright3}
\end{eqnarray}
where $n=O(\ell^{-1/2})$ is the number of zero-passings of the random
walk in the interval $[0, \ell^{-1}]$. The probability
(\ref{condright3}) can be evaluated by random walk methods (see,
e.g.,\cite{fellervol2}, Chap.~12, section~7), yielding
\begin{eqnarray}
  {\rm Prob}\! \left \{ \mbox{\sl Right} \right \} & \sim & {\rm
  Prob}\!\left\{ n \mbox{ first sums} \!>\! 0 \right \} \nonumber \\
  &\propto&  n^{-1/2} \propto \ell^{1/4}\!\!.
  \label{condrightfin}
\end{eqnarray}
By (\ref{leftboxright}), (\ref{box0}) and (\ref{condrightfin}), the
probability to have a regular point in a small interval of length
$\ell$ behaves as $\ell^{1/2}$ when $\ell\to0$. Thus, the regular
points have a box-counting dimension $1/2$.

This rigorous result on the fractal dimension of regular points served
as a basis in~\cite{afnb97} for a proof of the \emph{bifractality} of
the inverse Lagrangian map when the initial velocity is
Brownian. Namely, the moments $M_q(\ell) = \la (x_0(x+\ell) - x_0(x) )
\ra$ behave as $\ell^{\tau_q}$ at small separation $\ell$ and the
exponents $\tau_q$ experience the phase transition
\begin{eqnarray}
  &&\tau_q = 2q \mbox{ for } q\le 1/2\\ && \tau_q = 1 \mbox{ for }
  q\ge 1/2
\end{eqnarray}
At the moment, this is the only rigorous result on the bifractal
nature of the solutions to the Burgers equation in the case of
non-differentiable initial velocity. In particular, the case of
fractional Brownian motion is still opened.

%%%%%%%%%%%%%%%%%%%%%%%%%%%%%%%%%%%%%%%%%%%%%%
\section{Transport of mass in the Burgers/adhesion model}
\label{s:mass}

In the cosmological application of the Burgers equation, i.e.\ for the
adhesion model, it is of particular interest to analyze the behavior
of the density of matter, since the large-scale structures are
characterized as regions where mass is concentrated. This is done by
associating to the velocity field $\bv$ solution to the
$d$-dimensional decaying Burgers equation (\ref{eq:eqburgdecay}), a
continuity equation for the transport of a mass density field
$\rho$. In Eulerian coordinates, the mass density $\rho$ satisfies
\begin{equation}
  \partial_t \rho + \nabla \cdot (\rho \vec{v}) = 0\,,\quad
  \rho(\vec{x},0) = \rho_0(\vec{x})\,.
  \label{eq:continuity}
\end{equation}
A straightforward consequence of (\ref{eq:continuity}) and of the
formulation of Burgers dynamics in terms of characteristics
$\vec{X}(\vec{x}_0, t)$ is that, at the Eulerian locations where the
Lagrangian map is invertible, the mass density field $\rho$ can be
expressed as
\begin{eqnarray}
  \rho(\vec{x}, t) \!=\! \frac{\rho_0(\vec{x}_0)}{J(\vec{x}_0,t)},&&
  \mbox{where } \vec{X}(\vec{x}_0, t) \!=\! \vec{x},\nonumber \\
  \mbox{and} && J(\vec{x}_0, t) \!= \!\det\!  \left [(\partial
  X^i)/(\partial x_0^j) \right]\!\!.
  \label{eq:rholagmap}
\end{eqnarray}
Large but finite values of the density will be reached at locations
where the Jacobian $J$ of the Lagrangian map becomes very small. As we
will see in section \ref{subs:density}, they contribute a power-law
behavior in the tail of the probability density function of $\rho$.

The expression (\ref{eq:rholagmap}) is no more valid when the Jacobian
vanishes (inside shocks). Then the density field becomes infinite and
mass accumulates on the shock. We will see in section \ref{subs:mass}
that the evolution of the mass inside the singularities of the
solution can be obtained as the $\nu\to0$ limit of the well-posed
viscous problem. Finally, we will discuss in section
\ref{subs:optimization} some of the applications of the Burgers
equation to cosmology, and in particular how, assuming the dynamics of
the adhesion model, the question of reconstruction of the early
Universe from its present state can be interpreted as a convex optimal
mass transportation problem.

\subsection{Mass density and singularities}
\label{subs:density}

We give here the proof reported in \cite{fbv01} that in any dimension
large densities are localized near ``kurtoparabolic'' singularities
residing on space-time manifolds of co-dimension two. In any
dimension, such singularities contribute universal power-law tails
with exponent $-7/2$ to the mass density probability density function
(PDF) $p(\rho)$, provided that the initial conditions are smooth.

In one dimension, the mass density at regular points can be written as
\begin{equation}
  \rho(X(x_0,t),t) = \frac{\rho_0(x_0)}{1 -
  t[(\mathrm{d}^2\Psi_0)/(\mathrm{d}x_0^2)]}\,.
  \label{eq:rho1d}
\end{equation}
We suppose here that the initial density $\rho_0$ is strictly positive
and that both $\rho_0$ and $\Psi_0$ are sufficiently regular
statistically homogeneous random fields.  Large values of $\rho(x,t)$
are obtained in the neighborhood of Lagrangian positions with a
vanishing Jacobian, i.e.\ where
$\mathrm{d}^2\Psi_0(x_0)/\mathrm{d}x_0^2 = 1/t$.  Once mature shocks
have formed, the Lagrangian points with vanishing Jacobian are inside
shock intervals and thus not regular. The only points with a vanishing
Jacobian that are at the boundary of the regular points are obtained
at the \emph{preshocks}, that is when a new shock is just born at some
time $t_\star$. Such points, that we denote by $x_0^\star$, are local
minima of the initial velocity gradient which have to be negative, so
that the following relations are satisfied:
\begin{equation}
  \frac{\mathrm{d}^2\Psi_0}{\mathrm{d}x_0^2} (x_0^\star) =
  \frac{1}{t_\star}, \ \frac{\mathrm{d}^3\Psi_0}{\mathrm{d}x_0^3}
  (x_0^\star) = 0, \ \frac{\mathrm{d}^3\Psi_0}{\mathrm{d}x_0^3}
  (x_0^\star) < 0\,.
  \label{eq:defpreshocks}
\end{equation}
There is of course an extra global regularity condition that the
preshock Lagrangian location $x_0^\star$ has not been captured by a
mature shock at a time previous to $t_\star$. This global condition
affects only constants but not the scaling behavior of $p(\rho)$ at
large $\rho$.

\begin{figure}[ht]
  \centerline{\subfigure[\label{f:sketchlagpot1d}]{
      \includegraphics[width=0.45\textwidth]{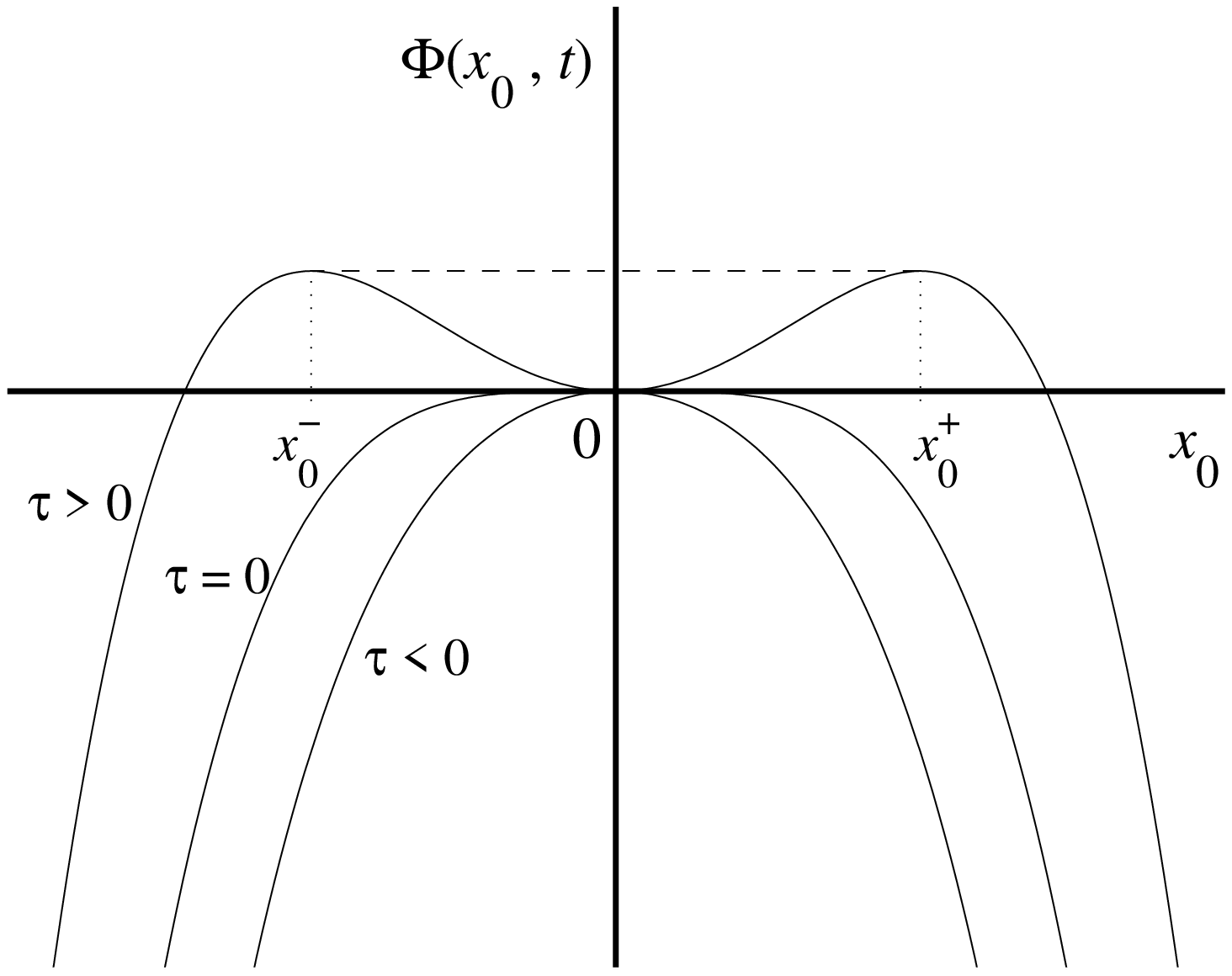}}}
  \vspace{-20pt} \centerline{\subfigure[\label{f:kurto2d}]{
  \includegraphics[width=0.4\textwidth]{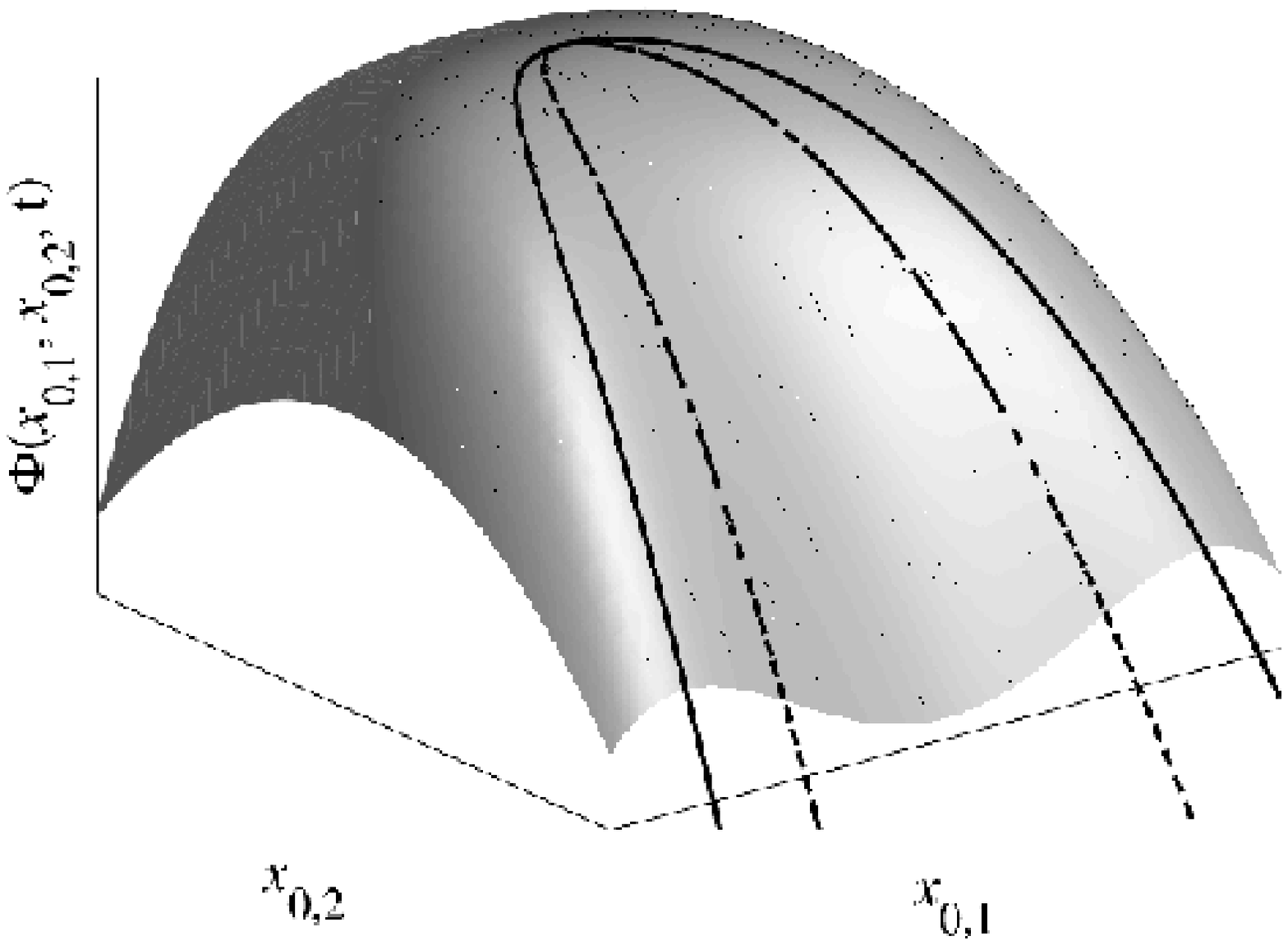}}}
  \caption{Normal form of the Lagrangian potential. (a) in one
  dimension, in the time-neighborhood of a preshock; at the time of
  the preshock ($\tau=0$), the Lagrangian potential changes from a
  single extremum to three extrema and develops a non-trivial convex
  hull (shown as a dashed line). (b) in two dimension, the space
  neighborhood of a shock ending point has a structure similar to the
  spatio-temporal normal form of a preshock in one dimension when
  replacing the $x_{0,2}$ variable by the time $\tau$; the continuous line
  is the separatrix between the regular part and the ruled surface of
  the convex hull; the dotted line corresponds to the locations where
  the Jacobian of the Lagrangian map vanishes.}
\end{figure}

We now Taylor-expand the initial density and the initial velocity
potential in the vicinity of $x_0^\star$. By adding a suitable
constant to the initial potential, shifting $x_0^\star$ to the origin
and making a Galilean transformation canceling the initial velocity at
$x_0^\star$, we obtain the following ``normal forms'' for the
Lagrangian potential (\ref{eq:deflagpot}) and for the density
\begin{equation}
  \Phi(x_0,t) \!\simeq\! \frac{1}{2} \tau x_0^2 \!+\! \zeta x_0^4, \ \
  \rho(X\!(x_0,t),t) \!\simeq\! \displaystyle
  \frac{-\rho_0}{\tau\!+\!12\zeta x_0^2},
  \label{eq:normforms}
\end{equation}
where
\begin{equation}
  \tau = \frac{t-t_\star}{t_\star} \mbox{ and } \zeta =
  \frac{t_\star}{24}\,\left.\frac{\mathrm{d}^4\Psi_0}{\mathrm{d}
  x_0^4} \right|_{x_0=0} < 0\,.
  \label{eq:deftauzeta}
\end{equation}
The Lagrangian potential bifurcates from a situation where it has a
single maximum at $\tau<0$ through a degenerate maximum with quartic
behavior at $\tau=0$, to a situation where convexity is lost and where
it has two maxima at $x_0^\pm = \pm \sqrt{-\tau/(4\zeta)}$ for
$\tau>0$. As a result of our choice of coordinates, the symmetry
implies that the convex hull contains a horizontal segment joining
these two maxima (see.\ figure\ \ref{f:sketchlagpot1d}).

We see from (\ref{eq:normforms}) that the Eulerian density $\rho$ is
proportional to $x_0^2$ in Lagrangian coordinates at $t=t_\star$.
Since $X = -\partial_{x_0} \Phi$, the relation between Lagrangian and
Eulerian coordinates is cubic, so that at $\tau=0$, the density has a
singularity $\propto x^{-2/3}$ in Eulerian coordinates.  At any time
$t\neq t_\star$, the density remains bounded except at the shock
position. Before the preshock ($\tau<0$), it is clear that
$\rho<-\rho_0/\tau$, while after ($\tau>0$), exclusion of the
Lagrangian shock interval $[x_0^-, x_0^+]$ implies that
$\rho<\rho_0/(2\tau)$. Clearly, large densities are obtained only in
the immediate space-time neighborhood of the preshock. More precisely,
it follows from~(\ref{eq:normforms}) that having $\rho(x,t) > \mu$
requires simultaneously
\begin{equation}
  |\tau| < \frac{\rho_0}{\mu} \mbox{ and } |x| < (-12\zeta)^{-1/2}
   \left(\frac{\rho_0}{\mu}\right)^{3/2}\!\!\!.
   \label{eq:condspcetimerho}
\end{equation}
The tail of the cumulative probability of the density can be
determined from the fraction of Eulerian space-time where $\rho$
exceeds a given value. This leads to
\begin{equation}
  P^{>}\!(\mu;\,x,t) = \mbox{Prob}\!\left\{ \rho(x,t) \!>\! \mu
  \right\} \simeq C(x,t) \mu^{-5/2}\!,
  \label{eq:psuprho}
\end{equation}
where the constant $C$ can be expressed as
\begin{equation}
  C(x,t) = A\,t\,\int_{-\infty}^0 |\zeta|^{-1/2}p_3(x,t,\zeta)\,
  \mathrm{d}\zeta,
  \label{eq:defconstpsuprho}
\end{equation}
$A$ is a positive numerical constant and $p_3$ designates the joint
probability distribution of the preshock space-time position and of
its ``strength'' coefficient~$\zeta$ (see \cite{fbv01} for
details). This algebraic law for the cumulative probability implies
that the PDF of the mass density has a power-law tail with exponent
$-7/2$ at large values.  Actually this law was first proposed in
\cite{ekms97} for the large-negative tail of velocity gradients in
one-dimensional forced Burgers turbulence, a subject to which we shall
come back in section \ref{s:1Dstatistics}.

In higher dimensions it was shown in \cite{fbv01} that the main
contribution to the probability distribution tail of the mass density
does not stem from preshocks but from ``kurtoparabolic'' points.  Such
singularities (called $A_3$ according to the classification of
\cite{gs84}, which is summarized in section \ref{subs:singularities})
correspond to locations which belong to the regular part of the convex
hull of the Lagrangian potential $\Phi(\vec{x}_0,t)$ and where its
Hessian vanishes. The name kurtoparabolic comes from the Greek
``kurtos'' meaning ``convex''. These points are located on the spatial
boundaries of shocks and generically form space-time manifolds of
co-dimension 2 (persisting isolated points for $d=2$, lines for $d=3$,
etc.).  As in one dimension, the normal form of such singularities is
obtained by Taylor-expanding in a suitable coordinate frame the
Lagrangian potential to the relevant order
\begin{equation}
  \Phi(\vec{x}_0,t) \simeq \zeta x_{0,1}^4 \!+\!\! \sum_{2\le j\le d}
  \left( -\frac{\mu_j}{2} x_{0,j}^2 \!+\! \beta_j x_{0,1}^2 x_{0,j}
  \right)\!,
  \label{eq:normformmultid}
\end{equation}
where the different coefficients satisfy inequalities that ensure that
the surface is below its tangent plane at $\vec{x}_0 = 0$.  The
typical shape of the Lagrangian potential in two dimensions is shown
in figure\ \ref{f:kurto2d}. The positions where the Jacobian of the
Lagrangian map vanishes can be easily determined from this normal
form. The convex hull of $\Phi$ and the area where the mass density
exceeds the value $\mu$ can also be constructed explicitly. An
important observation is that, in any dimension, the scalar product of
the vector $\vec{y}_0 = (x_{0,2},\dots,x_{0,d})$ with the vector
$\vec{\beta} = (\beta_2,\dots,\beta_d)$ plays locally the same role as
time does in the analysis of one-dimensional preshocks.

When $\mu\to\infty$, the cumulative probability can be estimated as
\begin{eqnarray}
  P^{>}(\mu;\,x,t) & \propto &
  \underbrace{\mu^{-3/2}}_{\mbox{\scriptsize from }x_{0,1}} \times
  \underbrace{\mu^{-1}}_{\mbox{\scriptsize from
  }\vec{\beta}\cdot\vec{y}_0} \nonumber \\ && \times\!\!\!\!\!
  \underbrace{1 \times \cdots \times 1}_{\mbox{\scriptsize from other
  components of } \vec{y}_0} \times \underbrace{1}_{\mbox{\scriptsize
  from time}}\!\!\!\!\!\!.
  \label{eq:estimpsup}
\end{eqnarray}
The only non-trivial contributions come from $x_{0,1}$ and from the
component of $\vec{y}_0$ along the direction of $\vec{\beta}$, all the
other components and time contributing order-unity factors.  Hence,
the cumulative probability $P^{>}(\mu)$ is proportional to
$\mu^{-5/2}$ in any dimension, so that the PDF of mass density has a
power-law behavior with the universal exponent $-7/2$.

As we have seen, the theory is not very different in one and higher
dimension even if kurtoparabolic points are persistent only in the
latter case. This is due to the presence of a time-like direction in
the case $d\ge 2$.

\subsection{Evolution of matter inside shocks}
\label{subs:mass}

As we have seen in the previous subsection, the mass density becomes
very large in the neighborhood of kurtoparabolic points ($A_3$
singularities) corresponding to the space-time boundaries of
shocks. Such singularities dominate the tail of the mass density
probability distribution and contribute a power-law behavior with
exponent $-7/2$. However the mass distribution depends strongly on
what happens inside the shocks where the density is infinite. Indeed,
after the formation of the first singularity a finite fraction of the
initial mass gets concentrated inside these low-dimensional
structures. Describing the mass distribution requires understanding
how matter evolves once concentrated in the shocks. But before it will
be useful to explain briefly the time evolution of the shock manifold.

\subsubsection{Dynamics of singularities}

Suppose that $\vec{X}(t)$ denotes the position of a shock at time
$t$. We suppose this singularity to be of type $A_1^n$ (see section
\ref{subs:singularities}), so that at this position, the velocity
field is discontinuous; we denote by $\vec{v}_1,\dots,\vec{v}_n$ the
$n$ different limiting values it takes at that point. At any time we
generically have $n\le d+1$ and occasionally $n=d+2$ at the space-time
positions of shock metamorphoses corresponding to instants when two
$A_1^d$ singularities merge. We first restrict ourselves to persistent
singularities, meaning that $n\le d+1$. In the neighborhood of
$\vec{X}(t)$, it is easily checked that the velocity potential can be
written as
\begin{eqnarray}
  \Psi(\vec{x}, t) & = & \Psi(\vec{X}(t), t) + \max_{j=1..n} \left[
  \vec{v}_j\cdot(\vec{X}(t)-\vec{x}) \right] \nonumber\\ && + {\rm
  o}(\|\vec{x}-\vec{X}(t)\|) \,.
  \label{eq:localpot}
\end{eqnarray}
This expansion divides locally the physical space in $n$ subdomains
$\Omega_j$ where $\vec{v}_j\cdot(\vec{X}(t)-\vec{x})$ is maximum, i.e.
\begin{equation}
  \vec{y}\in\Omega_j  \Leftrightarrow  (\vec{v}_i-\vec{v}_j)
  \cdot (\vec{y}-\vec{X}(t)) \ge 0, \ 1\le i\le n\,.
  \label{eq:defomegai}
\end{equation}
Writing the expansion (\ref{eq:localpot}) amounts to approximating the
velocity potential by a continuous function which is piecewise linear
on the subdomains $\Omega_j$. The boundaries between the $\Omega_j$'s
define the local shock manifold. The maximum in (\ref{eq:localpot})
ensures that we are focusing on entropic solutions to the Burgers
equation (solutions obtained in the limit of vanishing viscosity) and
results in the convexity of the local approximation of the
potential. Note also that the position $\vec{x} = \vec{X}(t)$ of the
reference singular point corresponds by construction to the unique
intersection of all subdomains $\Omega_j$. Remember that we have
assumed that locally, the solution does not have higher-order
singularity.

The approximation (\ref{eq:localpot}) fully describes the local
structure of the singularity. If $n=2$, corresponding to $\vec{X}(t)$
being the position of a simple shock, it is easily checked
from~(\ref{eq:localpot}) that there will actually exist a whole shock
hyper-plane given by the set of positions $\vec{y}$ satisfying
\begin{equation}
  (\vec{v}_1 - \vec{v}_2) \cdot (\vec{X}(t) -\vec{y}) = 0\,.
  \label{eq:locshockform}
\end{equation}
If $n>2$, meaning that $\vec{X}(t)$ is an intersection between
different shocks, all the singular manifolds of co-dimension $m\le n$
are present in the expansion and are given by the set of positions
$\vec{y}$ satisfying
\begin{equation}
  \vec{v}_{i_1} \cdot (\vec{X}(t) -\vec{y}) = \cdots = \vec{v}_{i_m}
  \cdot (\vec{X}(t) -\vec{y})\,,
  \label{eq:locintersecform}
\end{equation}
with $1\le i_1 < \cdots < i_m \le n$.

We next apply the variational principle (\ref{eq:varprincdecay2}) in
order to solve the decaying problem between times $t$ and $t+\delta t$
with the initial condition given by (\ref{eq:localpot}).  This yields
an approximation of the potential at time $t+\delta t$:
\begin{eqnarray}
  && \Psi(\vec{x}, t+\delta t) \simeq \Psi(\vec{X}(t), t) \nonumber \\
  && + \max_{\vec{y}} \, \max_{j=1..n} \left[
  \vec{v}_j\cdot(\vec{X}(t)-\vec{y}) -\frac{1}{2\delta t}\| \vec{x} -
  \vec{y} \|^2 \right]\,.
  \label{eq:locmaxprinc}
\end{eqnarray}
Note that here, $\delta t$ and $\|\vec{x}-\vec{X}(t)\|$ are chosen
sufficiently small in a suitable way to ensure that (i) any
singularity of higher co-dimension does not interfere with the
position of $\vec{X}(t)$ between times $t$ and $t+\delta t$ and that
(ii) the subleading terms are always dominated by the kinetic energy
contribution $\| \vec{x} - \vec{y} \|^2 / (2\delta t)$.

The two maxima in $\vec{y}$ and in $j$ of (\ref{eq:locmaxprinc}) can
be interchanged, under the condition that the maximum in $\vec{y}$ is
restricted to the domain $\Omega_j$ defined in
(\ref{eq:defomegai}). The potential at time $t+\delta t$ can thus be
written as
\begin{eqnarray}
  && \Psi(\vec{x}, t+\delta t) \simeq \Psi(\vec{X}(t), t) \nonumber \\
  && + \max_{j=1..n}\,\, \max_{\vec{y}\in\Omega_j} \left[
  \vec{v}_j\cdot(\vec{X}(t)-\vec{y}) -\frac{1}{2\delta t} \| \vec{x}
  -\vec{y} \|^2 \right]\,.
  \label{eq:reducemax}
\end{eqnarray}
We next remark that for all $\vec{x}$, $j$ and $\vec{y}$, one has
\begin{eqnarray}
  \vec{v}_j\cdot(\vec{X}(t)-\vec{y}) && -\frac{1}{2\delta t} \|
  \vec{x} -\vec{y} \|^2 \nonumber \\ && \le
  \vec{v}_j\cdot(\vec{X}(t)-\vec{x} +\delta t\vec{v}_j) -\frac{\delta
  t}{2} \| \vec{v}_j \|^2\,,
  \label{eq:trivialineg}
\end{eqnarray}
which gives an upper-bound to the maximum over $\vec{y}\in\Omega_j$ in
(\ref{eq:reducemax}). Suppose now that the maximum over the index $j$
is achieved for $j=j_0$. This means that for all $1\le i\le n$ and
$\vec{y}\in\Omega_i$
\begin{eqnarray}
  && \vec{v}_i\cdot(\vec{X}(t)-\vec{y}) -\frac{1}{2\delta t} \|
  \vec{x} -\vec{y} \|^2 \nonumber \\ && \ \ \le
  \max_{\vec{z}\in\Omega_{j_0}} \left[
  \vec{v}_{j_0}\cdot(\vec{X}(t)-\vec{z}) -\frac{1}{2\delta t} \|
  \vec{x} -\vec{z} \|^2 \right] \nonumber \\ && \ \ \le
  \vec{v}_{j_0}\cdot(\vec{X}(t)-\vec{x} +\delta t\vec{v}_{j_0})
  -\frac{\delta t}{2} \| \vec{v}_j \|^2\,.
  \label{eq:condmaxj_0}
\end{eqnarray}
Let $\Omega_{i_0}$ be the domain containing the vector
$(\vec{x}-\delta t\vec{v}_{j_0})$. Then, (\ref{eq:condmaxj_0}) applied
to $i=i_0$ and $\vec{y} = \vec{x} -\delta t\vec{v}_{j_0}$ trivially
implies that
\begin{equation}
  (\vec{v}_{i_0} - \vec{v}_{j_0})\cdot(\vec{x} -\delta t \vec{v}_{j_0}
  -\vec{X}(t)) \ge 0\,,
  \label{eq:ineqjj0}
\end{equation}
which together with the definition (\ref{eq:defomegai}) for
$\Omega_{i_0}$ leads to $i_0=j_0$. Hence, to summarize, if the first
maximum is reached for $j=j_0$ then the second maximum is necessarily
reached for $\vec{y} = \vec{x} -\delta t \vec{v}_{j_0}$. 

\begin{figure}[ht]
  \centerline{\includegraphics[width=0.25\textwidth]{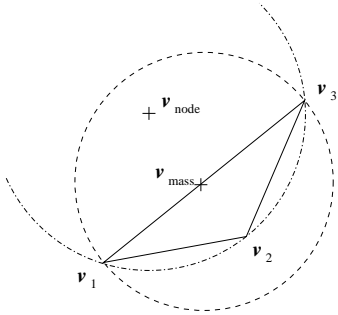}}
  \caption{Determination of the velocity of a triple point and of that
  of the mass inside it when the three limiting values of the velocity
  $\vec{v}_1$, $\vec{v}_2$, and $\vec{v}_3$ form an obtuse
  triangle. The dash-dotted circle is the circumcircle whose center
  gives the velocity of the singularity and the dashed circle is the
  smallest circle containing the triangle whose center gives the
  velocity of mass. }
  \label{fig:trianglevel}
\end{figure}

It is clear that the approximation (\ref{eq:locmaxprinc}) of the
velocity potential at time $t+\delta t$ preserves the local structure
of the singular manifold. Indeed, for $m\le n$, the positions
$\vec{y}$ satisfying
\begin{eqnarray}
  \vec{v}_1 \cdot (\vec{X}(t) -\vec{y}) & +& \frac{\delta t}{2} \|
  \vec{v}_1 \|^2 = \cdots \nonumber \\ \cdots& =& \vec{v}_m \cdot
  (\vec{X}(t) -\vec{y})+\frac{\delta t}{2} \| \vec{v}_m \|^2
  \label{eq:evolintersec}
\end{eqnarray}
form a $(d-m)$-dimensional shock manifold. The trajectory $\vec{X}(t)$
of the reference singular point satisfies
\begin{equation}
  \vec{v}_1 \cdot \frac{\mathrm{d}\vec{X}}{\mathrm{d}t}
  -\frac{1}{2}\|\vec{v}_1\|^2 =\cdots =\vec{v}_n \cdot
  \frac{\mathrm{d}\vec{X}}{\mathrm{d}t} -\frac{1}{2}\|\vec{v}_n\|^2\,,
  \label{eq:evolsingpos}
\end{equation}
which can be rewritten as
\begin{equation}
  \| {\mathrm{d}\vec{X}}/{\mathrm{d}t} -\vec{v}_1 \| =\cdots = \|
  {\mathrm{d}\vec{X}}/{\mathrm{d}t} -\vec{v}_n \|\,.
  \label{eq:evolsingpos2}
\end{equation}
This gives $n$ relations for the $d$ components of the vector
$\mathrm{d}\vec{X}/\mathrm{d}t$. These relations allow determining the
normal velocity of the singular manifold. The tangent velocity remains
undetermined. The velocity of the singularity located at $\vec{X}(t)$
is completely determined only if $n=d$, i.e.\ for point
singularities. For instance when $d=1$, the dynamics of shocks is
given by
\begin{equation}
  \frac{\mathrm{d} X}{\mathrm{d}t} = \frac{1}{2} (u_1+u_2)\,,
  \label{eq:evolshock}
\end{equation}
meaning that they move with a velocity equal to the half sum of their
right and left velocities. For $d=2$, only the positions of triple
points (singularities of type $A_1^3$ corresponding to the
intersection of three shock lines) are well determined. It is easily
checked that the two-dimensional velocity vector
$\mathrm{d}\vec{X}/\mathrm{d}t$ is the circumcenter of the triangle
formed by the three limiting values $(\vec{v}_1,\vec{v}_2,\vec{v}_3)$
that are achieved by the velocity field at this position (see
figure~\ref{fig:trianglevel}).

%Eventually:
% Comment on the (eventual) geometrical constraint on the vectors
% $(\vec{v}_1,\vec{v}_2,\vec{v}_3)$ for triple points in two dimension
% (can we exclude obtuse triangles)

\subsubsection{Dynamics of the mass inside the singular manifold}

One of the central themes of this review article is a connection
between Lagrangian particle dynamics and the inviscid Burgers
equation.  In the unforced case the velocity is conserved along
particle trajectories minimizing the Lagrangian action (see
section~\ref{s:basictools}). At a given moment of time, all particles
whose trajectories are not minimizers have been absorbed by the
shocks.  In the one-dimensional case when shocks are isolated points,
particles absorbed by shocks just follow the dynamics of a shock
point. However, in the multi-dimensional case the geometry of the
singular shock manifold can be rather complicated. This results in a
non-trivial particle dynamics inside the singular manifold. In other
words, the particle absorbed by shocks have a rich afterlife and the
main problem is to describe their dynamical properties inside the
singular manifold. This problem was addressed by I.\ Bogaevsky in
\cite{bog04}.

The basic idea is to consider first particle transport by the velocity
field given by smooth solutions to the viscous Burgers
equation. Indeed, let $\vec{v}^\nu(x,t)$ be a solution to the viscous
Burgers equation
$$ \partial_t\bv^\nu +(\bv^\nu\cdot\nabla)\bv^\nu= \nu \nabla^2\bv^\nu
- \nabla F(\bx,t).$$ Then the dynamics of a Lagrangian particle
labeled by its position $\vec{x}_0$ at time $t=0$ is described by the
system of ordinary differential equations
\begin{equation}
  \label{trpartnu}
  \dot{\vec{X}}^\nu(\vec{x}_0,t)=\bv^\nu(\vec{X}^\nu(\vec{x}_0,t),t),
  \quad \vec{X}^\nu(\vec{x}_0,0)=\vec{x}_0,
\end{equation}
where the dots stand for time derivatives.  It is possible to show
that in the inviscid limit $\nu \to 0$ solutions to (\ref{trpartnu})
converge to limiting trajectories $\{\vec{X}(\vec{x}_0,t)\}$. These
limiting trajectories are not disjoint anymore. In fact, two
trajectories corresponding to different initial positions
$\vec{x}_0^1$ and $\vec{x}_0^2$ can merge:
$\vec{X}(\vec{x}_0^1,t^*)=\vec{X}(\vec{x}_0^2,t^*)$. This corresponds
to absorption of particles by the shock manifold. Of course, two
trajectories coincide after they merge:
$\vec{X}(\vec{x}_0^1,t)=\vec{X}(\vec{x}_0^2,t)$ for $t\geq t^*$.
Particles which until time $t$ never merged with any other particles
correspond to minimizers. Such trajectories obviously satisfy the
limiting differential equation:
\begin{equation}
\label{trpart}
\dot{\vec{X}}(\vec{x}_0,t)=\bv(\vec{X}(\vec{x}_0,t),t), \quad
\vec{X}(\vec{x}_0,0)=\vec{x}_0,
\end{equation}
where $\bv(x,t)$ is the entropic solution of the inviscid Burgers
equation which is well defined outside of the shock manifold. However,
we are mostly interested in the dynamics of particles which have
merged with other particles and thus were absorbed by shocks.  One can
prove that for such trajectories one-sided time derivatives exist
\begin{equation}
\label{onesidedder}
\frac{\mathrm{d}^+}{\mathrm{d}t}{\vec{X}}(t)=\lim_{\Delta t \to 0+}
\frac{\vec{X}(t+\Delta t) - \vec{X}(t)}{\Delta t}
\end{equation}
and satisfy a ``one-sided" differential equation:
\begin{equation}
\label{trpart2}
\frac{\mathrm{d}^+}{\mathrm{d}t}{\vec{X}}(t)=\bv^{(\mathrm{s})}(\vec{X}(t),t).
\end{equation}
Here $\bv^{(\mathrm{s})}(\cdot,t)$ is the velocity field on the shock
manifold (index $\mathrm{s}$ stands for shocks). It turns out that
$\bv^{(\mathrm{s})}(\bx,t)$ and the corresponding shock trajectories
satisfy a variational principle, described hereafter.  Denote by
$\Psi(\bx,t)$ a potential of the viscous velocity field $\bv(\bx,t)$:
$\bv(\bx,t)= - \nabla \Psi(\bx,t)$. As we have pointed out many times
before, $-\Psi(\bx,t)$ corresponds to a minimum Lagrangian action
among all the Lagrangian trajectories which pass through point $\bx$
at time $t$.  Shocks correspond to a situation where the minimum is
attained for several different trajectories.  Correspondingly, one has
several smooth branches such that $\Psi(\bx,t) = \Psi_i(\bx,t), \,
1\leq i\leq k$. Suppose a particle moves from a point of shock
$(\bx,t)$ with a velocity $\bv$. Then at infinitesimally close time
$t+\delta t$ its position will be $\bx + \bv \delta t$.  In linear
approximation (see previous subsection) the Lagrangian action of this
infinitesimal piece of trajectory is equal to $[|\bv|^2/2 -
F(\bx,t)]\delta t$. Of course, the action minimizing trajectory at the
point $(\bx + \bv \delta t, t +\delta t)$ does not pass through a
shock point $(\bx,t)$. Hence, the minimum action $-\Psi(\bx + \bv
\delta t, t +\delta t)$ is smaller than $-\Psi(\bx,t) + [\|\bv \|^2/2
- F(\bx,t)]\delta t$ for any velocity $\bv$.  However, we can put a
variational condition on $\bv$ which requires the difference between
$-\Psi(\bx,t) + [\|\bv \|^2/2 - F(\bx,t)]\delta t$ and $-\Psi(\bx +
\bv \delta t, t +\delta t)$ to be as small as possible. This is
exactly the variational principle which determines the velocity $\bv=
\bv^{(\mathrm{s})}(\bx,t)$ at a shock point. It is easy to see that in
linear approximation
\begin{eqnarray}
  &&\Psi(\bx \!+\! \bv \delta t, t\!+\!\delta t) =\max_{1\leq i\leq
  k}[\Psi_i(\bx \!+\! \bv \delta t, t \!+\!\delta t)]\nonumber \\ &&=
  \Psi(\bx,t) -\! \min_{1\leq i\leq k}[-\nabla \Psi_i(\bx,t)\cdot \bv
  -\partial_t\Psi_i(\bx,t)]\,\delta t .
  \label{psi1}
\end{eqnarray}
Let us denote by $\bv_i$ the limiting velocities $-\nabla
\Psi_i(\bx,t)$ at the shock point $(\bx,t)$.  Then, using
Hamilton--Jacobi equation for the velocity potential
\begin{eqnarray}
\partial_t\Psi_i(\bx,t) &=& \frac{1}{2}\left \|\nabla
  \Psi_i(\bx,t)\right \|^2 + F(\bx,t)\nonumber\\ &=&
  \frac{1}{2}\|\bv_i\|^2+F(\bx,t) \label{condt}
\end{eqnarray}
we have
\begin{eqnarray}
  \Psi(\bx + \bv \delta t,&& t +\delta t)= \Psi(\bx,t) - \nonumber \\
  &&-\min_{1\leq i\leq k}\left[\bv_i\!\cdot\! \bv - \frac{1}{2}\|\bv_i\|^2
  \right]\!\delta t \!- \! F(\bx,t)\delta t .\label{psi2}
\end{eqnarray}
Hence, the difference of actions can be written as
\begin{eqnarray}
\label{psi3}
\Delta \mathcal A &=&-\Psi(\bx,t) \!+\! \frac{1}{2}\|\bv\|^2 \delta t
+ \Psi(\bx\!+\! \bv \delta t, t \!+\!\delta t) \nonumber \\ &=&
\frac{1}{2} \|\bv\|^2\delta t - \min_{1\leq i\leq k}\left[\bv_i\cdot
\bv - \frac{1}{2}\|\bv_i\|^2\right]\delta t \nonumber\\ &=&
\frac{1}{2}\max_{1\leq i\leq k}\|\bv-\bv_i\|^2 \delta t.
\end{eqnarray}
Obviously minimization of $\Delta \mathcal A$ over $\bv$ corresponds
to a center of a minimum ball covering $\bv_i$. It implies that such a
center gives the velocity $\bv^{(\mathrm{s})}(\bx(t),t)$ of particles
concentrated at a shock point $(\bx,t)$.  It is interesting that this
variational principle implies that a particle absorbed by a shock
cannot leave the singular shock manifold in the future.

Let us now consider the first nontrivial generic example of a shock
point, namely a triple point in two dimensions $d=2$. The point
$(\vec{X}(t),t)$ is thus the intersection of three shock lines. In
this case there are exactly three smooth branches $\Psi_i(\cdot,t)$
with limiting velocities $\bv_i= -\nabla \Psi_i, \, 1\leq i \leq 3$.
As we have seen in previous section the motion of the triple point is
determined by continuity of the velocity potential at
$(\vec{X},t)$. The ``geometrical velocity''
$\mathrm{d}\vec{X}/\mathrm{d}t$ of the triple point is then the
circumcenter of the triangle formed by the three velocities $\bv_1,
\bv_2, \bv_3$. It is easy to see that
$\mathrm{d}\vec{X}/\mathrm{d}t=\bv^{(s)}$ only in the case when the
vectors $\bv_1$, $\bv_2$, and $\bv_3$ form an acute triangle. If so, a
cluster of particles follows the triple point. In the opposite case
when the triangle is obtuse, the particles leave the node.  Such a
situation is presented in figure~\ref{fig:trianglevel}, where the mass
leaves the node along the shock line delimiting the values $\bv_1$ and
$\bv_3$ of the velocity.

The analysis presented above was carried out for the Burgers equation
jointly with A. Sobolevski{\u\i} as a part of ongoing work on a
similar theory for the case of a general Hamilton--Jacobi equation,
with a Hamiltonian that is convex in the momentum variable.  The
formal extension of this analysis to the latter case is
straightforward and can be left to the interested reader; however at
present a rigorous justification of it, employing methods
of~\cite{bog04}, is known only for the case of $H(x, \dot{x}, t) =
a(x, t) |\dot{x}|^2$, with $a(x, t) > 0$.

\subsection{Connections with convex optimization problems}
\label{subs:optimization}

As discussed in section~\ref{ssec:adhesion}, Burgers dynamics is known
in cosmology as the adhesion model and frequently used to understand
the formation of the large-scale structures in the Universe. Recently,
this model was used as a basis for developing new techniques for one
of the most challenging questions in modern cosmology, namely
\emph{reconstruction}. This problem aims at reconstructing the
dynamical history of the Universe through the mass density initial
fluctuations that evolved into the distribution of matter and galaxies
which is nowadays observed (see, e.g., \cite{p93}). The main
difficulty encountered is that the velocities of galaxies (the
peculiar velocities) are usually unknown, so that most approaches lead
to non-unique solutions to this ill-posed problem. The reconstruction
technique we present here, which was proposed
in~\cite{fmms02,bfhlmms03}, is based on the observation that, to the
leading order, the mass is initially uniformly distributed in space
(see, e.g., \cite{p93}). This observation, together with the Zeldovich
approximation, leads to a reformulation of the problem as a well-posed
instance of an optimal mass transportation problem between the initial
(uniform) and the present (observed) distributions of mass. More
precisely it amounts to a convex optimization problem related to the
Monge--Amp{\`{e}}re equation and dually, as found by
Kantorovich~\cite{k42}, to a linear programming problem. This is the
reason why the name MAK (Monge--Amp{\`{e}}re--Kantorovich) has been
proposed for this method in~\cite{fmms02}. Namely, one has to find the
transformation from initial to current positions (the Lagrangian map)
that maps the initial density $\rho(\vec{x}_0,0) = \rho_0$ to the
field $\rho(\vec{x},t)$ which is nowadays observed. One then use a
well-known fact in cosmology: because of the expansion of the
Universe, the initial velocity field of the self-gravitating matter is
\emph{slaved} to the initial gravitational field (see,
e.g.,~\cite{bfhlmms03}). This observation implies that the initial
velocity field is potential and allows one to deduce from it the
subleading fluctuations of the mass density.

The MAK reconstruction technique is based on two crucial
assumptions. First the Lagrangian map $\vec{x}_0 \mapsto \vec{x} =
\vec{X}(\vec{x}_0,t)$ is assumed to be potential, i.e.\ $\vec{X} =
\nabla_{x_0} \Phi (\vec{x}_0)$. Second, the Lagrangian potential
$\Phi(\vec{x}_0)$ is assumed to be a convex function. As explained in
\cite{bfhlmms03} these two hypotheses are motivated by the adhesion
model (and thus inviscid Burgers dynamics) where they are trivially
satisfied. As we will see later the reverse is actually true: the
potentiality of the Lagrangian map and the convexity of the potential
is equivalent to assuming that the latent velocity field is a solution
to the Burgers equation. We will now see how, under these hypotheses,
the reconstruction problem relates to Monge--Amp{\`{e}}re
equation. Conservation of mass trivially implies that $\rho(\vec{x},t)
\mathrm{d}^3 x= \rho_0 \mathrm{d}^3 x_0$, which can be rewritten in
terms of the Jacobian matrix $(\partial X^i)/(\partial x_0^j)$ as
\begin{equation}
  \mathrm{det}\left[ \frac{\partial X^i}{\partial x_0^j} \right]=
  \frac{\rho_0}{\rho(\vec{X}(\vec{x}_0,t),t)}.
\end{equation}
Potentiality of the Lagrangian map leads to 
\begin{equation}
  \mathrm{det}\left[ \frac{\partial^2 \Phi}{\partial x_0^i \partial
  x_0^j} \right]= \frac{\rho_0}{\rho(\nabla_{x_0} \Phi,t)}.
  \label{eq:conservmass}
\end{equation}
The problem with this formulation is that the unknown potential $\Phi$
enters the right-hand side of the equation in a non-trivial way.
Convexity of the Lagrangian potential $\Phi$ is next used to
reformulate the problem in term of the inverse Lagrangian map. Indeed,
if $\Phi$ is convex, the inverse Lagrangian map is also potential,
i.e.\ $\vec{x}_0 = \vec{X}_0(x,t) = \nabla_{x} \Theta (\vec{x})$ with
the potential $\Theta$ itself convex. The two potentials $\Phi$ and
$\Theta$ are moreover related by Legendre transforms:
\begin{eqnarray}
  && \Theta(\vec{x}) = \max_{\vec{x}_0} [\vec{x}\cdot\vec{x}_0 -
  \Phi(\vec{x}_0)], \\ && \Phi(\vec{x}_0) = \max_{\vec{x}}
  [\vec{x}\cdot\vec{x}_0 - \Theta(\vec{x})].
\end{eqnarray}
In terms of the inverse Lagrangian potential $\Theta$ the conservation
of mass (\ref{eq:conservmass}) reads
\begin{equation}
  \mathrm{det}\left[ \frac{\partial^2 \Theta}{\partial x^i \partial
  x^j} \right]= \rho(\vec{x} ,t),
  \label{eq:monge-ampere}
\end{equation}
which is exactly the elliptic Monge-Amp\`{e}re equation. This time,
the difficulty expressed above has disappeared since the unknown
potential $\Theta$ does not enter the right-hand side of the
equation. Note that we have implicitly assumed here that the present
distribution of mass has no singularity. The case of a singular
distribution could actually be treated using a weak formulation of the
Monge-Amp\`{e}re equation, which amounts to applying conservation of
mass on any subdomain but requires allowing the inverse Lagrangian map
to be multivalued. The next step in the design of the MAK method is to
reformulate (\ref{eq:monge-ampere}) as an optimal transport problem
with quadratic cost. Indeed, as shown in \cite{b91}, the map
$\vec{X}(\vec{x}_0,t)$ (and its inverse $\vec{X}_0(\vec{x},t)$)
minimizing the cost
\begin{eqnarray}
  \mathcal{I} &=& \int \| \vec{X}(\vec{x}_0,t) - \vec{x}_0\|^2 \rho_0
  \, \mathrm{d}^3 x_0 \nonumber\\ &=& \int \| \vec{x} -
  \vec{X}_0(\vec{x},t)\|^2 \rho(\vec{x},t) \, \mathrm{d}^3 x,
\end{eqnarray}
is a potential map whose potential is convex and is the solution to
the Monge--Amp\`{e}re equation (\ref{eq:monge-ampere}). This can be
understood using a variational approach as proposed
in~\cite{fmms02}. Suppose we perform a small displacement $\delta
\vec{X}_0(\vec{x})$ of the inverse Lagrangian map
$\vec{X}_0(\vec{x},t)$ solution of the optimal transport problem. On
the one hand the only admissible displacement are those satisfying the
constraint to map the initial density field $\rho_0$ to the final one
$\rho(\vec{x},t)$. It is shown in~\cite{bfhlmms03} that this is
equivalent to require that $\nabla_x \cdot [\rho(\vec{x},t)\delta
\vec{X}_0(\vec{x})] =0$. On the other hand one easily see that the
variation of the cost function corresponding to the variation $\delta
x$ reads
\begin{equation}
  \delta\mathcal{I} = -2\int [ \vec{x} - \vec{X}_0(\vec{x},t)]\cdot
  [\rho(\vec{x},t) \delta \vec{X}_0(\vec{x}) ] \, \mathrm{d}^3 x.
\end{equation}
This integral can be interpreted as the scalar product (in the $L_2$
sense) between $\vec{x} - \vec{X}_0(\vec{x}_0,t)$ and $\rho(x) \delta
\vec{X}(\vec{x}_0)$. Hence the optimal solution, which should satisfy
$\delta I = 0$ for all $\delta \vec{X}_0$, is such that the
displacement $\vec{x} - \vec{X}_0(\vec{x}_0,t)$ (or equivalently
$\vec{X}(\vec{x}_0) - \vec{x}_0$) is orthogonal to all divergence-free
vector fields. This means that it is necessarily the gradient of a
potential, from which it follows that $\vec{X}(\vec{x}_0,t) =
\nabla_{x_0} \Phi (\vec{x}_0)$. Convexity follows from the observation
that the Lagrangian map $\vec{x}_0\mapsto\vec{X}$ has to
satisfy
\begin{equation}
  (\vec{x}_0 - \vec{x}_0^\prime)\cdot
  [\vec{X}(\vec{x}_0)-\vec{X}(\vec{x}_0^\prime)] \ge 0.
  \label{eq:monotonicity}
\end{equation}
Indeed, if that was not the case, one can easily check that any map
where the Lagrangian pre-image of a neighborhood of $\vec{x_0}$ and of
one of $\vec{x_0}^\prime$ are inverted would lead to a smaller
cost. Formulated in terms of potential maps, the relation
(\ref{eq:monotonicity}) straightforwardly implies convexity of
$\Phi$. This finishes the proof of equivalence between
Monge--Amp\`{e}re equation and the optimal transport problem with
quadratic cost.

The goal of reformulating reconstruction as an optimization problem is
mostly algorithmic. Once discretized, the problem of finding the
optimal map between initial and final positions amounts is equivalent
to solving a so-called assignment problem. An efficient method to deal
numerically with such problems is based on the auction algorithm
\cite{b81} and was used in \cite{bfhlmms03} with data stemming from
$N$-body cosmological simulations.
\begin{figure}[ht]
  \centerline{\includegraphics[width=0.35\textwidth]{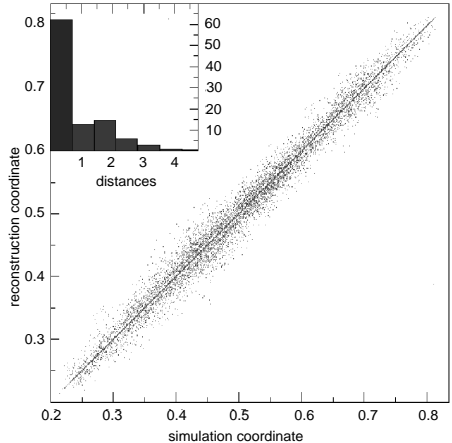}}
  \caption{Test of the MAK reconstruction for a sample of $N' =17,178$
    points from a $N$-body simulation (from~\cite{bfhlmms03}). The
    scatter diagram plots reconstructed versus true initial
    positions. The histogram inset gives the distribution (in
    percentages) of distances between true and reconstructed initial
    positions; the horizontal unit is the distance between two sampled
    points. The width of the first bin is less than unity to ensure
    that only exactly reconstructed points fall in it. More than sixty
    percent of the points are exactly reconstructed.}
  \label{fig:test_mak}
\end{figure}
As summarized in figure~\ref{fig:test_mak}, the MAK reconstruction
method leads to very promising results. More than 60\% of the discrete
points are assigned to their actual Lagrangian pre-image. Such a
number has to be compared with other reconstruction methods for which
the success rate barely exceed 40\% for the same data set.

Even if the mapping from initial to final positions is unique, the
peculiar velocities are not well defined except if we have some extra
knowledge of what is happening at intermediate times $0\le t^\prime
\le t$. Of course the density field $\rho(\vec{x}^\prime,t^\prime)$ is
unknown. However, there are trivial physical requirements. First the
two mass transport problems between $0$ and $t^\prime$ and between
$t^\prime$ and $t$ have both to be optimal. This means that one looks
for two Lagrangian maps, $\vec{X}_1$ from $0$ to $t^\prime$ and
$\vec{X}_2$ from $t^\prime$ to $t$ which are minimizing the respective
costs
\begin{eqnarray}
\mathcal{I}_1 &=& \int \| \vec{X}_1(\vec{x}_0) - \vec{x}_0\|^2
\rho_0\, \mathrm{d}^3x_0,\nonumber\\ \mathcal{I}_2 &=& \int \| \vec{x}
- \vec{X}_2^{-1} (\vec{x})\|^2 \rho(\vec{x},t) \,\mathrm{d}^3x.
\end{eqnarray}
The second physical requirement is that the composition of these two
optimal maps have to give the Lagrangian map between times $0$ and
$t$, namely $\vec{X}(\vec{x}_0,t) = \vec{X}_2(\vec{X}_1)$. Under these
two conditions there is equivalence between the optimal transport with
a quadratic cost and the Burgers dynamics supplemented by the
transport of a density field (see~\cite{bb00} for details).

%%%%%%%%%%%%%%%%%%%%%%%%%%%%%%%%%%%%%%%%%%%%%%
\section{Forced Burgers turbulence}
\label{s:force}

\subsection{Stationary r\'{e}gime and global minimizer}
\label{subs:stationary}

We consider in this section solutions to the forced Burgers
equation. As we have seen in section \ref{s:basictools}, the solution
in the limit of vanishing viscosity can be expressed at any time $t$
in terms of the initial condition at time $t_0$ through a variational
principle which consists in minimizing an action along particle
trajectories.  The statistically stationary r\'{e}gime toward which
the solution converges at large time can be studied assuming that the
by rejecting the initial time $t_0$ is at minus infinity. The solution
is then given by the variational principle
\begin{eqnarray}
  \Psi(\vec{x}, t) \!=\! -\!\inf_{\vec{\gamma}(\cdot)} \!\left\{
  \int_{-\infty}^t\! \left [\frac{1}{2} \|\dot{\vec{\gamma}}(s) \|^2
  \!-\! F(\vec{\gamma}(s), s) \right ] \! \mathrm{d}s\! \right \}\!,
  \label{eq:statvarprinc}
\end{eqnarray}
where the infimum is taken over all (absolutely continuous) curves
$\vec{\gamma}: (-\infty, t] \to \Omega$ such that $\vec{\gamma}(t) =
\vec{x}$.  In this setting, the action is computed over the whole half
line $(-\infty, t]$ and the argument of the infimum does not depend
anymore on the initial condition.  Of course, (\ref{eq:statvarprinc})
defines $\Psi$ up to an additive constant. This means that only the
differences $\Psi(\vec{x}, t) - \Psi(0, t)$ can actually be defined. A
trajectory $\vec{\gamma}$ minimizing (\ref{eq:statvarprinc}) is called
a \emph{one-sided minimizer}. It is easily seen from
(\ref{eq:statvarprinc}) that all the minimizers are solutions of the
Euler--Lagrange equation
\begin{equation}
  \ddot{\vec{\gamma}}(s) = -\nabla F(\vec{\gamma(s)}, s)\,,
  \label{eq:eullag2}
\end{equation}
where the dots denote time derivatives. This equation defines a
$2d$-dimensional (possibly random) dynamical system in the
position-velocity phase space $(\vec{\gamma},\dot{\vec{\gamma}})$. The
Lagrangian one-sided minimizers $\vec{\gamma}$ defined over the
half-infinite interval $(-\infty,t]$ play a crucial role in the
construction of the global solution and of the stationary r\'{e}gime.
Namely, a global solution to the randomly forced inviscid Burgers
equation is given by $\vec{v}(\vec{x}, t) = \dot{\vec{\gamma}}(t)$
where $\vec{\gamma}(t) = \vec{x}$. To prove that such half-infinite
minimizers exist, one has to take the limit $t_0\to-\infty$ for
minimizers defined on the finite time interval $[t_0,t]$. The
existence of this limit follows from a uniform bound on the absolute
value of the velocity $|\dot{\vec{\gamma}}|$ (see, e.g.,
\cite{ekms00}). Obtaining such a bound becomes the central problem for
the theory, as we shall now see.

When the configuration space $\Omega$ where the solutions live is
compact (bounded), one can expect the velocity of a minimizer to be
uniformly bounded. Indeed, in this case the displacement of a
minimizer for any time interval is then bounded by the diameter of the
domain $\Omega$, so that action minimizing trajectories cannot have
large velocities. For forcing potential that are delta-correlated in
time, it has been shown by E \textit{et al.}\/ \cite{ekms00} in one
dimension and by Iturriaga and Khanin \cite{ik01,ik03} in higher
dimensions that the minimizing problem (\ref{eq:statvarprinc}) has a
unique solution $\Psi$ with the following properties:
\begin{itemize}
\item $\Psi$ is the unique statistically stationary solution to the
  Hamilton--Jacobi equation (\ref{eq:visckpz}) in the inviscid limit
  $\nu\to0$;
\item $\Psi$ is almost everywhere differentiable with respect to the
  space variable $\vec{x}$;
\item $-\nabla\Psi$ uniquely defines a statistically stationary
  solution to the Burgers equation in the inviscid limit;
\item there exists a unique one-sided minimizer at those Eulerian
  positions $\vec{x}$ where the potential $\Psi$ is differentiable;
  the locations where $\Psi$ is not differentiable correspond to
  shocks.
\item There exists a unique minimizer $\vec{\gamma}^{\rm (g)}$ that
  minimizes the action calculated from $-\infty$ to any time $t$. It
  is called the \emph{global minimizer} (or \emph{two-sided
  minimizer}) and corresponds to the trajectory of a fluid particle
  that is never absorbed by shocks. Moreover, all one-sided minimizers
  are asymptotic to it as $s\to-\infty$.
\end{itemize}
All the properties above follow from the variational approach.  In
fact, the variational principle (\ref{eq:varprincvisc}) imply similar
statements in the viscous case.  Of course, when viscosity is positive
the unique statistically stationary solution is smooth. However, one
can show that the stationary distribution corresponding to such
solutions converges to inviscid stationary distribution in the limit
$\nu \to 0$~\cite{gikp05}. Although the variational proofs are
conceptual, general and simple, they are based on the fluctuation
mechanism and therefore do not give a good control of the rate of
convergence to the statistically stationary regime.  Exponential
convergence would follow from the hyperbolicity of the global
minimizer. Although one expects hyperbolicity holds in any dimension,
mathematically it is an open problem. At present a rigorous proof of
hyperbolicity is only available in dimension one~\cite{ekms00}.

The assumption of compactness of the configuration space $\Omega$ is
essential in the construction of the stationary r\'{e}gime. As we will
see in subsection \ref{subs:extended}, the situation is much more
complex in the non-compact case when for instance the solution is
defined on the whole space $\Omega=\rset^d$.

\subsection{Topological shocks}
\label{subs:topological}

\begin{figure}[ht]
  \centerline{\subfigure[\label{f:unwrap1d}]{
      \includegraphics[width=0.3\textwidth]{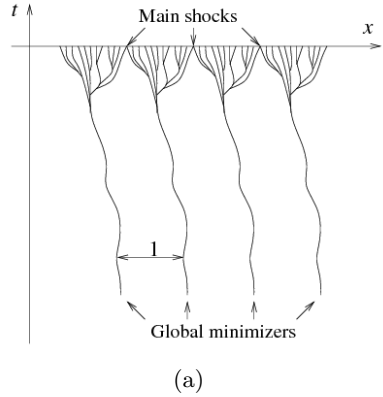}}}
   \centerline{\subfigure[\label{f:unwrap2d}]{
      \includegraphics[width=0.3\textwidth]{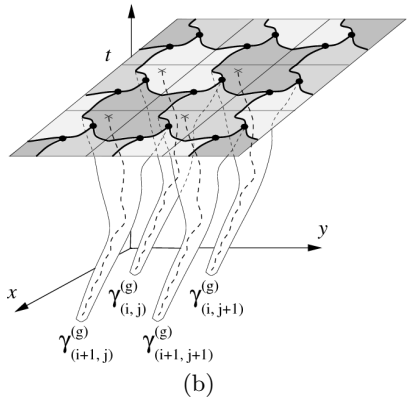}}}
  \caption{Space-time sketch of the unwraping of the periodic domain
  $\tset^d$ to the whole space $\rset^d$ for $d=1$ (a) and $d=2$ (b).}
\end{figure}

To introduce the notion of topological shock we first focus on the
one-dimensional case in a periodic domain, i.e.\ in $\Omega = \tset =
\rset/\zset$. If we ``unwrap'' at a given time $t$ the configuration
space to its universal cover $\rset$ (see figure~\ref{f:unwrap1d}), we
then obtain an infinite number of global minimizer $\gamma^{\rm
(g)}_k$, which at all time $s\le t$ satisfy $\gamma^{\rm (g)}_{k+1}(s)
= \gamma^{\rm (g)}_{k}(s) + 1$. All the one-sided minimizers converge
backward in time to one of these global minimizers. The
\emph{topological shock} (or \emph{main shock}) is defined as the set
of $x$ positions giving rise to several minimizers approaching two
successive replicas of the global minimizer. This particular shock is
also the only shock that has existed for all times.

This construction can easily be extended to higher dimensions
(see~\cite{bik02}). For this we unwrap the $d$-dimensional torus
$\tset^d$ to its universal cover, the full space $\rset^d$ (see
figure~\ref{f:unwrap2d} for $d=2$). Then, the different replicas of
the periodic domain define a lattice of global minimizers
$\vec{\gamma}^{\rm (g)}_{\vec{k}}$ parameterized by integer vectors
$\vec{k}$. The backward-in-time convergence on the torus of the
one-sided minimizers to the global minimizer implies that a minimizer
associated to a location $\vec{x}$ in $\rset^d$ at time $t$ will be
asymptotic to one of the global minimizer $\vec{\gamma}^{\rm
(g)}_{\vec{k}}$ of the lattice.  Hence, every position $\vec{x}$ which
has a unique one-sided minimizer is associated to an integer vector
$\vec{k}(\vec{x})$. This defines a tiling of space at time $t$. The
tiles $\mathcal{O}_{\vec{k}}$ are the sets of points whose associated
one-sided minimizers are asymptotic to the $\vec{k}$-th global
minimizer.  The boundaries of the $\mathcal{O}_{\vec{k}}$'s are the
\emph{topological shocks}. They are the locations from which at least
two one-sided minimizers approach different global minimizers on the
lattice.  Indeed, a point where two tiles $\mathcal{O}_{\vec{k}_1}$
and $\mathcal{O}_{\vec{k}_2}$ meet, has at least two one-sided
minimizers, one of which is asymptotic to $\vec{\gamma}^{(\rm
g)}_{\vec{k}_1}$ and another to $\vec{\gamma}^{(\rm
g)}_{\vec{k}_2}$. Of course, there are also points on the boundaries
where three or more tiles meet and thus where more than two one-sided
minimizers are asymptotic to different global minimizers.  For $d=2$
such locations are generically isolated points corresponding to the
intersections of three or more topological shock lines, while for
$d=3$, they form edges and vertices where shock surfaces meet. Note
that, generically, there exist other points inside
$\mathcal{O}_{\vec{k}}$ with several minimizers.  They correspond to
shocks of ``local'' nature because at these locations, all the
one-sided minimizers are asymptotic to the same global minimizer
$\vec{\gamma}^{(\rm g)}_{\vec{k}}$ and hence, to each other. In terms
of Lagrangian dynamics, the topological shocks play a role dual to
that of the global minimizer. Indeed, all the fluid particles are
converging backward-in-time to the global minimizer and are absorbed
forward-in-time by the topological shocks. For the transportation of
mass when we assume that the Burgers equation is supplemented by a
continuity equation for the mass density, all the mass concentrate at
large times in the topological shocks.

\begin{figure}[t]
  \centerline{\subfigure[\label{f:torus}]{
      \includegraphics[width=0.15\textwidth]{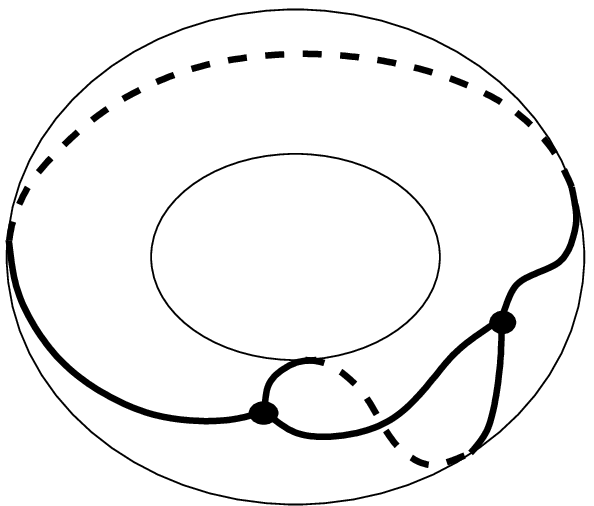}}}
  \vspace{-10pt}\centerline{\subfigure[\label{f:figpot}]{
      \includegraphics[width=0.4\textwidth]{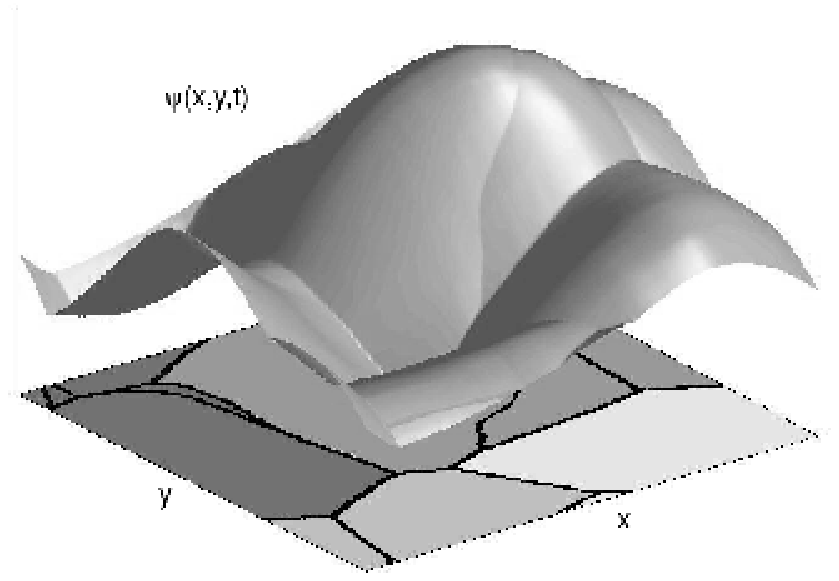}}}
  \caption{(a) Position of the topological shock on the torus; the two
    triple points are represented as dots. (b) Snapshot of the
    velocity potential $\psi(x,y,t)$ for $d=2$ in the statistical
    steady state, obtained numerically with $256^2$ grid points. Shock
    lines, corresponding to locations where $\psi$ is not
    differentiable, are represented as black lines on the bottom of
    the picture; the four gray areas are different tiles separated by
    the topological shocks; the other lines are local shocks.}
\end{figure}
The global structure of the topological shocks is related to the
various singularities generically present in the solution to the
Burgers equation that were detailed in section
\ref{subs:singularities}. Generically there are no locations
associated to more than $(d+1)$ minimizers. As one expects to see only
generic behavior in a random situation, the probability to have points
with more than $(d+1)$ one-sided minimizers is zero.  It follows that
there are no points where $(d+2)$ tiles $\mathcal{O}_{\vec{k}}$ meet,
which is an important restriction on the structure of the tiling. For
$d=2$ it implies that the tiling is constituted of curvilinear
hexagons. Indeed, suppose each tile $\mathcal{O}_{\vec{k}}$ is a
curvilinear polygon with $s$ vertices corresponding to triple points.
For a large piece of the tiling that consists of $N$ tiles, the total
number of vertices is $n_v \sim sN/3$ and the total number of edges is
$n_e \sim sN/2$.  The Euler formula implies that $1 = n_v - n_e + N
\sim (6-s)N/6$, and we necessarily have $s=6$, corresponding to an
hexagonal tiling. As shown in figure~\ref{f:torus}, this structure
corresponds on the periodicity torus $\tset^2$, to two triple points
connected by three shock lines that are the curvilinear edges of the
hexagon $O_{\vec{0}}$. The connection between the steady-state
potential and the topological shocks is illustrated numerically on
figure~\ref{f:figpot}. The different tiles covering the periodic
domain were obtained by tracking backward in time fluid particle
trajectories and by determining to which periodic image of the global
minimizer they converge.

\begin{figure}[htbp]
  \centerline{\includegraphics[height=2.5cm,width=2.5cm]
    {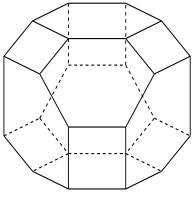}}
  \caption{Sketch of the simplest configuration of the topological
    shock in dimension $d=3$.}
  \label{f:tile3d}
\end{figure}

In dimensions higher than two, the structure of topological shocks is
more complicated. For instance it is not possible to determine in a
unique manner the shape of the polyhedra forming the tiling. However,
it has been shown by Matveev \cite{ma90} that for $d=3$ the minimum
polyhedra forming such tiling has 24 vertices and 36 edges and is
composed of 8 hexagons and 6 rectangles (see
figure~\ref{f:tile3d}). It is of interest to note that the structure
of topological shocks is in direct relation with the notions of
complexity and minimum spines of manifolds introduced by Matveev from
a purely topological viewpoint.

\bigskip
\noindent\textit{Algebraic characterization of the topological shock}

\begin{figure}[htbp]
  \centerline{\includegraphics[width=0.35\textwidth]{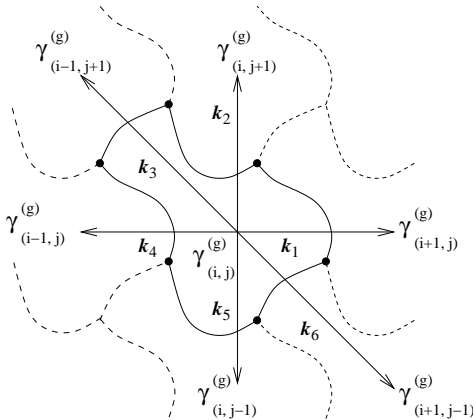}}
  \caption{The algebraic structure of the topological shock in
    dimension $d=2$ is determined by the indexes corresponding to
    immediate neighbors of the tiling considered.}
  \label{f:tile2d}
\end{figure}

In two dimensions, when periodic boundary conditions are considered,
very strong constraints are imposed on the structure of the solution.
In particular, the topology of the torus $\tset^2$ imply that the
topological shocks generically form a periodic tiling of $\rset^2$
with curvilinear hexagons. However, this tiling can be of various
algebraic types. Consider the tile $O_{\vec{0}}$ surrounded by its six
immediate neighbors $O_{\vec{k}_i}$, where the integer vectors
$\vec{k}_i$ are labeled in anti-clockwise order, $\vec{k}_1$ having
the smallest polar angle (see figure~\ref{f:tile2d}). It is easily
seen that the periodicity of the tiling implies
\begin{eqnarray}
  \vec{k}_3 = \vec{k}_2 - \vec{k}_1, && \ \vec{k}_4 = -\vec{k}_1, \ \
  \vec{k}_5 = -\vec{k}_2 \nonumber \\ && \mbox{and } \ \vec{k}_6 =
  \vec{k}_1 - \vec{k}_2,
  \label{eq:implyperiod}
\end{eqnarray}
so that the whole information on the algebraic structure of the tiling
is contained in the vectors $\vec{k}_1$ and $\vec{k}_2$ which form a
matrix $\mathcal{S}$ from the group ${\rm SL}(2,\zset)$ of $2\times2$
integer matrices with unit determinant. The matrix $\mathcal{S}$ gives
information on the number of times each shock line turns around the
torus before reconnecting to another triple point.  Figure~\ref{f:torus}
corresponds to the simplest case when $\mathcal{S}$ is the identity
matrix. When the forcing is stochastic, the matrix $\mathcal{S}$ is
random and stationary solutions to the two-dimensional Burgers
equation define a stationary distribution on ${\rm SL}(2,\zset)$.

\begin{figure}[htbp]
  \centerline{\includegraphics[width=0.45\textwidth]{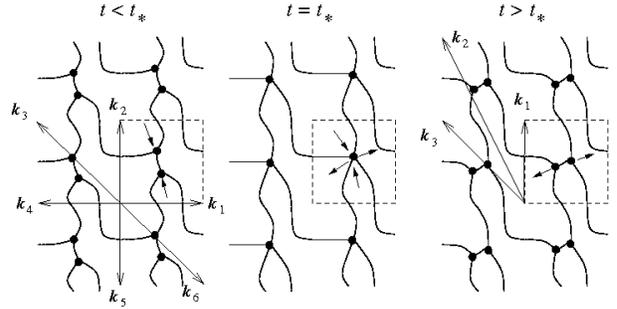}}
  \caption{\label{f:fliptile} Sketch of the tiling before, at the
    flipping time $t_*$ and after it. This example corresponds to a
    bifurcation from the matrix $\mathcal{S}_1=[^1_ 0~^0_1]$ to
    $\mathcal{S}_2=[^0_1~^{-1}_{\;\;\: 2}]$. The dashed boxes
    represent the periodicity domain $[0,1]^2$.}
\end{figure}

Certainly, topological shocks evolve in time and may change their
algebraic structure.  This happens through bifurcations (or
metamorphoses) described in section \ref{subs:singularities}. In two
dimensions, the generic mechanism which transforms the algebraic
structure of topological shocks is the merger of two triple
points. This metamorphosis is called the flipping bifurcation and
corresponds to the appearance at time $t_\star$ of an $A_1^4$
singularity in the solution associated to a position with four
minimizers. The mechanism transforming the algebraic structure of the
topological shock is illustrated in figure~\ref{f:fliptile}. Issues
such as the minimum number of flips needed to transform the matrix
$\mathcal{S}_1$ associated to the algebraic structure of the
topological shock to another matrix $\mathcal{S}_2$ are discussed in
in~\cite{an01}.

\subsection{Hyperbolicity of the global minimizer}

The nature of the convergence to a statistical steady state is
determined by the local properties of the global minimizer. The
hyperbolicity of this action-minimizing trajectory implies an
exponential convergence, so that the global picture of the solution is
reached very rapidly, after just a few turnover times.

Since the trajectory of the global minimizer is unique and can be
extended to arbitrary large times, it corresponds to an ergodic
invariant measure for the stochastic flow defined by the
Euler--Lagrange equation (\ref{eq:eullag2}). Conditioned by the random
force, this measure is simply the delta measure sitting at the
location
$(\vec{\gamma}^{\rm(g)}(0),\dot{\vec{\gamma}}^{\rm(g)}(0))$. By the
Oseledets ergodic theorem (see, e.g.\ \cite{p93}), $2d$ non-random
Lyapunov exponents can be associated to the global minimizer
trajectory. Since the flow is symplectic these non-random exponents
come in pairs with opposite signs. That is
\begin{equation}
 \lambda_1 \ge \cdots \ge \lambda_d \ge 0 \ge -\lambda_d \ge \cdots
 \ge -\lambda_1\,.
 \label{eq:deflyap}
\end{equation}
Hyperbolicity is defined as the non-vanishing of all these
exponents. Thus, the issue of hyperbolicity can be addressed in terms
of the backward-in-time convergence of the one-sided minimizers to the
global one or, better, in terms of forward-in-time dynamics. In the
latter case, this amounts to looking how fast Lagrangian fluid
particles are absorbed by shocks. For this we consider the set
$\Omega_{\rm reg}(T)$ of locations $\vec{x}$ such that the fluid
particle emanating from $\vec{x}$ at time $t=0$ survives, i.e.\ is not
absorbed by any shock, until the time $t=T$. The long-time shrinking
of $\Omega_{\rm reg}$ as a function of time is asymptotically governed
by the Lyapunov exponents. To ensure the absence of vanishing Lyapunov
exponents, it is sufficient to show that the diameter of $\Omega_{\rm
reg}(T)$ decays exponentially as $T\to+\infty$.

In one dimension, it has been shown in \cite{ekms00} that this is
indeed the case, and particularly that there exists positive constants
$\alpha$, $\beta$, $A$ and $B$ such that
\begin{equation}
  {\rm Prob}\left\{ {\rm diam}\, \Omega_{\rm reg}(T) \ge A{\rm
    e}^{-\alpha T} \right\} \le B{\rm e}^{-\beta T}\,.
  \label{eq:hyperbol}
\end{equation}
Unfortunately this proof of hyperbolicity is purely one-dimensional
and at present time there is no extension of this result to higher
dimensions.

\begin{figure}[ht]
  \centerline{\includegraphics[width=0.4\textwidth]
    {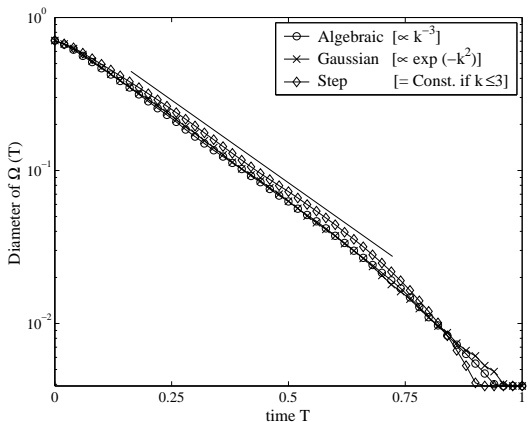}}
  \caption{\label{f:exponentialdiameter} Time evolution of the
    diameter of the Lagrangian set $\Omega(T)$ (points corresponding
    to the regular region) for three different types of forcing
    spectrum; average over 100 realizations and with $256^2$ grid
    points (from \cite{bik02}).}
\end{figure}

In two dimensions, the behavior of ${\rm diam}\, \Omega_{\rm reg}(T)$
at large times was studied numerically in \cite{bik02} by using the
fast Legendre transform described in section \ref{subs:numerics} and a
forcing that is a sum of independent random impulses concentrated at
discrete times. The ideas of this numerical method are related to the
Lagrangian structure of the flow. This easily permits to track
numerically the set $\Omega_{\rm reg}$ of regular Lagrangian
locations. As seen from figure~\ref{f:exponentialdiameter}, the
diameter of this set decays exponentially fast in time for three
different types of forcing, providing good evidence of the
hyperbolicity of the global minimizer for $d=2$.

Hyperbolicity of the global minimizer implies existence at any time
$t$ of two $d$-dimensional smooth manifolds in phase space
$(\vec{\gamma},\dot{\vec{\gamma}})$ that are invariant by the
Euler--Lagrange dynamics (\ref{eq:eullag2}): a stable (attracting)
manifold $\Gamma^{\rm (s)}(t)$ and an unstable (repelling) manifold
$\Gamma^{\rm (u)}(t)$, defined as the instantaneous location of
trajectories converging to the global minimizer forward in time and
backward in time, respectively.  Since all the minimizers converge
backward in time to the global minimizer, the graph in the
position-velocity phase space $(\vec{x}, \vec{v})$ of the solution in
the statistical steady state is made of pieces of the unstable
manifold $\Gamma^{\rm (u)}(t)$ with discontinuities along the shocks
lines or surfaces. In other words, shocks appear as jumps between two
different folds of the unstable manifold. The smoothness of the
unstable manifold is an important property; for instance, it implies
that when $d=2$, the topological shock lines are smooth curves.

\begin{figure}[t]
  \centerline{\includegraphics[width=0.4\textwidth]
    {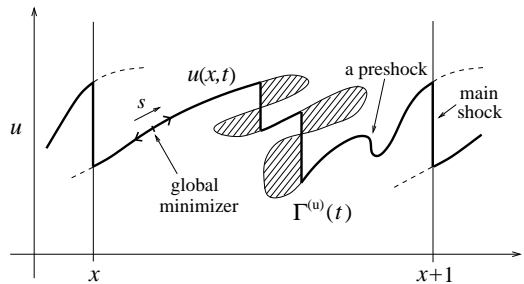}}
  \caption{\label{f:vinstable} Sketch of the unstable manifold for
  $d=1$ in the $(x,v)$ plane. Shock locations ($A_1^2$ singularities)
  are obtained by applying Maxwell rules to the loops. A preshock
  ($A_3$ singularity) is represented; it corresponds to the formation
  of a loop in the manifold. The velocity profile which is the actual
  solution to the Burgers equation is represented as a bold line.}
\end{figure}

In one dimension, where hyperbolicity is ensured, the main shock
corresponds to a jump between the right branch and the left branch of
the unstable manifold. Its position can be obtained geometrically
after observing that the area $b$ covered by the unstable manifold,
once the latter is cut by the main shock, should be equal to the first
integral of motion which is conserved, i.e.\
\begin{equation}
  b = \int v(x,t)\,\mathrm{d}x = \int v_0(x)\,\mathrm{d}x\,.
  \label{eq:firstconserved}
\end{equation}
The other shocks (or secondary shocks that have existed only for a
finite time) cut through the double-fold loops of the unstable
manifold (see figure~\ref{f:vinstable}). Their locations can be
obtained by a Maxwell rule applied to those loops. Indeed, the
difference of the two areas defined by cutting such a loop at some
position $x$ is equal to the difference of actions of the two
trajectories emanating from the upper and lower locations and, thus,
vanishes at the shock location. We will see in section
\ref{s:timeperiodic} that this construction of the solution is also
valid when the forcing is periodic in time, problem which can be
related to Aubry--Mather theory relative to
commensurate-incommensurate phase transitions.

The above geometrical construction of the solution has much in common
with that appearing in the unforced problem. Indeed, as we have seen
in section \ref{subs:geometrical}, when $F=0$ the solution can be
obtained geometrically by considering in the $(\vec{x}, \vec{v})$
space, the Lagrangian manifold defined by the position and the
velocity of the fluid particles at a given time. This analogy gives
good ground predicting that some universal properties associated to
the unforced problem will still hold in the forced case, as we will
indeed see in section \ref{s:1Dstatistics}. Another instance concerns
transport of mass in higher dimension. We have seen in section
\ref{subs:density} that, for the unforced case, large but finite mass
densities are localized near boundaries of shocks (``kurtoparabolic''
singularities) contributing power-law tails with the exponent $-7/2$
to the probability density function of the mass density. When a force
is applied the smoothness of the unstable manifold associated to the
global minimizer should lead to the same universal law.

\subsection{The case of extended systems}
\label{subs:extended}

So far, we have discussed the global structure of the solution to the
forced Burgers equation with periodic boundary conditions.  Is is
however of physical interest to understand instances when the size of
the domain is much larger than the typical length scale of the
forcing. In this section, we will focus on describing, in the
one-dimensional case, the singular structure of the solution in
unbounded domains. Based on the formalism of \cite{bk03}, we achieve
this goal by considering a spatially periodic forcing with a
characteristic scale $L_f$ much smaller than the system size $L$. More
precisely, for a fixed size $L$ we consider the stationary r\'{e}gime
corresponding to the limit $t\to\infty$ and then study the limit
$L\to\infty$ by keeping constant the energy injection rate (i.e.\ the
$\mathcal{L}^2$ norm of the forcing grows like $L$).

\begin{figure}[t]
  \centerline{\subfigure[\label{f:replicas}]{
      \includegraphics[width=0.45\textwidth]{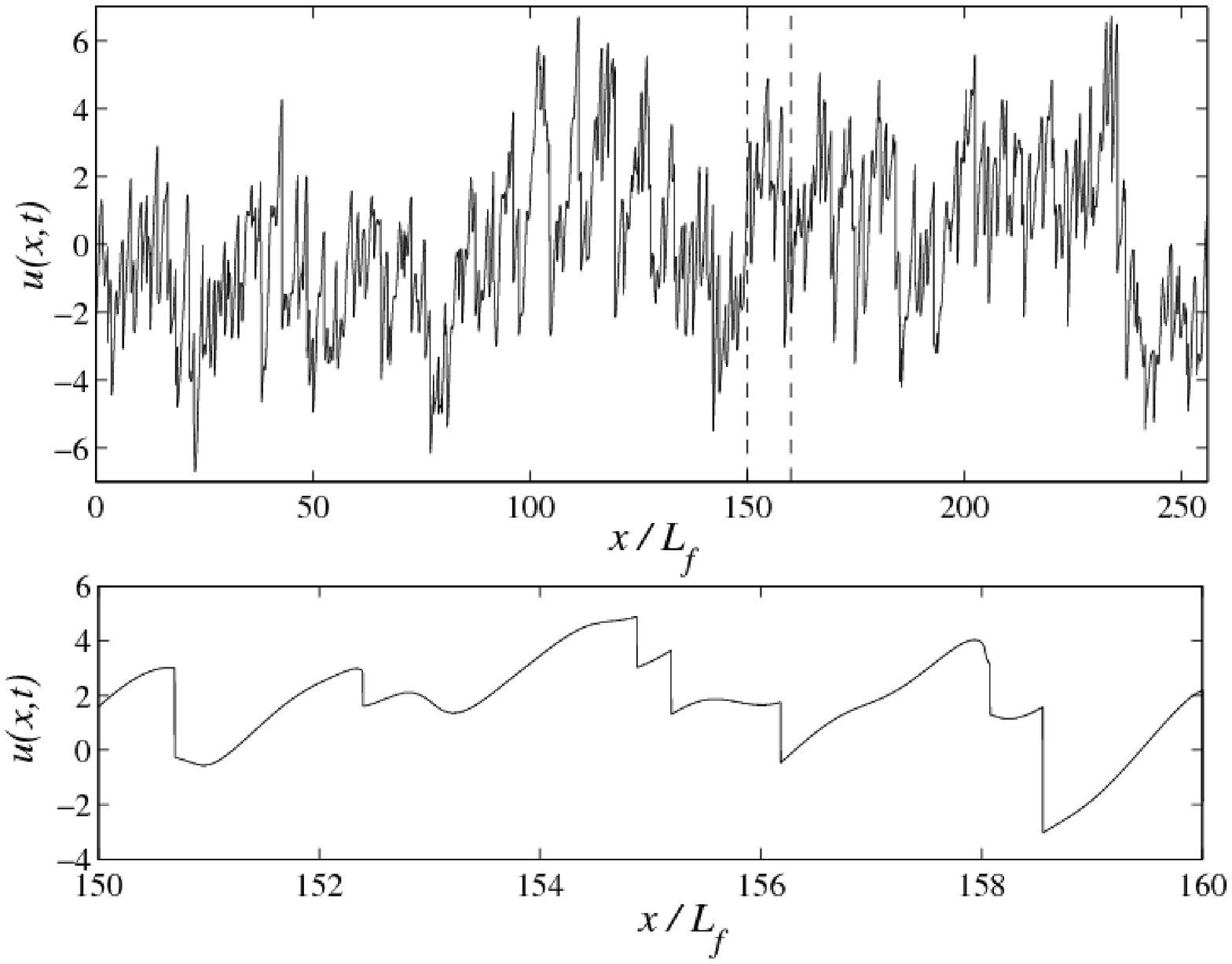}}}
  \vspace{-10pt}\centerline{\subfigure[\label{f:minimizersextended}]{
      \includegraphics[width=0.35\textwidth]
      {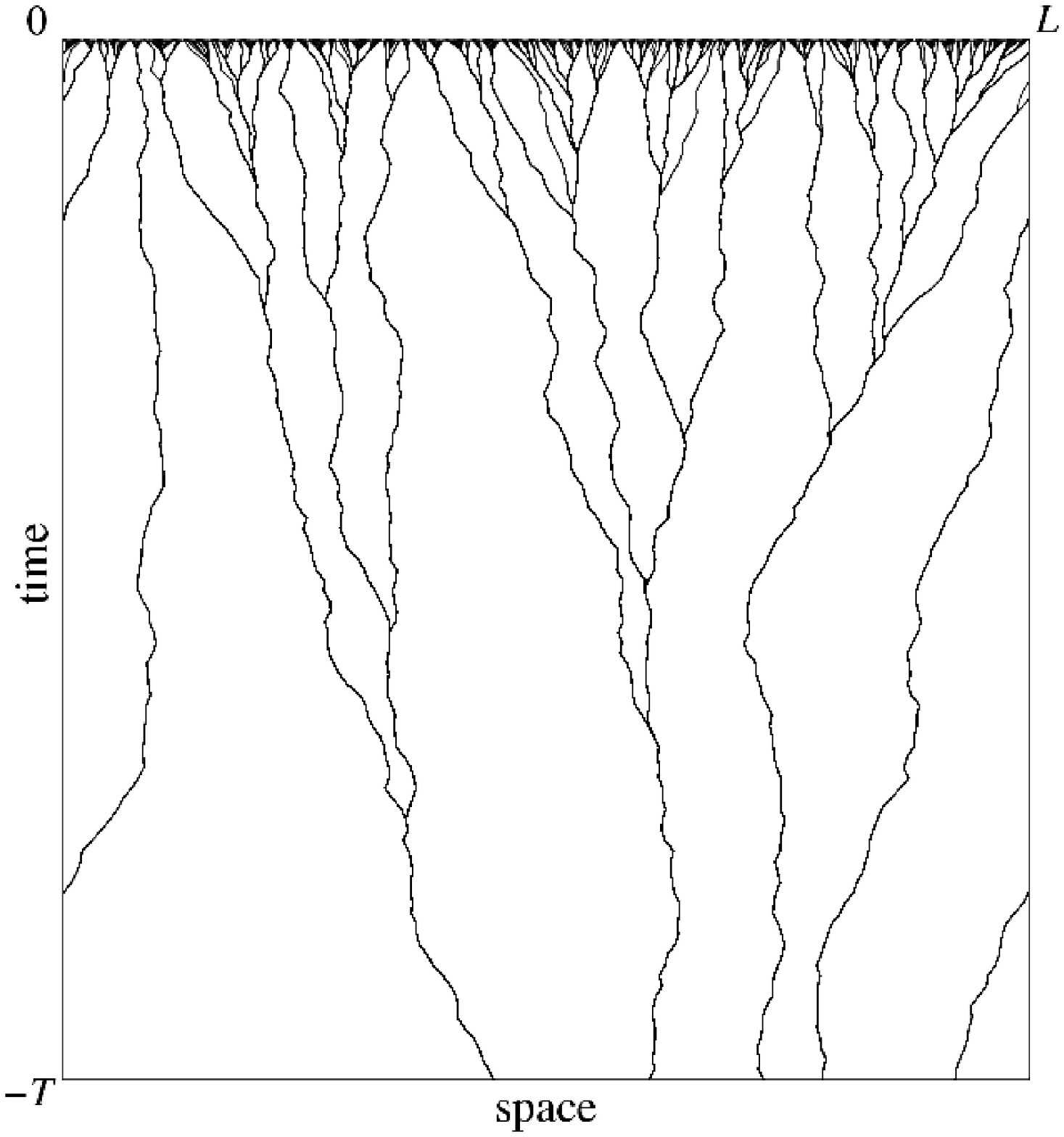}}}
  \caption{(a) Upper: snapshot of the velocity field for $L =
    256\,L_f$. Lower: zoom of the field in a interval of length
    $10\,L_f$. (b) Minimizing trajectories in space time for $L =
    256\,L_f$ and over a time interval of length $T=100$}
\end{figure}
\begin{figure}[t]
  \centerline{\subfigure[\label{f:shocksextended}]{
      \includegraphics[width=0.4\textwidth]{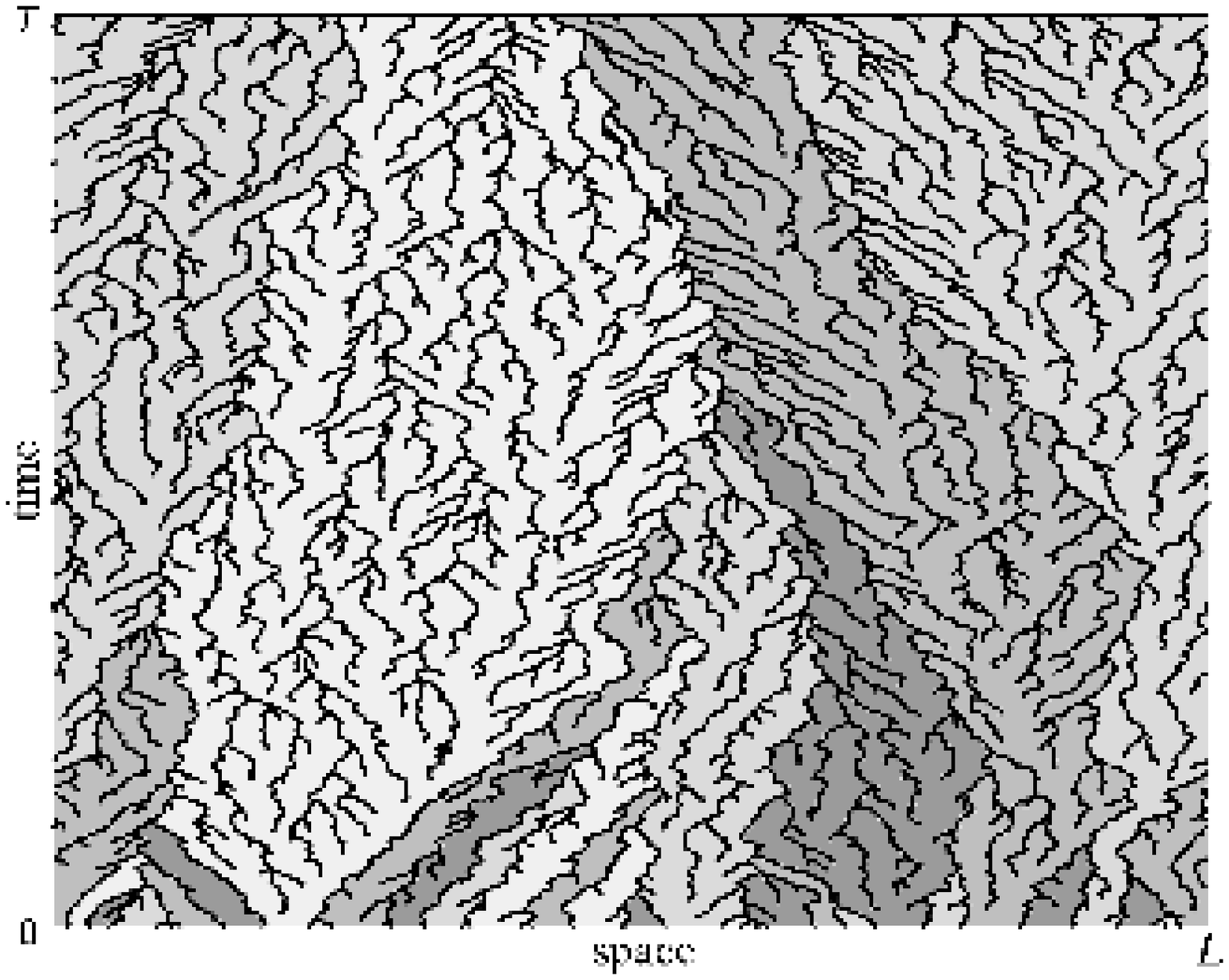}}}
  \centerline{\subfigure[\label{f:defTmini}]{
      \includegraphics[width=0.3\textwidth] {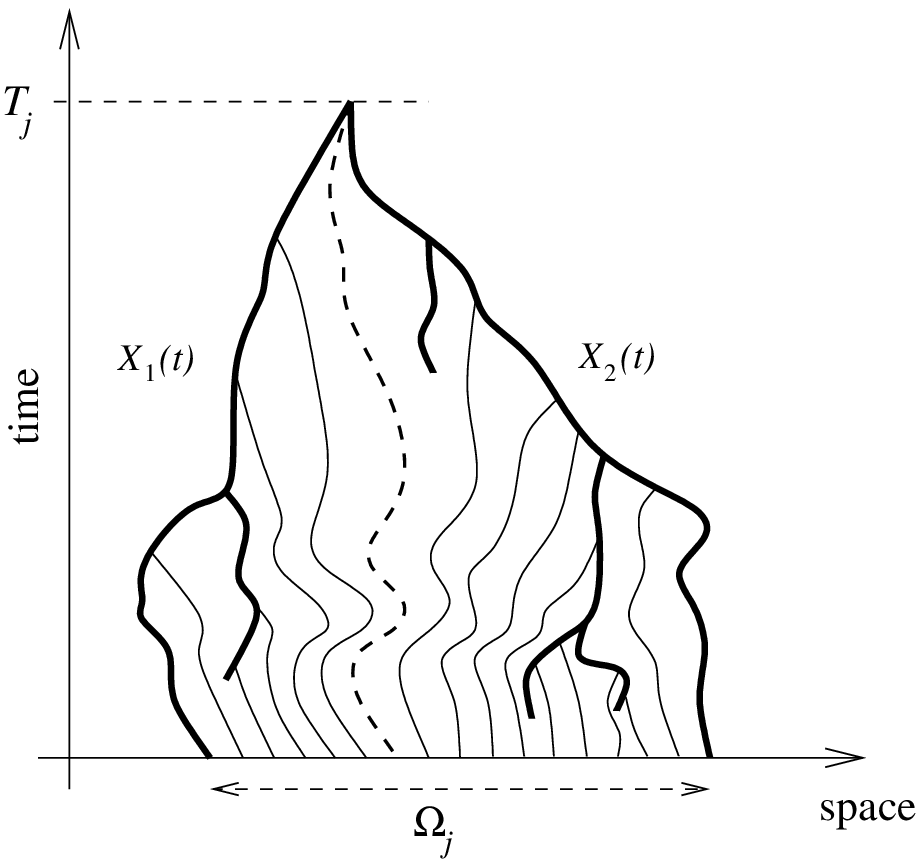}}}
  \caption{(a) Shock trajectories for aspect ratio $L/L_f = 32$ and
  with $T=10$. The different gray areas correspond to the space-time
  domains associated to the different smooth pieces $\Omega_j$ of the
  velocity field at time $t=0$. (b) Sketch of the space-time evolution
  of a given smooth piece $\Omega_j$ located between two shock
  trajectories $X_1(t)$ and $X_2(t)$ that merge at time $T_j$.}
\end{figure}

In order to get an idea of the behavior of the solution, the limit of
infinite aspect ration $L/L_f$ was investigated numerically in
\cite{bk03}. As seen from figure~\ref{f:replicas} numerical
observations suggest that at any time in the statistical steady state,
the shape of the velocity profile is similar to the order-unity aspect
ratio problem, duplicated over independent intervals of size $L_f$. In
particular, when tracking backward in time the trajectories of fluid
particles the minimizers converge to each other in a very non-uniform
way. Figure~\ref{f:minimizersextended} shows that the minimizers form
different branches, which are converging to each other backward in
time; in space time a tree structure is obtained. As shown in
figure~\ref{f:shocksextended} a similar behavior is observed for
shocks.

The velocity field at a given time $t$, consists of smooth pieces
separated by shocks. Let us denote by $\{\Omega_j\}$ the set of
intervals in $[0,L)$, on which the solution $u(\cdot,t)$ is smooth.
The boundaries of the $\Omega_j$'s are the shocks positions. Each of
these shocks is associated to a root-like structure formed by the
trajectories of the various shocks that have merged at times less than
$t$ to form the shock under consideration (see
figure~\ref{f:shocksextended}). This root-like structure contains the
whole history of the shock and in particular its age (i.e.\ the length
of the deeper branch of the root structure).  Indeed, if the root has
a finite depth, the shock considered has only existed for a finite
time. A \emph{$T$-global shock} is defined as a shock whose associated
root is deeper than $-T$. They can alternatively be defined
geometrically by considering the leftmost and the rightmost minimizer
associated to it. After tracing them backward for a sufficiently long
time, these two minimizers are getting close and eventually converge
to each other exponentially fast (see
figure~\ref{f:minimizersextended}). For a $T$-global shock, the time
when the two minimizers are getting within a distance smaller than the
forcing correlation length $L_f$ is larger than $T$. As we have seen
in section~\ref{subs:topological}, the existence in one dimension of a
main shock in the spatially periodic situation follows from a simple
topological argument.  The main shock can also be defined as the only
shock that has existed forever in the past. It is hence infinitely
old, contrary to all other shocks, all of them being created at a
finite time and having a finite age. When the periodicity condition is
dropped, the main shock disappears and it is useful to consider the
$T$-global shocks that mimic the behavior of a main shock over time
scales larger than $T$.

One can dually define \emph{$T$-global minimizers}. All the smoothness
intervals $\Omega_j$ defined above, except that which contains the
global minimizer, will be entirely absorbed by shocks after a
sufficient time (see figure\ \ref{f:defTmini}).  For each of these
pieces, one can define a life-time $T_j$ as the time when the last
fluid particle contained in this piece at time $t$ enters a shock. It
corresponds to the first time for which the shock located on the left
of this smooth interval at time $t$ merges with the shock located on
the right. When the life-time of such an interval is greater than $T$,
the trajectory of the last surviving fluid particle is here called a
$T$-global minimizer. Note that, when $T\to\infty$, the number of
$T$-main shocks and of $T$-global minimizers is one, recovering
respectively the notions of main-shock and of two-sided minimizer.

\begin{figure}[htbp]
  \centerline{\includegraphics[width=0.4\textwidth] {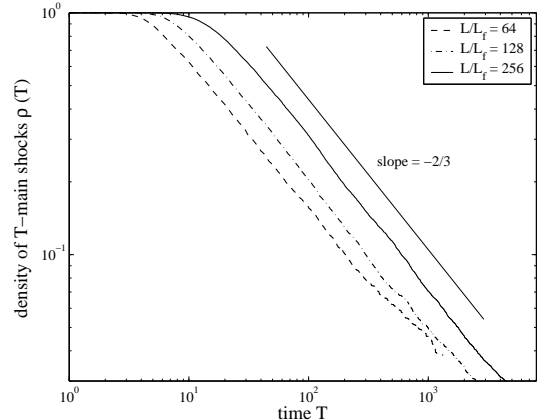}}
  \caption{\label{f:Tdensity} Density of $T$-main shocks as a function
    of $T$ for three different system sizes $L/L_f=64$, $128$ and
    $256$; average over 100 realizations. Lower inset: local scaling
    exponent.}
\end{figure}
Hence, at a given instant $t$, and for any timelag $T$, the spatial
domain $[0,L)$ contains a certain number of $T$-objects. We define
their spatial density as being the number of such objects, averaged
with respect to the forcing realizations, divided by the size of the
domain $L$. The density $\rho(T)$ of $T$-main shocks was investigated
numerically in~\cite{bk03} for the kicked case by using a two-step
method: first, the simulation was run until a large time $t$ for which
the statistically stationary r\'{e}gime is reached; secondly, each
shock present at time $t$ was tracked backward-in-time down to the
instant of its creation, giving an easy way to characterize the
density $\rho(T)$. It is seen in figure~\ref{f:Tdensity} that, for
three different aspect ratios $L/L_f$, the density $\rho(T)$ displays
a power-law behavior $\rho(T)\propto T^{-2/3}$ for the intermediate
time asymptotics $L_f/u_{\rm rms} \ll T \ll L/u_{\rm rms}$.

We now present a simple phenomenological theory aiming to explain the
scaling exponent $2/3$. We consider the solution at a fixed time
($t=0$, for instance). Denote by $\ell(T)$ the typical spatial
separation scale for two nearest $T$-global shocks. Obviously,
$\ell(T) \sim 1/{\rho(T)}$.  The mean velocity of the spatial segment
of length $\ell$ is given by
\begin{equation}
  b_\ell= \frac{1}{\ell}\int_{[y,\,y+\ell]}u(x,0)\,\mathrm{d}x
  \label{eq:minvelL}
\end{equation}
Since the expected value $\langle u(x,0) \rangle=0$, and that the
integral in (\ref{eq:minvelL}) is over an interval of size much larger
than the forcing correlation length, it is equivalent to a sum of
independent centered random variables and scales as the Brownian
motion. Hence, for large $\ell$ one has the following asymptotics
\begin{equation}
  \int_{[y,\,y+\ell]}u(x,0)\,\mathrm{d}x \sim \sqrt{\ell},
  \label{eq:fluctmeanvelocity}
\end{equation}
which gives $b_\ell \sim \ell^{-1/2}$ for mean velocity fluctuations.
Consider now the rightmost minimizer corresponding to the left
$T$-global shock and the leftmost minimizer related to the right one.
Since there are no $T$-global shocks in between, it follows that the
two minimizers we selected get close to each other backward-in-time
around times of the order of $-T$. This means that the
backward-in-time displacement of a spatial segment of length $O(\ell)$
is itself $O(\ell)$ for time intervals of the order of $T$. The
corresponding displacement is given as the sum of two competing
behaviors: the first, which can be understood as a drift induced by
the local mean velocity $b_\ell$, is due to the mean velocity
fluctuations and is responsible for a displacement $\propto b_\ell T$;
the second contribution is due to a standard diffusive scale $\propto
T^{1/2}$ expressing the diffusive behavior of the minimizing
trajectories. Taking into account both terms we obtain
\begin{equation}
  \ell \sim B_1 T\ell^{-1/2} + B_2 T^{1/2},
  \label{eq:Tglob}
\end{equation}
where $B_1$ and $B_2$ are numerical constants. It is easy to see that
the dominant contribution comes from the first term. Indeed, if the
second term were to dominate, then $\ell$ would be much larger than
$T$, which contradicts~(\ref{eq:Tglob}). Hence, one has $\ell \sim B_1
T\ell^{-1/2}$, leading to the scaling behavior
\begin{equation}
  \ell(T) \propto T^{2/3}, \qquad \rho(T) \propto T^{-2/3}.
  \label{eq:scales}
\end{equation}
As we have already discussed, $T$-global shocks are shocks older than
$T$. Denote by $p(A)$ the probability density function (PDF) for the
age of shocks. More precisely, $p(A)$ is a density in the stationary
r\'{e}gime of a probability distribution of the age $A(t)$ of a shock,
say the nearest to the origin. It follows from~(\ref{eq:scales}) that
the probability of shocks whose age is larger than $A$ decays like
$A^{-2/3}$; this implies the following asymptotics for the PDF $p(A)$:
\begin{equation}
  p(A) \propto A^{-5/3}.
  \label{eq:scales1}
\end{equation}

\begin{figure}[htbp]
  \centerline{\includegraphics[width=0.45\textwidth]
    {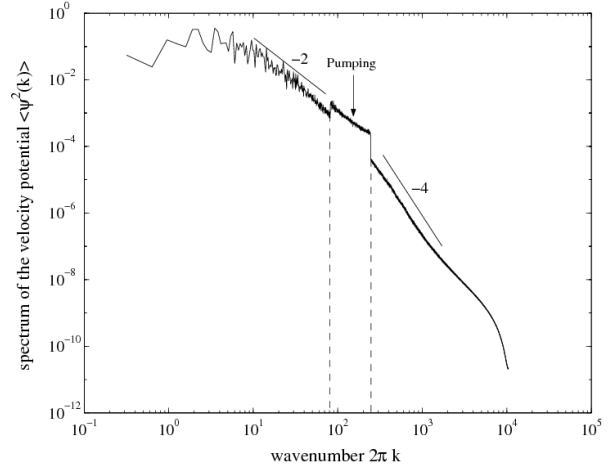}}
  \caption{Spectrum $\langle\hat\psi^2(k)\rangle$ of the velocity
    potential in the stationary r\'{e}gime for the aspect ratio
    $L/L_f=128$. This spectrum contains two power-law ranges: at
    wavenumbers $k\gg L/L_f$, the traditional $\propto k^{-4}$
    inertial range connected to the presence of shocks in the solution
    and, for $k\ll L/L_f$, an ``inverse cascade'' $\propto k^{-2}$
    associated to the large-scale fluctuations of $\psi$}
  \label{fig:spectrum128}
\end{figure}

Actually, the power-law behavior of the density $\rho(T)$ of
$T$-global shocks can be interpreted in term of an inverse cascade in
the spectrum of the solution (although there is no conserved
energy-like quantity). Indeed, the fluctuations
(\ref{eq:fluctmeanvelocity}) of the mean velocity suggest that, for
large-enough separations $\ell$, the velocity potential increment
scales like
\begin{equation}
  \left | \psi(x+\ell,\,t)-\psi(x,\,t)\right| \propto \ell^{1/2}.
  \label{potincr}
\end{equation}
This behavior is responsible for the presence of an intermediate
power-law range with exponent $-2$ in the spectrum of the velocity
potential at wavenumbers smaller than the forcing scale (see figure\
\ref{fig:spectrum128}).
In order to observe the $k^{-2}$ range at small wavenumbers, the
spectrum of the forcing potential must decay faster than $k^{-2}$;
otherwise the leading behavior is non-universal but depends on the
functional form of the forcing correlation.

The one-dimensional randomly forced Burgers equation in an unbounded
domain has been studied in \cite{hk03} with a different type of
forcing: it was assumed that the forcing potential has at any time its
global maximum and its global minimum in a prescribed compact region
of space. It was proven that with these particular settings the
statistically stationary r\'{e}gime exists and is very similar to that
arising in compact domains. In particular, there exists a unique
global minimizer located in a finite spatial interval for all times
and all other minimizers are asymptotic to it in the limit
$t\to-\infty$. The main idea behind considering such type of forcing
potential is to ensure that the potential energy plays a dominant role
in comparison with the kinetic (elastic) term in the action. This
leads to effective compactification and allows estimates on the
velocities of fluid particles. As we already mentioned in
section~\ref{subs:stationary}, these estimates are very important and
pave the way to the construction of the whole theory of the
statistically stationary r\'{e}gime.

Note finally that it was shown in \cite{kks03} that for special cases
of forcing potentials $F(x,t)$, the velocity of a minimizers can be
arbitrarily large. More specifically, one can construct pathological
forcing potentials such that minimizers are accelerated and reach
infinite velocities. Randomness is of course expected to prevent such
a type of non-generic blow-up.

%%%%%%%%%%%%%%%%%%%%%%%%%%%%%%%%%%%%%%%%%%%%%%
\section{Time-periodic forcing}
\label{s:timeperiodic}

This section is devoted to the study of the solutions to the
one-dimensional Burgers equation with time-periodic forcing.  In this
case many of the objects we have discussed above can be constructed
almost explicitly: the global minimizer, the main shock etc.  Also, a
mathematical analysis is then much simpler.  For instance,
hyperbolicity of the global minimizer follows immediately from first
principles. Finally, the case of time-periodic forcing is directly
related to the Aubry-Mather theory as we explain below.

\subsection{Kicked Burgers turbulence}

We shall be concerned here with the initial-value problem for the
one-dimensional Burgers equation when the force is concentrated in
Dirac delta functions at discrete times:
\begin{equation}
f(x,t) = \sum_j f_j(x)\,\delta(t-t_j),
\label{kickforce}
\end{equation}
where both the ``impulses'' $f_j(x)$ and the ``kicking times'' $t_j$
are prescribed (deterministic or random). The kicking times are
ordered and form a finite or infinite sequence. The impulses $f_j(x)$
are always taken spatially smooth, i.e.\ acting only at large
scales. The general scheme we are presenting below holds for any
sequence of impulses $f_j(x)$ and kicking time. Later on we shall
assume that they define a time-periodic forcing. The precise meaning
we ascribe to the Burgers equation with such forcing is that at time
$t_j$, the solution $u(x,t)$ changes discontinuously by the amount
$f_j(x)$
\begin{equation}
u(x,t_{j+})= u(x,t_{j-})+f_j(x),
\label{discon}
\end{equation}
while, between $t_{j+}$ and $t_{(j+1)-}$ the solution evolves according to
the unforced Burgers equation.

We shall also make use of the formulation in terms 
of the velocity potential $\psi(x,t)$ and the force potentials $F_j(x)$
\begin{equation}
u(x,t) = - \partial_x \psi(x,t),\qquad f_j(x)= -
\frac{\mathrm{d}}{\mathrm{d}x} F_j(x).
\label{psiF}
\end{equation}
The velocity potential satisfies
\begin{eqnarray}
&&\partial_t \psi = \frac{1}{2} (\partial_x \psi)^2 +
\nu \partial_{xx} \psi + \sum_j F_j(x)\,\delta(t-t_j),
\label{BEP}\\
&&\psi(x,t_0)=\psi_0(x),
\label{initpsi}
\end{eqnarray}
where $\psi_0(x)$ is the initial potential.

Using the variational principle we obtain the following ``minimum
representation'' for the potential in the limit of vanishing viscosity
which relates the solutions at any two times $t>t'$ between which no
force is applied:
\begin{equation}
\psi (x,t) = -\min_y \left[\frac{(x-y)^2 } {2(t-t')}- \psi(y,t')\right].
\label{MAXunforced}
\end{equation}
As before, when $t'$ is the initial time, the position $y$ which
minimizes (\ref{MAXunforced}) is the Lagrangian coordinate associated
to the Eulerian coordinate $x$. The map $y\mapsto x$ is called the
Lagrangian map. By expanding the quadratic term it is easily shown
that the calculation of $\psi(\cdot,t)$ from $\psi(\cdot,t')$ is
equivalent to a Legendre transformation. For details, see
\cite{saf92,vdfn94}.

We now turn to the {\em forced\/} case with impulses applied at the
kicking times $t_j$. Let $t_{J(t)}$ be the last such time before $t$.
Using (\ref{MAXunforced}) iteratively between kicks and changing the
potential $\psi(y,t_{j+1})$ discontinuously by the amount $F_{j+1}(y)$
at times $t_{j+1}$, we obtain
\begin{equation} 
\psi(x,t) = -\!\!\!\!\min_{\{y_j\}_{j_0\le j\le J}}\!\!\!\!
\left[\mathcal{A}(\{y_j\};x,t;j_0)) - \psi_0(y_{j_0})\right],
\label{MAXkicked}
\end{equation}
\begin{eqnarray} 
\mathcal{A}(\{y_j\};x,t;j_0)&& \equiv \frac{(x-y_J)^2}{2(t-t_J)}
\nonumber \\ +&& \!\sum_{j=j_0}^{J-1}\left[
\frac{(y_{j+1}-y_j)^2}{2(t_{j+1}-t_j)} - \!F_{j+1}(y_{j+1}) \right]\!,
\label{defA}
\end{eqnarray}
where $A(j_0;x,t;\{y_j\})$ is called the action.  We shall assume that
the force potential and the initial condition are periodic in the
space variable and the period is taken to be unity. This assumption is
very important for the discussion below.
 
For a given initial condition at $t_{j_0}$ we next define a
``minimizing sequence'' associated to $(x,t)$ as a sequence of $y_j$'s
($j=j_0,j_0+1,\ldots, J(t)$) at which the right-hand side of
(\ref{MAXkicked}) achieves its minimum.  Differentiating the action
(\ref{defA}) with respect to the $y_j$'s one gets necessary conditions
for such a sequence, which can be written as a sequence of
(Euler--Lagrange) maps
\begin{eqnarray}
v_{j+1} &&= v_j + f_j(y_j),\label{map1}\\ y_{j+1} &&= y_j +
v_{j+1}(t_{j+1}-t_j) \nonumber \\ &&= y_j + (v_j +
f_j(y_j))(t_{j+1}-t_j),
\label{map2} 
\end{eqnarray}
where 
\begin{equation}
v_j \equiv \frac{y_j-y_{j-1}}{t_j-t_{j-1}}.
\label{defvj}
\end{equation}
These equations must be supplemented by the initial and final conditions: 
\begin{eqnarray}
v_{j_0} &=& u_0(y_{j_0}),\label{mapinit}\\
x &=& y_J + v_{J+1}(t-t_J).
\label{mapfin}
\end{eqnarray}
It is easily seen that $u(x,t)= v_{J+1} = (x-y_J)/(t-t_J)$. Observe
that the ``particle velocity'' $v_j$ is the velocity of the fluid
particle which arrives at $y_j$ at time $t_j$ and which, of course,
has remained unchanged since the last kick (in Lagrangian
coordinates). Equation (\ref{map1}) just expresses that the particle
velocity changes by $f_j(y_j)$ at the the kicking time $t_j$.

Note that (\ref{map1})-(\ref{map2}) define an area-preserving and
(explicitly) invertible map.

As in the case of continuous-in-time forcing we can formulate the
Burgers equation in the half-infinite time interval $(-\infty, t]$
without fully specifying the initial con\-di\-tion $u_0(x)$ but only
its (spatial) mean value $\la u\ra \equiv \int _0 ^1
u_0(x)\mathrm{d}x$.

The construction of the solution in a half-infinite time interval is
done by extending the concept of minimizing sequence to the case of
dynamics starting at $t_0=-\infty$. For a half-infinite sequence
$\{y_j\}$ ($j\le J$), let us define the action
$\mathcal{A}(\{y_j\};x,t;-\infty)$ by (\ref{defA}) with
$j_0=-\infty$. Such a half-infinite sequence will be called a
``minimizer'' (or ``one-sided minimizer'') if it minimizes this action
with respect to any modification of a finite number of
$y_j$'s. Specifically, for any other sequence $\{{\hat y_j}\}$ which
coincides with $\{y_j\}$ except for finitely many $j$'s (i.e.\ ${\hat
y_j}=y_j$, $j\le J-k, k\ge 0$), we require
\begin{equation}
A(\{\hat y_j\};x,t;J-k)\ge A(\{y_j\};x,t;J-k).
\label{inegalaction}
\end{equation}

Of course, the Euler--Lagrange relations (\ref{map1})-(\ref{map2})
still apply to such minimizers. Hence, if for a given $x$ and $t$ we
know $u(x,t)$, we can recursively construct the minimizer 
$\{y_j\}$ backwards in time by using the inverse of
(\ref{map1})-(\ref{map2}) for all $j<J$ and the final condition -- now
an initial condition -- (\ref{mapfin}) with $v_{J+1} = u(x, t)$. This is
well defined except where $u(x,t)$ has a shock and thus more than one
value. 

One way to construct minimizers is to take a sequence of initial
conditions at different times $t_0\to -\infty$. At each such time some
initial condition $u_0(x)$ is given with the only constraint that it
have the same prescribed value for $\la u\ra$. Then, (finite)
minimizing sequences extending from $t_0$ to $t$ are constructed for
these different initial conditions. This sequence of minimizing
sequences has limiting points (sequences themselves) which are
precisely minimizers (E {\it et al}.\ 1998). The uniqueness of such
minimizers, which would then imply the uniqueness of a solution to the
Burgers equation in the time interval $]-\infty,t]$, can only be shown
by using additional assumptions, for example for the case of random
forcing or when the forcing is time-periodic.

If $\la u\ra=0$, the sequence $\{y_j\}$ minimizes the action
$\mathcal{A}(\{y_j\};x,t;-\infty)$ in a stronger sense. Consider any
sequence $\{{\hat y_j}\}$ such that, for some integer $P$ we have
${\hat y_j} = y_j+P$, $j\le J-k, k\ge0$ and which differs arbitrarily
from $\{y_j\}$ for $j>J-k$. (In other words, in a sufficiently remote
past the hatted sequence is just shifted by some integer multiple of
the spatial period.)  We then have
\begin{equation}
\mathcal{A}(\{\hat y_j\};x,t;-\infty)\ge
\mathcal{A}(\{y_j\},x,t;-\infty).
\label{inegalperiodic}
\end{equation}
Indeed, for $\la u\ra=0$, the velocity
potential for any initial condition is itself periodic.  In this case
a particle can be considered as moving on the circle $S^1$ and its
trajectory is a curve on the space-time cylinder. The $y_j$'s are now
defined modulo~1 and can be coded on a representative $0\le y_j<
1$. The Euler--Lagrange map (\ref{map1})-(\ref{map2}) is still valid
provided (\ref{map2}) is defined modulo~1. 

The condition of minimality implies now that $y_j$ and $y_{j+1}$ are
connected by the shortest possible straight segment. It follows that
$\vert v_{j+1}\vert = \rho(y_j, y_{j+1})/(t_{j+1}-t_j)$, where $\rho$
is the distance on the circle between the points $y_j, y_{j+1}$,
namely $\rho(a,b)\equiv \min \{|a-b|, 1 -|a-b|\}$.  Hence, the action
$\mathcal{A}$ can be rewritten in terms of cyclic variables:
\begin{eqnarray}
\mathcal{A}(\{y_j\};x,t;-\infty) && = \frac{\rho ^2(x,y_J)}{2(t-t_J)}
\nonumber \\ + \sum _{j<J}&& \left [\frac{\rho ^2(y_{j+1}, y_j)}
{2(t_{j+1}-t_j)} - F_{j+1}(y_{j+1})\right ].
\label{actionz}
\end{eqnarray}
The concept of ``global minimizers'' can be defined in a usual
way. Namely, global minimizers correspond to one-sided minimizers that
can be continued to a bilateral sequence $\{y_j, -\infty < j <+\infty
\}$ while keeping the minimizing property. Such global minimizers
correspond to trajectories of fluid particles that, from $t=-\infty$
to $t=+\infty$, have never been absorbed in a shock.  As before we
define a ``main shock'' as a shock which has always existed in the
past.

From now on we shall consider exclusively the case where the kicking
is periodic in both space and time. Specifically, we assume that 
the force in the Burgers equation is given by
\begin{eqnarray}
f(x,t) &=& g(x)\sum_{j=-\infty}^{+\infty}
 \delta(t-jT),\label{forceperiodic}\\ g(x)&\equiv&
 -\frac{\mathrm{d}}{\mathrm{d}x}G(x),
\label{defg}
\end{eqnarray}
where $G(x)$, the kicking potential, is a deterministic function of
$x$ which is periodic and sufficiently smooth (e.g.\ analytic) and
where $T$ is the kicking period. The initial potential $\psi_{\rm
init}(x)$ is also assumed smooth and periodic. This implies that the
initial velocity integrates to zero over the period. The case where
this assumption is relaxed will be considered later in connection with
the Aubry--Mather theory.

The numerical experiments of~\cite{bfk00} reported here have been made with
the kicking potential
\begin{equation}
G(x) = \frac{1}{3}\sin 3x +\cos x,
\label{defG}
\end{equation}
and a kicking period $T\!=\!1$. Other experiments were done with (i)
$G(x)= -\cos x$ and (ii) $G(x)= (1/2) \cos (2x) -\cos x$. The former
potential produces a single shock and no preshock. As a consequence it
displays no $-7/2$ law in the PDF of gradients.  The latter potential
gives essentially the same results as reported hereafter but has an
additional symmetry.  To avoid non-generic behaviors that could result
from this symmetry, it was chosen to focus on the forcing potential
given by~(\ref{defG}).

The number of collocation points chosen for such simulations was
mostly $N_x=2^{17}\approx 1.31\times 10^5$, with a few simulations
done at $N_x=2^{20}$ (for the study of the relaxation to the periodic
r\'{e}gime presented below). Since the numerical method allows going
directly to the desired output time (from the nearest kicking time)
there is no need to specify a numerical time step. However, in order
to perform temporal averages, e.g.\ when calculating PDF's or
structure functions, without missing the most relevant events (which
can be sharply localized in time) sufficiently frequent temporal
sampling is needed. The total number of output times $N_t\approx
1000$, is thus chosen such that the increment between successive
output times is roughly the two-thirds power of the mesh (this is
related to the cubic structure of preshocks, see section
\ref{subs:singularities}).

\begin{figure}
\centerline{\includegraphics[width=0.45\textwidth]{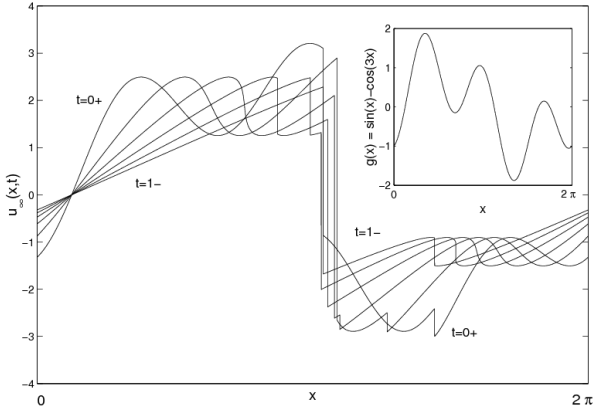}}
\caption{Snapshots of the velocity for the unique time-periodic
solution corresponding to the kicking force $g(x)$ shown in the upper
inset; the various graphs correspond to six output times equally
spaced during one period. The origin of time is taken at a
kick. Notice that during  each period, two new shocks are born  and two
mergers occur. (From~\cite{bfk00}.)}
\label{f:global-evol}
\end{figure}

Figure~\ref{f:global-evol} shows snapshots of the time-periodic
solution at various instants. 
\begin{figure}
\centerline{\includegraphics[width=0.45\textwidth]{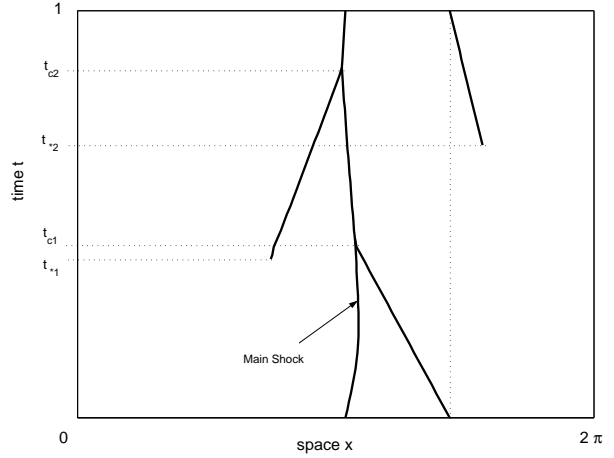}}
\caption{Evolution of shock positions during one period. The
beginnings of lines correspond to births of shocks (preshocks) at
times $t_{\star 1}$ and $t_{\star 2}$; shock mergers take place at
times $t_{c1}$ and $t_{c2}$. The ``main shock'', which survives for
all time, is shown with a thicker line.}
\label{f:shock-position}
\end{figure}
It is seen that shocks are always present (at least two) and that at
each period two new shocks are born at $t_{\star 1}\approx 0.39$ and
$t_{\star 2}\approx 0.67$. There is one main shock which remains near
$x=\pi$ and which collides with the newborn shocks at $t_{c1}\approx
0.44$ and $t_{c2}\approx 0.86$.  Figure~\ref{f:shock-position} shows
the evolution of the positions of shocks during one period.

\begin{figure}
\centerline{\includegraphics[width=0.45\textwidth]{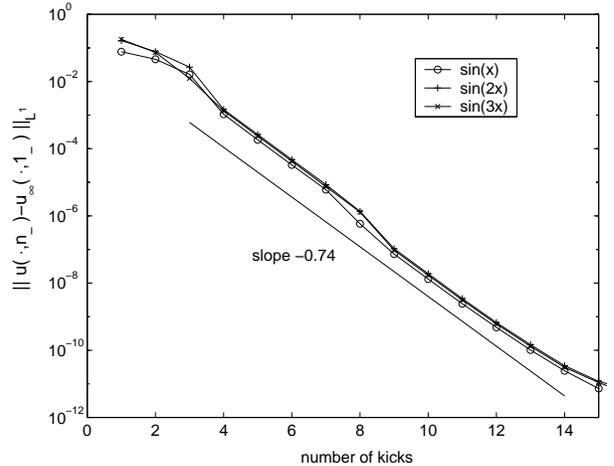}}
\caption{Exponential relaxation to a time-periodic solution for three
different initial velocity data as labeled. The horizontal axis gives
the time elapsed since $t=0$. (From~\cite{bfk00}.)}
\label{f:relax-expo}
\end{figure}

It was found that, for all initial conditions $u_0(x)$ used, the
solution $u(x,t)$ relaxes exponentially in time to a unique function
$u_\infty(x,t)$ of period 1 in time.  Figure~\ref{f:relax-expo} shows
the variation of $\int_0^{2\pi}|u(x,n_-)-
u_\infty(x,1_-)|\,\mathrm{d}x/(2\pi)$ for three different initial
conditions as a function of the discrete time $n$.

The phenomenon of exponential convergence to a unique space- and
time-periodic solution is something quite general: whenever the
kicking potential $G(x)$ is periodic and analytic and the initial
velocity potential is periodic (so that the mean velocity $\la u\ra$
=0 at all times), there is exponential convergence to a unique
piecewise analytic solution. This can be proved rigorously (see
Appendix to \cite{bfk00}) in the case when the functions $G(x)$ have a
unique point of maximum with a non-vanishing second derivative (Morse
generic functions). Here, we just explain the main ideas of the proof
and give some additional properties of the unique solution.

One very elementary property of solutions is that, for any initial
condition of zero mean value, the solution after at least one kick
satisfies
\begin{equation}
  |u(x,t)|\le (1/2) + \max_x |\mathrm{d}G(x)/\mathrm{d}x|.
\end{equation}
Indeed, at a time $t=n_-$ just before any kick we have $x= y+
u(x,n_-)$ where $y$ is the position just after the previous kick of
the fluid particle which goes to $x$ at time $n_-$. It follows from
the spatial periodicity of the velocity potential that the location
$y$ which minimizes the action is within less than half a period from
$x$. Thus, $|u(x,n_-)|\le 1/2$. The additional $\max_x
|\mathrm{d}G(x)/\mathrm{d}x|$ term comes from the maximum change in
velocity from one kick. Hence the solution is bounded. Note that if
the spatial and temporal periods are $L$ and $T$, respectively, the
bound on the velocity becomes $L/(2T) + \max_x
|\mathrm{d}G(x)/\mathrm{d}x|$.

The convergence at large times to a unique solution can be understood
in terms of the two-dimensional conservative (area-preserving)
dynamical system defined by the Euler--Lagrange map
(\ref{map1})-(\ref{map2}). By construction, we have
$u(x,1_+)=\hat{u}(x) -\mathrm{d}G(x)/\mathrm{d}x$, where $\hat{u}(x)$
is the solution of the unforced Burgers equation at time $t=1_-$ from
the initial condition $u(x)$ at time $t=0_+$. The map $u\mapsto
\hat{u}(x) +g(x)$, where $g(x)\equiv -\mathrm{d}G(x)/\mathrm{d}x$,
here denoted $B_g$, solves the kicked Burgers equation over a time interval
one.  The problem is to show that the iterates $B_g^{n}u_0$ converge
as $n\to \infty$ to a unique solution.

If it were not for the shocks it would suffice to consider the
two-dimensional Euler--Lagrange map.  Note that, for the case of
periodic kicking, this map has an obvious fixed point $P$, namely
$(x=x_c,v=0)$, where $x_c$ is the unique point maximizing the kicking
potential.  It is easily checked that this fixed point is an unstable
(hyperbolic) saddle point of the Euler--Lagrange map with two
eigenvalues $\lambda = 1 + c +\sqrt{c^2 + 2c}$ and $1/\lambda$, where
$c=-\partial ^2_{xx} G(x_c)/2$.

\begin{figure}[t]
\centerline{\subfigure[\label{f:stable-instable}]
  {\includegraphics[width=0.2\textwidth]{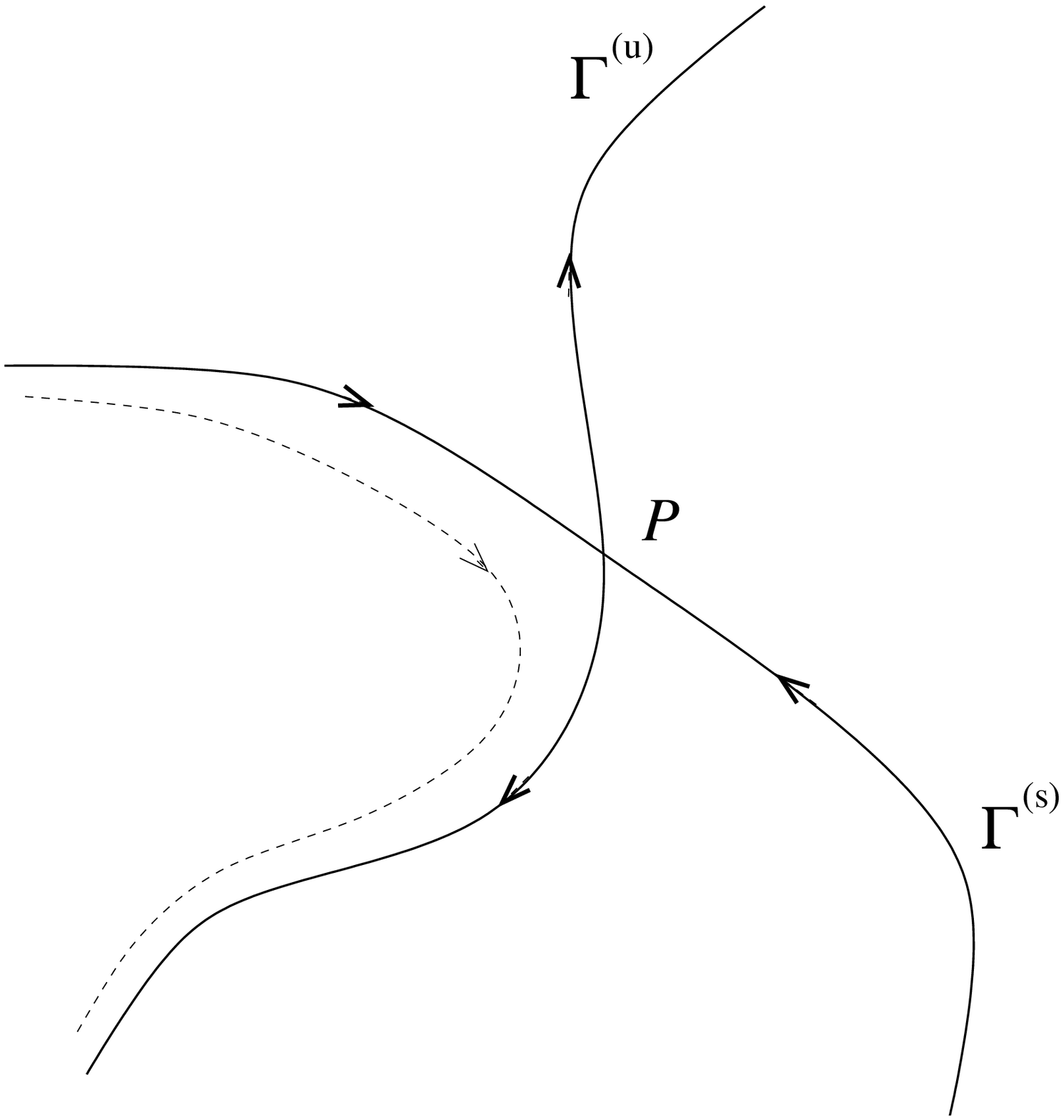}}}
\centerline{\subfigure[\label{f:unstable}]{
  \includegraphics[width=0.4\textwidth]{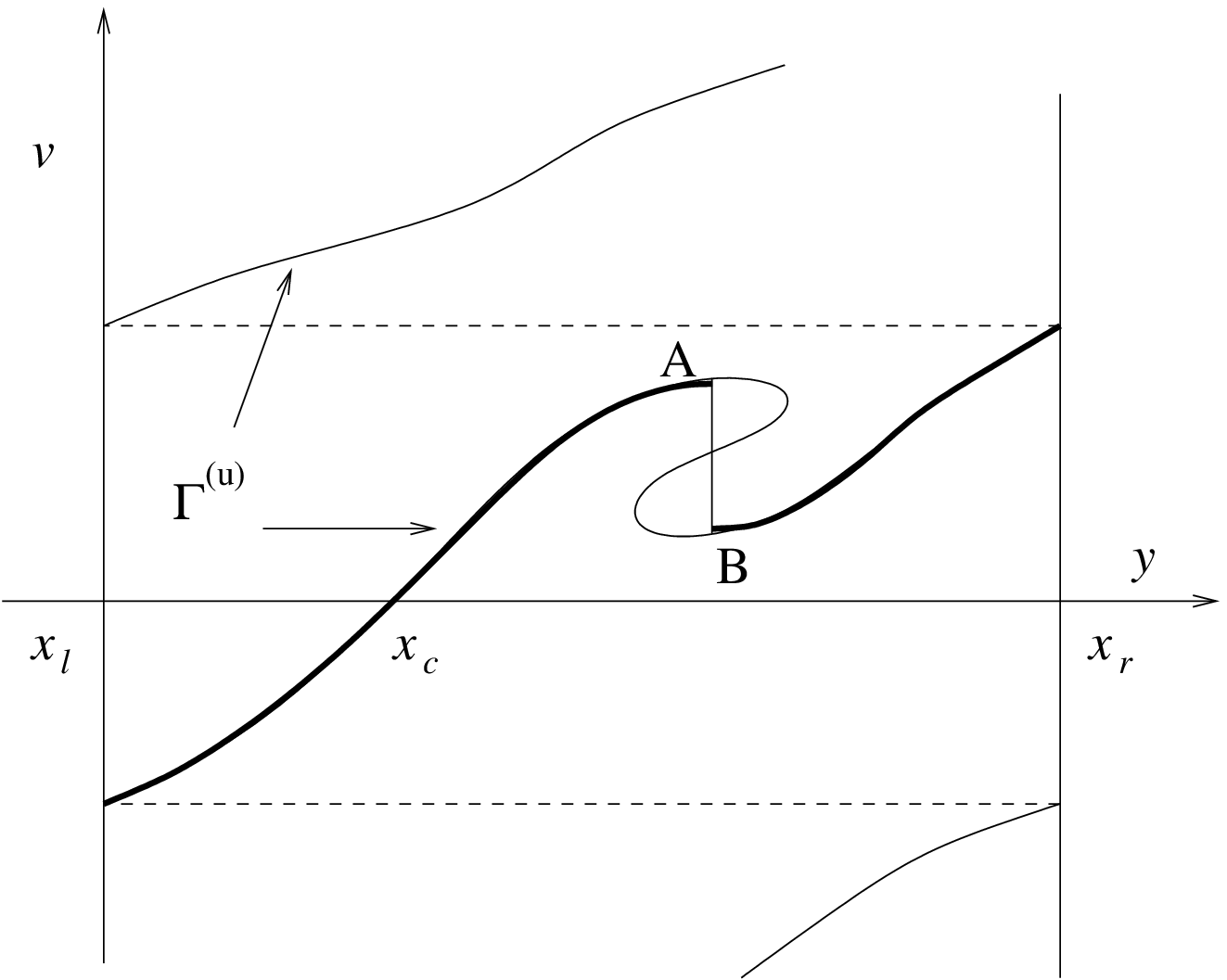}}}
\caption{(a) Sketch of a hyperbolic fixed point $P$ with stable
  ($\Gamma^{\rm (s)}$) and unstable ($\Gamma ^{\rm (u)}$)
  manifolds. The dashed line gives the orbit of successive iterates of
  a point near the stable manifold. (b) Unstable manifold $\Gamma
  ^{\rm (u)}$ on the $(x,v)$-cylinder (the $x$-coordinate is defined
  modulo~1) which passes through the fixed point $P=(x_c,0)$. The bold
  line is the graph of $u_\infty(x,1_-)$. The main shock is located at
  $x_l=x_r$. Another shock at $x_1$ corresponds to a local zig-zag of
  $\Gamma ^{\rm (u)}$ between A and B.}
\end{figure}
Like for any two-dimensional map with a hyperbolic fixed point, there
are two curves globally invariant by the map which intersect at the
fixed point. The first is the stable manifold $\Gamma ^{\rm (s)}$,
i.e.\ the set of points which converge to the fixed point under
indefinite iteration of the map; the second is the unstable manifold
$\Gamma ^{\rm (u)}$, i.e.\ the set of points which converge to the
fixed point under indefinite iteration of the inverse map, as
illustrated in figure~\ref{f:stable-instable}.  Any curve which
intersects the stable manifold transversally (at the intersection
point, the two curves are not tangent to each other) will, after
repeated applications of the map, be pushed exponentially against the
unstable manifold at a rate determined by the eigenvalue $1/\lambda$.
In the language of Burgers dynamics, the curve in the $(x,v)$ plane
defined by an initial condition $u_0(x)$ will be mapped after time $n$
into a curve very close to the unstable manifold. In fact, for the
case studied numerically, $1/\lambda\approx 0.18$ is within one
percent of the value measured from the exponential part of the graph
shown in figure~\ref{f:relax-expo}. Note that if the initial condition
$u_0(x)$ contains the fixed point, the convergence rate becomes
$\left(1/\lambda\right)^2$ (even higher powers of $1/\lambda$ are
possible if the initial condition is tangent to the unstable
manifold).

The fixed point $P$ is actually a very simple global minimizer: $(y_j
=x_c,\, v_j=0)$ for all positive and negative $j$'s. It follows indeed
by inspection of (\ref{actionz}) that any deviation from this
minimizer can only increase the action; actually, this trajectory
minimizes both the kinetic and the potential part of the action. Note
that the corresponding fluid particle is at rest forever and will
never be captured by a shock (it is actually the only particle with
this property).  It is easy to see that any minimizer is attracted
exponentially to such a global minimizer as $t\to -\infty$. Thus, any
point $(y_j,v_j)$ on a minimizer belongs to the {\em unstable
manifold\/} $\Gamma ^{\rm (u)}$ and, hence, any regular part of the
graph of the limiting solution $u_\infty(x)$ belongs to the unstable
manifold $\Gamma ^{\rm (u)}$. This unstable manifold is analytic but
can be quite complicated. It can have several branches for a given $x$
(see figure~\ref{f:unstable}) and does not by itself define a
single-valued function $u_\infty(x)$. The solution has shocks and is
only piecewise analytic. Consideration of the minimizers is required
to find the position of the shocks in the limiting solution: two
points with the same $x$ corresponding to a shock, such as A and B on
figure~\ref{f:unstable} should have the same action.

\begin{figure}[t]
\centerline{\includegraphics[width=0.3\textwidth]{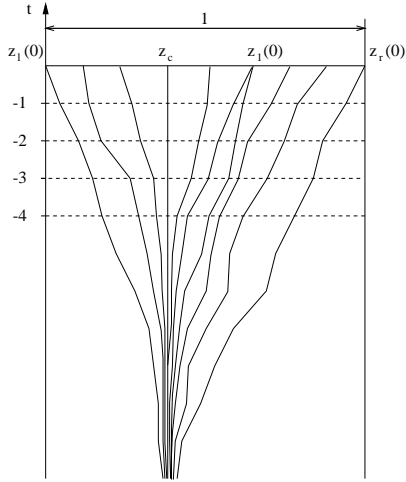}}
\caption{Minimizers (trajectories of fluid particles) on the
$(x,t)$-cylinder. Time starts at $-\infty$. Shock locations at $t=0_-$
are characterized by the presence of two minimizers (an instance is at
$x_1$).  The main shock is at $x_l=x_r$. The fat line $x=x_c$ is the
global minimizer.}
\label{f:main}
\end{figure}
Finally, we give the geometric construction of the main shock, the
only shock which exists for an infinite time.  Since the eigenvalue
$\lambda$ is positive, locally, minimizers which start to the right of
$x_c$ approach the global minimizer from the right, and those which
start to the left approach it from the left. Take the rightmost and
leftmost points $x_r$ and $x_l$ on the periodicity interval such that
the corresponding minimizers approach the global minimizer from the
right and left respectively (see figure~\ref{f:main}). These points
are actually identical since there cannot be any gap between them that
would have minimizers approaching the global minimizer neither from
the right nor the left. The solution $u_\infty(x)$ has then its main
shock at $x_l=x_r$.

%%%%%%%%%%%%%%%%%%%%%%%%%%%%%%%%%%%%%%%%%%%%%%
\subsection{Connections with Aubry--Mather theory}
\label{subs:aubry-mather}

\begin{figure}[htbp]
  \centerline{\includegraphics[width=0.4\textwidth]
    {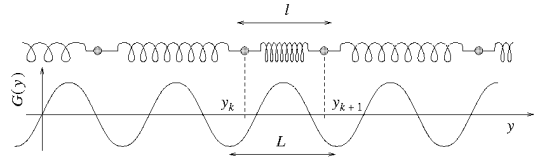}}
  \caption{Sketch of the Frenkel--Kontorova model for the equilibrium
  states of an atom chain in a periodic potential.}
  \label{fig:frenkelkontorova}
\end{figure}

In the previous subsection, the study of the solutions to the
periodically kicked Burgers equation was limited to initial conditions
with a vanishing spatial average $b$. With a non-vanishing mean
velocity $b$, which in the forced case cannot be eliminated by a
Galilean invariance, many of the properties of the solutions described
above are still valid. However the action now depends on $b$. Global
minimizers $\{y_j^{\rm (g)},\, j\in\zset\}$ exist in this case as
well. However generically they are not unique and do not correspond to
fixed points of the Euler--Lagrange map (\ref{map1})-(\ref{map2}). A
global minimizer now minimizes the action
\begin{eqnarray}
  \mathcal{A}_\infty (\{y_k\}) &= &\mathcal{A}
  (\{y_k\};+\infty;-\infty) \nonumber \\ &=&\!\!
  \sum_{k=-\infty}^{+\infty} \!\!\left[
  \frac{1}{2T}(y_{k+1}\!-\!y_k\!-\!b)^2 \!\!-\! G(y_{k+1})
  \right]\!. \label{eq:defactionbneq0}
\end{eqnarray}
This action is exactly the potential energy associated to an infinite
chain of atoms linked by elastic springs and embedded in a periodic
potential, problem known as the Frenkel--Kontorova model \cite{fk39}.
The parameter $b$ represents the equilibrium length $l$ of the springs
and the spatial period $L$ of the external potential (see
figure~\ref{fig:frenkelkontorova}) is equal to 1.
A global minimizer of (\ref{eq:defactionbneq0}) represents an
equilibrium configuration of this system. The properties of this
equilibrium, or ground states are determined by the competition
between two tendencies: on the one hand the atoms tend to stabilize at
those locations where the potential is minimum; on the other hand, the
springs tend to maintain them at a fixed distance of each other. When
$b=0$ this competition disappears and the equilibrium is given by $y_k
= x_c$, where $x_c$ is the location at which $G$ attains its global
minimum. For $b\neq0$, the situation is more delicate and the
structure of the ground states involves, as we shall now see, a
problem of commensurate-incommensurate transition. The properties of
ground states were studied in great details by Aubry \cite{a83} and
Mather \cite{m82}.

\begin{figure}[htbp]
  \centerline{\includegraphics[width=0.4\textwidth]
    {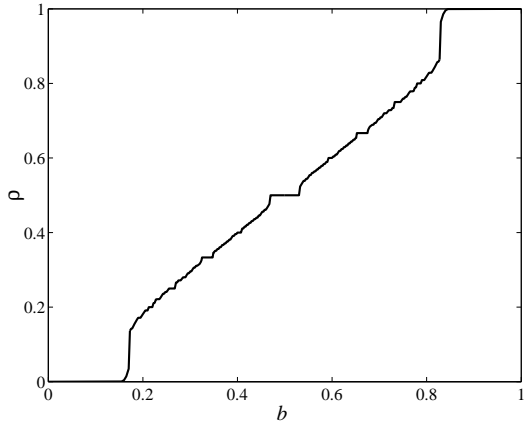}}
  \caption{Rotation number $\rho$ as a function of the spatial mean of
  the velocity $b$ for the standard map. }
  \label{fig:devilstaircase}
\end{figure}

The relations between the Burgers equation with a time-periodic
forcing and Aubry--Mather theory were discussed for the first time
in~\cite{jkm99} and in \cite{ekms00}. The theory was further developed
in \cite{e99,s99}.  For integer values of $b$, the global minimizer is
trivially associated to the fixed point $(x,v) = (x_c, b)$ of the
Euler--Lagrange map (\ref{map1})-(\ref{map2}), which corresponds to a
fluid trajectory located at integer times at $x=x_c$ and which moves
on distance of $b$ spatial periods during one temporal period. A
similar argument implies that it is enough to study values of $b$ in
the interval $[0,1)$.To each global minimizer $\{y_j^{\rm (g)},\,
j\in\zset\}$ is associated a \emph{rotation number} defined as
\begin{equation}
  \rho \equiv \lim_{J\to\infty}\frac{1}{J} \sum_{j=0}^J \left(
  y_{j+1}^{\rm(g)} - y_{j}^{\rm(g)} \right)\,,
  \label{eq:defrotationnumber}
\end{equation}
which represents the time-average velocity of the minimizer.  For a
fixed value of the spatial average $b$ of the velocity, all global
minimizers associated to the solution of the Burgers equation have the
same rotation number $\rho$. Indeed, as the dynamics is restricted to
a compact domain of the configuration space (in our case $\tset$), two
minimizers with different rotation numbers necessarily cross each
other; this is an obvious obstruction to the action minimization
property.  In the case of rational rotation numbers the global
minimizers correspond to periodic orbits of the dynamical system
defined by the Euler--Lagrange map. An important feature is that for
rational $\rho$, the rotation number does not change when varying $b$
over a certain closed interval $[b_{\min}, b_{\max}]$, called the
\emph{mode-locking interval}. On the contrary, irrational $\rho$
correspond to a unique value of the parameter $b$. Such ``irrational"
values of $b$ form a Cantor set of zero Lebesgue measure. In
particular, the graph of $\rho$ as a function of the parameter $b$ is
a ``Devil staircase'' (see figure~\ref{fig:devilstaircase}).

When $\rho$ is rational ($\rho = p/q$ in irreducible form), global
minimizers correspond to a periodic orbit of period $q$. It is easy to
see that such an orbit generates $q$ different but closely related
global minimizers. Of course each of these global minimizer is the
image of another one by the Euler--Lagrange map and is mapped back to
itself after $q$ iterations. This procedure generates a periodic
orbit, which turns out to be hyperbolic one. Hence, each of the $q$
global minimizers has a one-dimensional unstable manifold associated
to it.  The solution to the Burgers equation is formed by branches of
these various manifolds with jumps between them defining $q$ global
shocks.

The picture is very different for values of $b$ corresponding to
irrational rotation numbers.  Consider velocities and positions of all
global minimizers at a fixed moment of time, say $t=0$. They form a
subset $\mathcal G$ of the phase space $\tset \times \rset$. Then two
cases have to be distinguished:
\begin{itemize}
\item The set $\mathcal G$ forms a closed invariant curve for the
  Euler--Lagrange map. This invariant curve has a one-to-one
  projection onto the base $\tset$ and dynamics on the curve is
  conjugated to a rigid rotation by angle $\rho$. The limiting
  solution of the Burgers equation is given by the invariant curve and
  does not contain any shocks.
\item The set $\mathcal G$ forms an invariant Cantor set and the
  limiting solution of the Burgers equation contains an infinite
  number of shocks, none of which is a main shock.
\end{itemize}
The Kolmogorov \cite{k57}, Arnold~\cite{a63} and Moser~\cite{m62}
theory (frequently referred to as KAM) describes invariant curves (or
invariant tori) for small analytic perturbations of integrable
Hamiltonian systems, and thus the various types of dynamical
trajectories. The KAM theory ensures that for sufficiently small
perturbations, most of the invariant curves associated to Diophantine
irrational rotation numbers are stable with respect to small analytic
perturbations of the system. Diophantine irrational numbers possess
fast converging approximations by rational numbers (in a suitable
technical sense).  However, these invariant curves may disappear from
the perturbed system when an interaction corresponding to a
non-integrable perturbation gets sufficiently strong. Aubry--Mather
theory provides another variational description for the KAM invariant
curves.  But even more importantly, it describes the invariant Cantor
sets that appear instead of invariant curves in the case of strong
nonlinear interactions. We have mentioned already that these Cantor
sets correspond to global minimizers. Thus Aubry--Mather theory
provides information about the global minimizers and, hence, allows
one to study in such a situation the properties of limiting entropic
solutions and, in particular, the structure of shocks.

A numerical study of the Burgers equation in the inviscid limit, with
periodic forcing and a non-vanishing spatial average of the velocity,
reveals the appearance of shock accumulations. Such events occur for
the values of the mean velocity $b$ near the end-points of the
mode-locking intervals, corresponding to rational rotation
numbers. The shock accumulation phenomenon is due to the fact that the
end-points $b_{\min}, b_{\max}$ of the mode-locking intervals can be
approximated by convergent sequences of ``irrational" values of the
parameter $b$.  This implies accumulation of shocks, since for
irrational rotation numbers the number of shocks is infinite.

The limiting solution $u_\infty(x,t)$ is completely determined by the
function $\hat u(x)$ defined in the previous subsection. The function
$\hat u(x)$ corresponds to a stroboscopic section of $u_\infty$ right
after each impulse. The regular parts of $\hat u$ are made of
single-valued functions related to the unstable manifolds.  The shocks
correspond to jumps, either between different branches of the same
manifold (secondary shocks), or between the manifolds associated to
different global minimizers (main shocks).

\begin{figure}[htbp]
  \centerline{\includegraphics[width=0.45\textwidth]
    {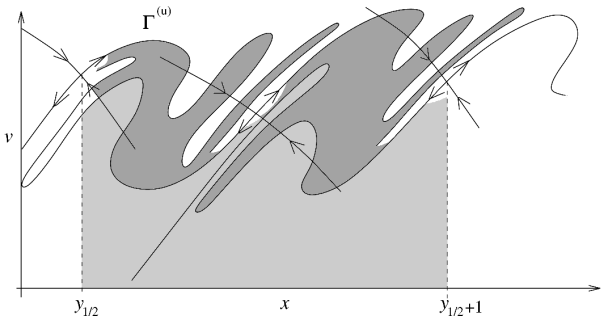}}
  \caption{Sketch of the unstable manifolds of the two global
  minimizers associated to the rotation number $\rho=1/2$. The values
  $b_{\min}$ and $b_{\max}$ given by this configurations are
  represented as grey areas.}
  \label{fig:periode2unstable}
\end{figure}

\begin{figure}[ht]
  \centerline{\subfigure[\label{f:homocline}]{
      \includegraphics[width=0.35\textwidth]{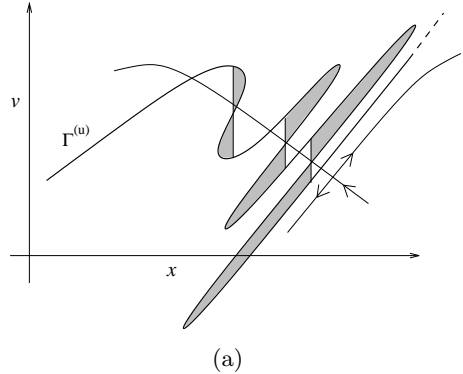}}}
    \centerline{\subfigure[\label{f:accumulation}]{
      \includegraphics[width=0.4\textwidth]{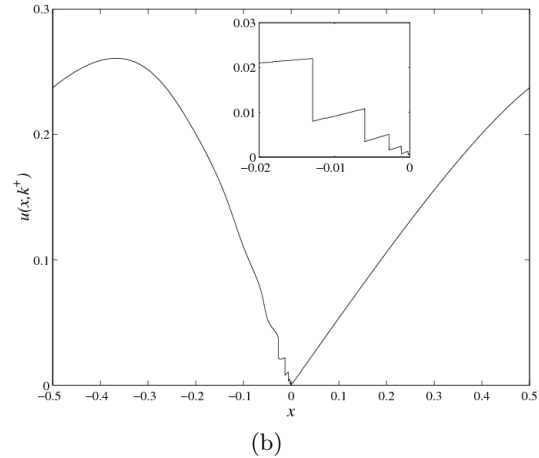}}}
  \caption{(a) Accumulations of shocks occurring for $b=b_{\min}$ or
    $b=b_{\max}$, due to the presence of an infinite number of loops
    of the unstable manifold in the homocline or heterocline
    tangle. (b) Shock accumulation at the fixed point $(0,0)$ of the
    standard map. Here, $\lambda=0.1$ and $b=0.15915$. The latter
    value is close to the upper bound of the interval associated to
    $\rho=0$. The upper inset is a zoom near $(0,0)$, illustrating the
    accumulation of shocks.}
\end{figure}

When the rotation number is rational ($\rho = p/q$), there are $q$
global minimizers. The positions of the $q$ main shocks of $\hat u$
are determined by a requirement that the area defined by the graph of
the solution is equal to the conserved quantity $b$. The latter
constraint shows that the values of $b$ compatible with the rotation
number $p/q$ belong to an interval $[b_{\min}, b_{\max}]$ bounded by
the minimum and maximum areas defined by the unstable manifolds, as
illustrated in figure~\ref{fig:periode2unstable}.
The detailed shape of the manifolds can actually not be sketched on a
figure. Generically the unstable manifold of a global minimizer
corresponding to a particular point of the basic periodic orbit of
period $q$ intersects transversally with the stable manifold of
another minimizer corresponding to another point of the periodic
orbit.  Such an intersection leads to formation of a
\emph{heteroclinic tangle}, a notion which can be traced back to the
work of Poincar\'{e}. The heteroclinic intersection results in the
formation of an infinite number of zig-zags of the unstable
manifolds. These zig-zags are accumulating along the stable manifold
and come arbitrary close to the corresponding point of the periodic
orbit.  The zig-zags contract exponentially in one direction (along
the stable manifold) and are stretched exponentially in the other
direction. It is easy to see that the accumulation of zig-zags
generates an infinite number of ``potential" shocks of smaller and
smaller size which also accumulate near the periodic orbit. When the
parameter $b$ is located well inside the mode-locking interval, the
position of the main shock cuts off the accumulated shocks of small
size so that the total number of shocks is of the order of
unity. However, when $b$ gets closer and closer to $b_{\max}$ or
$b_{\min}$, the main shocks move closer to the periodic points and a
larger number of the small accumulating shocks appears in the
solution. This mechanism leads to an infinite number of shocks in the
solution when $b$ is equal to $b_{min}$ or $b_{\max}$ (see
figure~\ref{f:homocline}).
Both the distances between two consecutive shocks and the sizes of the
shocks decrease exponentially fast with the number of shocks; the rate
is given by the stable eigenvalue associated to the hyperbolic
periodic orbit. It is interesting to mention that when $b=b_{min}$ or
$b=b_{max}$ the main shocks merge with the periodic orbit associated
to the global minimizers. Hence, for the end-points of the
mode-locking interval the main shocks disappear.

To illustrate numerically the change in behavior of the solution to
the Burgers equation when the mean velocity $b$ changes, we focus here
on the simple periodic kicking potential $G(x) = (\lambda/2\pi)\,
\cos(2\pi x)$ where $\lambda$ is a free parameter. The associated
Euler--Lagrange map then reads
\begin{eqnarray}
  && \mathcal{T}\!\!: (y,v) \!\mapsto\!
  \left(y\!+\!v\!+\!\lambda\sin(2\pi y), v\!+\!\lambda \sin(2\pi
  y)\right)\!.
  \label{eq:lagmapstdmap}
\end{eqnarray}
This transformation is usually called the \emph{standard map} (or
Chirikov--Taylor map). It is one of the simplest model for studying
the presence of chaos in Hamiltonian dynamical systems and in
particular particularly to study the KAM theory.

Figure~\ref{f:accumulation} illustrates the accumulation of shocks due
to the homoclinic or heteroclinic tangling for the first transition
(starting from $b=0$). This transition corresponds to a rotation
number of the global minimizer changing value from $\rho = 0$ to
$\rho>0$.  When $b$ is increased and gets close to the critical value,
shocks accumulate on the left-hand side of the global minimizer
located at $(y,v) = (0,0)$.

\begin{figure}[t]
  \centerline{\subfigure[\label{f:different-rotations1}]{
      \includegraphics[width=0.4\textwidth]{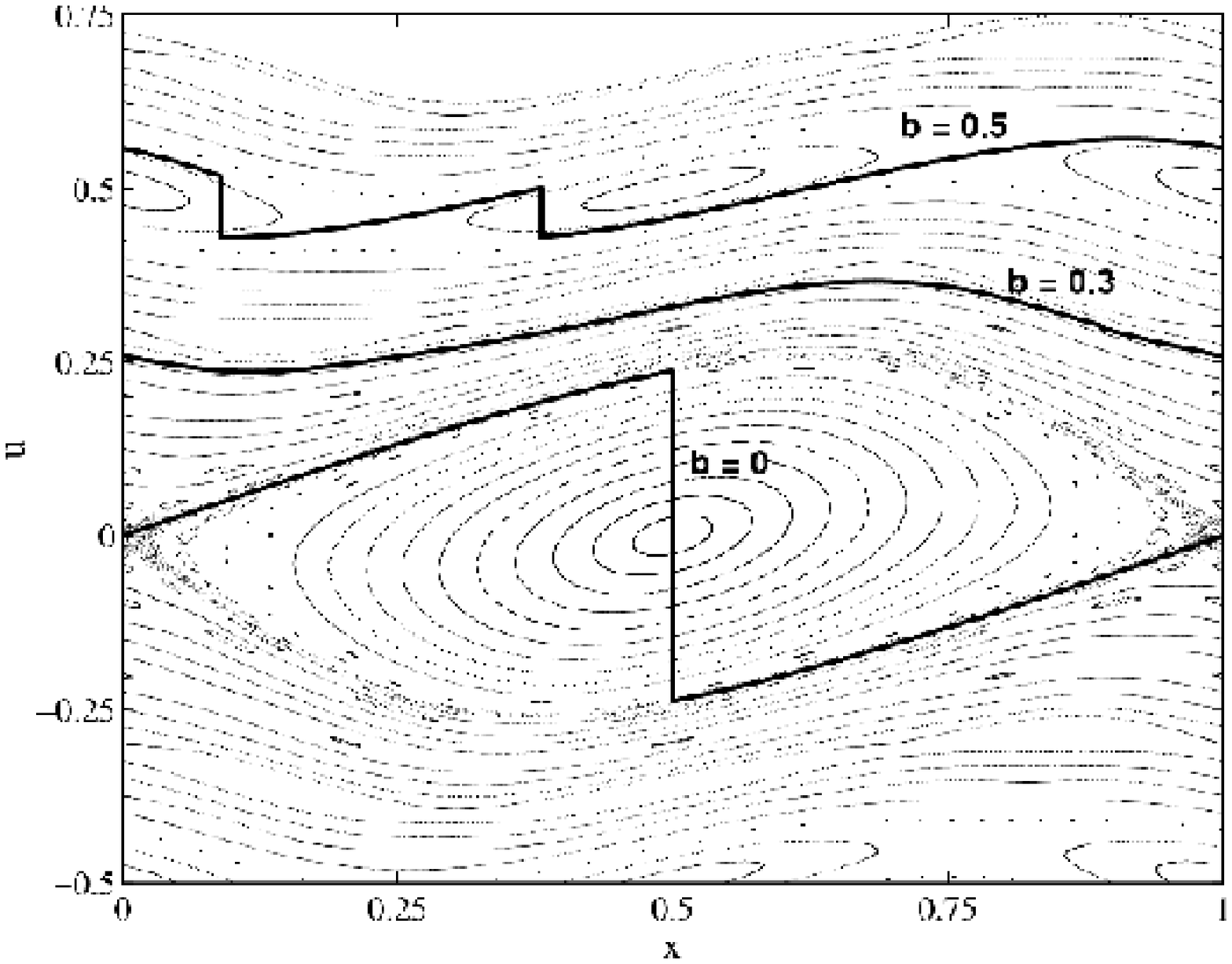}}}
      \centerline{\subfigure[\label{f:different-rotations2}]{
      \includegraphics[width=0.4\textwidth]{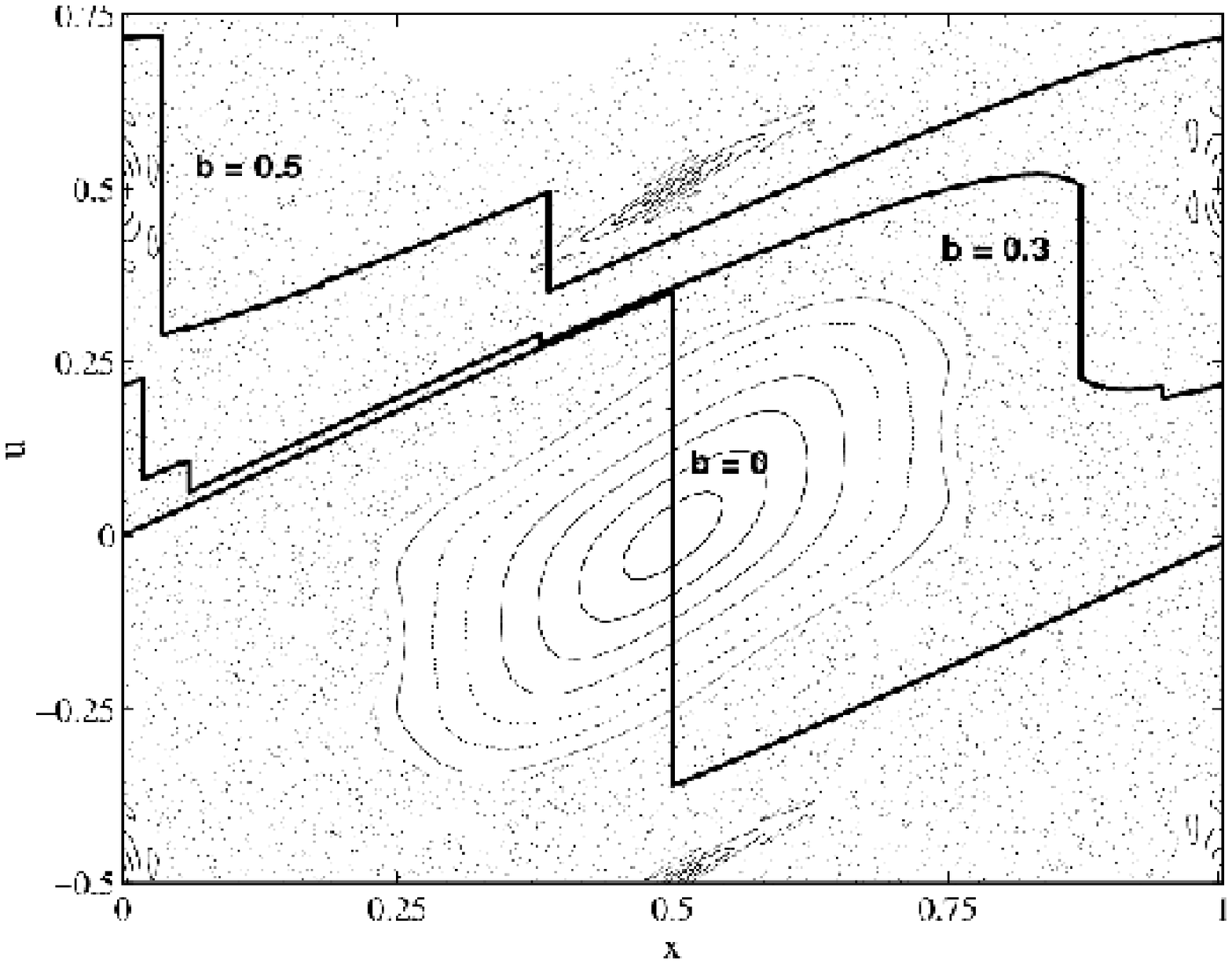}}}
  \caption{General aspect in position-velocity phase space of the
    dynamical system defined by the standard map
    (\ref{eq:lagmapstdmap}) for two different values of the parameter
    (a) $\lambda = 0.1$ and (b) $\lambda = 0.3$. The corresponding
    time-periodic solutions to the kicked Burgers equation are
    represented as bold lines in both cases. The results are presented
    for the spatial mean velocities $b=0$, $b=0.3$ and $b=0.5$.
    \label{fig:different-rotations}}
\end{figure}
\begin{figure}[ht]
  \centerline{\includegraphics[width=0.4\textwidth]{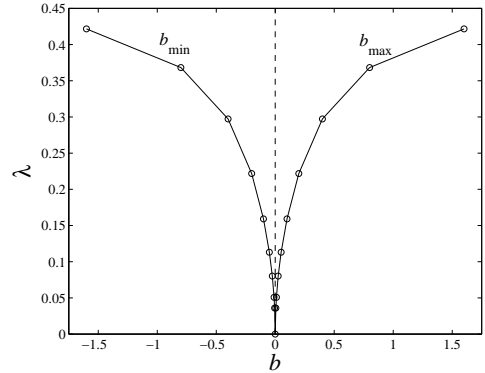}}
  \caption{Evolution as a function of the parameter $\lambda$ of the
  mode-locking interval $[b_{\min},b_{\max}]$ associated to the
  rotation number $\rho=0$. Such a graph is frequently referred to as
  an \emph{Arnold tongue}.\label{f:arnoldtongue}}
\end{figure}

Other numerical experiments were performed in order to observe the
destruction of invariant curves and the accumulation of shocks on
Cantor sets for irrational rotation numbers. It is of course
impossible numerically to set the rotation number to an irrational
value. Indeed, the values of $b$ for which $\rho$ is irrational are in
a Cantor set. It is however possible to be very close to irrational
rotation numbers. Figure~\ref{fig:different-rotations} illustrates the
changes in the behavior of the solutions to the periodically kicked
Burgers equation when varying the parameter $\lambda$. The
time-asymptotic solutions associated to various values of the mean
velocity $b$ are shown for $\lambda=0.1$ and $\lambda=0.3$. For the
latter value, all KAM invariant curves have already disappeared. For
$b=0$ and for all values of $\lambda$ the global minimizer trivially
corresponds to the fixed point $(0,0)$ with a vanishing rotation
number. For $b=0.5$ there are two global minimizers associated to the
rational rotation number $\rho=1/2$. For $\lambda=0.1$ and $b=0.3$ the
rotation number is much closer to an irrational than in previous
cases. The solution is then very close to the invariant curve
associated to this value. Note that the main shock is actually located
close to $x\approx 0.85$. It is so small that it can hardly be
seen. When $\lambda=0.3$ the value $b=0.3$ of the mean velocity no
more corresponds to a rotation number close to an irrational value; it
is now in the mode-locking interval associated to $\rho = 1/3$. This
change in the rotation number reflects the dependence of the
mode-locking intervals $[b_{\min},b_{\max}]$ on the parameter
$\lambda$.  The interval of values of $b$ associated to $\rho=0$ is
represented as a function of $\lambda$ in figure~\ref{f:arnoldtongue}.
Such a structure is frequently called an \emph{Arnold tongue} (see,
e.g., \cite{js98}).

Finally, we discuss the structure of shocks in the case when the
global minimizers form a Cantor set. There are then infinitely many
gaps with no global minimizers.  It is known in this case that all the
gaps can be split into the finite number of images of the \emph{main
gaps}. For the standard map there is only one main gap. Its end-points
$(x^1,v^1)$ and $(x^2,v^2)$ belong to the Cantor set associated to the
global minimizers. All other gaps can be obtained by iterating this
main gap with the Euler-Lagrange map (Standard map) for both positive
and negative times: $(x^1_i,v^1_i) = \mathcal{T}^i (x^1,v^1), \,
(x^2_i,v^2_i) = \mathcal{T}^i (x^2,v^2), \, i \in \zset$. One can show
that the length of the $i$th gap tends to zero as $i\to \pm
\infty$. Since global minimizers are hyperbolic trajectories one can
connect the end-points of the main gap by two smooth curves: the
stable manifold $\Gamma^{\rm (s)}$ and the unstable manifold
$\Gamma^{(\rm u)}$.  As $i\to \infty$ the iterates of the stable
manifold $\mathcal{T}^i\Gamma^{(\rm s)}$ tend to a straight segment
connecting the $i$-th gap with end-points at $(x^1_i,v^1_i)$ and
$(x^2_i,v^2_i)$. The same is true for iterates of the unstable
manifold $\mathcal{T}^i \Gamma^{(\rm u)}$ in the limit $i \to
-\infty$. On the contrary, negative iterates of the stable manifold
and positive of the unstable one form exponentially long curves
connecting corresponding gaps. As usual we are interested in the
iterates of the unstable manifold since they appear in the
time-periodic solution of the Burgers equation. Such a solution is
formed by the iterates of the unstable manifold connecting all the
gaps. Note that in the case of large negative $i$, the unstable
manifold is close to a straight segment; hence there are no shocks
located inside the corresponding gaps. Conversely, for large positive
$i$, the unstable manifold is exponentially long and possesses large
zig-zags.  Hence, the solution to the Burgers equation has one or
several shocks inside such gaps. Since there are no shocks for gaps
with large enough negative $i$, it follows that all the shocks have a
finite age. In other words, the time-periodic solution has no main
shocks.

At the moment it was not possible to study numerically the strange
behavior of the solutions to the Burgers equation corresponding to
global minimizers living on Cantor-like sets.  Looking for such cases
requires a very high spatial resolution in order to minimize the
numerical error in the approximation of the solution. Moreover, a
large number of values for the parameters $b$ and $\lambda$ has to be
investigated in order to observe such a phenomenon.  This would
require heavy computer ressources. However, many other aspects of the
Aubry--Mather theory for Hamiltonian systems can be studied
numerically using the Burgers equation with periodic kicks. For
instance it could be very useful for analyzing the higher dimensional
versions.

%%%%%%%%%%%%%%%%%%%%%%%%%%%%%%%%%%%%%%%%%%%%%%
\section{Velocity statistics in randomly forced Burgers
  turbulence}
\label{s:1Dstatistics}

The universality of small-scale properties in fully developed
Navier--Stokes turbulence has frequently been investigated, assuming
that a steady state is maintained by an external large-scale
forcing. It is generally conjectured that the velocity increments have
universal statistical properties with respect to such a force.
Understanding this issue in simpler models of turbulence has motivated
much work for over ten years.  A toy model which has been extensively
studied is the passive transport of a scalar field by random flows
(see, e.g., \cite{fgv01}).  Tools borrowed from statistical physics
and field theory were used to describe and explain the anomalous
scaling laws observed in the scalar spatial distribution.  It was
shown that the scale invariance symmetry is broken by geometrical
constraints on tracer configurations that are statistically conserved
by the dynamics.  Universality of the intermittent scaling exponents
with respect to the forcing was proven for the case where energy is
injected at large scales \cite{cfkl95,gk95,ss95,bgk96}.

Issues of universality for the nonlinear Burgers turbulence model has
also been very much on the focus.  The possibility to solve exactly a
hydrodynamical problem displaying the same kind of quadratic
nonlinearity as Navier--Stokes turbulence constitutes of course the
central motivation.  Three independent approaches were published
almost simultaneously in 1995 and were at the origin of the growing
interest in Burgers turbulence.  First, an analogy was made
in~\cite{bmp95} between forced Burgers turbulence and the problem of a
directed polymer in a random medium. This analogy was used to show
that the shocks appearing in the solution lead to anomalous scaling
laws for the structure functions. The strong intermittency could be
related to the replica-symmetry-breaking nature of the disordered
system associated to Burgers turbulence. This approach is discussed in
subsection~\ref{subs:intermittency}.  Second, ideas using operator
product expansions borrowed from quantum field theory were proposed
in~\cite{p95}. The goal was to close in the inertial range the
equations governing the correlations of the velocity field in one
dimension. This treatment of the dissipative anomaly is described in
subsection~\ref{subs:anomaly}.  It yields a prediction for the
probability density function (PDF) of velocity increments and
gradients and in particular to a power-law behavior for the PDF of
$\partial_x v$ at large negative values~\cite{p95}. However, the value
of the exponent of this algebraic tail has been a matter of
controversy.  An overview of the various works related to this issue
is given in subsection~\ref{subs:-7/2}.  Finally, the turbulent model
of the one-dimensional Burgers equation with a self-similar forcing
was proposed in~\cite{cy95} as one of the simplest nonlinear
hydrodynamical problem displaying multiscaling of the velocity
structure function.  As stressed in subsection~\ref{subs:selfsimilar}
this problem is easily tractable numerically and some of the numerical
observations can be confirmed by a one-loop renormalization group
expansion.

In what follows we consider the solutions to the Burgers equation with
a homogeneous Gaussian random forcing that is delta-correlated in
time. Namely, the spatio-temporal correlation of the forcing potential
is taken to be
\begin{equation}
  \left\langle F(\vec{x},t)\,F(\vec{x}^\prime,{t}^\prime)
  \right\rangle = B(\vec{x}-\vec{x}^\prime)\,\delta(t-{t}^\prime)\,.
\label{eq:correlforce}
\end{equation}
The function $B$ contains information on the spatial structure of the
forcing. It can be either smooth (i.e.\ concentrated at large spatial
scales) or asymptotically self-similar (i.e.\ behaving as a power law
at small separations). In the former case the solution reaches
exponentially fast a statistically stationary r\'{e}gime in any space
dimension.  The construction of the solution in this r\'{e}gime in
terms of global minimizer and main shock is described in detail in
section~\ref{s:force}.  When $B$ does not decrease sufficiently fast
at small separations (e.g.\ $B(r)\sim r^{2h}$ with $h<1$ as $r\to0$ in
one dimension), there is no rigorous proof of the existence of a
statistically stationary r\'{e}gime. However we assume in the sequel
that such a stationary r\'{e}gime exists in order to perform a
statistical analysis of the solutions to Burgers equation.

\subsection{Shocks and bifractality -- a replica variational approach}
\label{subs:intermittency}
The replica solution for Burgers turbulence proposed in~\cite{bmp95}
is based on its analogy with the problem of a directed polymer in a
random medium.  As already stated in the Introduction, the
\emph{viscous} Burgers equation forced by the potential $F$ is
equivalent to finding the partition function $\mathcal{Z}$ of an
elastic string in the quenched spatio-temporal disorder $V(\vec{x},t)
\!=\! F(\vec{x},t)/2\nu$ (remember that $t$ has to be interpreted as
the space direction in which the polymer is oriented). This relation
is obtained by applying to the velocity potential $\Psi$ the
Hopf--Cole transformation $\mathcal{Z}(\vec{x},t) \!=\!
\exp(\Psi(\vec{x},t)/2\nu)$.  The solution of the problem can be
written in terms of the path integral
\begin{eqnarray}
  && \mathcal{Z}(\vec{x},t) = \displaystyle \int_{\vec{\gamma}(t) =
  \vec{x}} \exp(-\mathcal{H}(\vec{\gamma}))\,\,
  \mathrm{d}[\vec{\gamma}(\cdot)]\,, \nonumber \\ && \mbox{with }
  \displaystyle \mathcal{H}(\vec{\gamma}) = \frac{1}{2\nu}
  \int_{-\infty}^t \left[ \left\|\dot{\vec{\gamma}}(s)\right\|^2 +
  F(\vec{\gamma}(s),s) \right]\mathrm{d}s. \label{eq:formvisc}
\end{eqnarray}
In the analogy between Burgers turbulence and directed polymers, the
polymer temperature is assumed to be unity and its elastic modulus is
$1/(2\nu)$. The strength of the potential fluctuations applied to the
polymer depends on the viscosity and is $\propto \varepsilon^{1/2}
L_f/(2\nu)$ (where $\varepsilon$ is the energy injection rate and
$L_f$ is the spatial scale of forcing). In order to calculate the
various moments of the velocity field $\vec{v} = -\nabla \Psi$, one
needs to average the logarithm of the partition function
$\mathcal{Z}$, a celebrated problem in disordered systems.

\begin{figure}[ht]
  \centerline{\subfigure[\label{fig:replica}]{
      \includegraphics[width=0.4\textwidth]{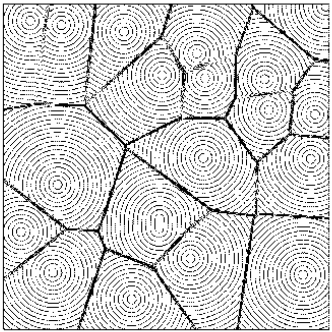}}}
    \centerline{\subfigure[\label{fig:scalingexponents}]{
      \includegraphics[width=0.3\textwidth]{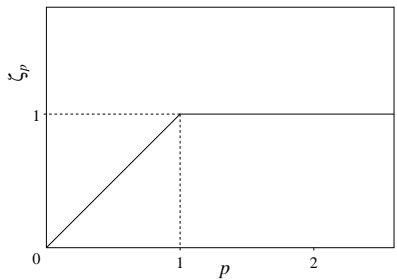}}}
  \caption{(a) Typical shape of the velocity field given by the
    replica approximation in dimension $d=2$ obtained from
    (\ref{eq:approxLin}) for $\mbox{\it Re} = 10^3$. The contour lines
    represent the velocity modulus. Note the cell structure of the
    domain. (b) Scaling exponents of the $p$th order structure
    function.}
\end{figure}

Bouchaud, M{\'{e}}zard and Parisi proposed in~\cite{bmp95} the use of
a replica trick in order to estimate the average free energy $\langle
\ln \mathcal{Z} \rangle$. The first step is to write the zero-replica
limit $\ln \mathcal{Z} = \lim_{n\to 0}\, (\mathcal{Z}^n -1)/n$. Then,
the moments $\langle \mathcal{Z}^n \rangle$ are used to generate an
effective attraction between replicas: they are written as the
partition functions of the disorder-averaged Hamiltonian
$\mathcal{H}_n(\vec{\gamma}_1, \dots, \vec{\gamma}_n)$ associated to
$n$ replicas of the same system~\cite{mpv87}
\begin{eqnarray}
  && \mathcal{H}_n \!=\!\! \sum_{i=1}^n
  \!\!\int_{-\infty}^t\!\!\!\!\!\!\!\mathrm{d}s\!\! \left[
  \!\frac{1}{2\nu}\!  \left\|\dot{\vec{\gamma}}_i(s)\!\right\|^2
  \!\!\!\!- \!\frac{1}{4\nu^2} \!\!\sum_{j=1}^n\!
  B(\vec{\gamma}_i(s)\!-\!\vec{\gamma}_j(s))\!\right]\!\! ,
  \label{eq:replicaHamil}
\end{eqnarray}
where $B$ denotes the spatial part of the forcing potential
correlation.  The next step is to study this problem by a variational
approach.  The Hamiltonian $\mathcal{H}_n$ is replaced by an effective
Gaussian quadratic Hamiltonian that can be written as
\begin{eqnarray}
  && \mathcal{H}_{\rm eff} \!=\! \frac{1}{2}\! \sum_{i=1}^n
  \sum_{j=1}^n \!\!\int_{-\infty}^{t}\!\int_{-\infty}^{t}\!\!\!\!\!\!
  \vec{{\gamma}}_i(\tau)\mathcal{G}_{ij}(\tau\!-\!\tau^{\prime})
  \vec{{\gamma}}_j(\tau^{\prime})\mathrm{d}\tau
  \mathrm{d}\tau^{\prime}\!.
  \label{eq:effHamil}
\end{eqnarray}
The kernel $\mathcal{G}_{ij}$ is then chosen in such a way that it
minimizes the free energy. It is shown in~\cite{bmp95} that the
optimal Gaussian Hamiltonian is the solution of a system of equations
that can be solved following the ansatz proposed in~\cite{mp90}. When
$d>2$ this approach singles out two r\'{e}gimes depending on the
Reynolds number $\mbox{\it Re} =
\varepsilon^{1/3}L_f^{3/4}/\nu$. These r\'{e}gimes are separated by
the critical value $\mbox{\it Re}_c = [2(1-2/d)^{1-d/2}]^{1/3}$. When
$\mbox{\it Re} < \mbox{\it Re}_c$ the optimal solution is of the form
$\mathcal{G}_{ij} = \mathcal{G}_0 \,\delta_{ij} + \mathcal{G}_1$ and
obeys the replica symmetry. In finite-size systems it corresponds to a
linear velocity profile. When $\mbox{\it Re} > \mbox{\it Re}_c$ the
correct solution is given by the \emph{one-step
replica-symmetry-breaking scheme} (see \cite{mp90}). The off-diagonal
elements of $\mathcal{G}_{ij}$ are then parameterized with two
functions depending on whether the indices $i$ and $j$ belong to the
same block or to different blocks. Qualitatively, the one-step
replica-symmetry-breaking approach amounts to the assumption that the
instantaneous velocity potential can be written as a weighted sum of
Gaussians, leading to an approximation of the velocity field as
\begin{equation}
  \vec{v} (\vec{x},t) \simeq \frac{2\nu}{\sigma}\, \frac{\sum_{\alpha}
      (\vec{x}-\vec{r}_\alpha)\,{\rm e}^{-\mbox{\scriptsize\it
      Re}\,(f_\alpha +
      \|\vec{x}-\vec{r}_\alpha\|^2/2L_f^2)}}{\sum_{\alpha} {\rm
      e}^{-\mbox{\scriptsize\it Re}\,(f_\alpha +
      \|\vec{x}-\vec{r}_\alpha\|^2/2L_f^2)}}\,,
  \label{eq:approxLin}
\end{equation}
where the $f_\alpha$'s are independent variables with a Poisson
distribution of density $\exp(-f)$. The $\vec{r}_\alpha$ are uniformly
and independently distributed in space. In (\ref{eq:approxLin}) the
sum over $\alpha$ is running from 1 to a large-enough integer $M$.
The typical shape of the approximation of the velocity field given by
(\ref{eq:approxLin}) is represented in figure~\ref{fig:replica} in the
two-dimensional case. In the limit of large Reynolds numbers the
random velocity field given by (\ref{eq:approxLin}) typically contains
cells of width $\propto L_f$. The width of a shock separating two
cells is of the order of $L_f/\mbox{\it Re}$.

The replica approximation (\ref{eq:approxLin}) leads to an estimate of
the PDF $p(\Delta v, r)$ of the longitudinal velocity increment
$\Delta v = (\vec{v}(\vec{x}+r\,\vec{e}, t)- \vec{v}(\vec{x}, t))\cdot
\vec{e}$, where $\vec{e}$ is an arbitrary unitary vector.  When
$\mbox{\it Re}\gg1$ and $r\ll L_f$ this approximation takes the
particularly simple asymptotic form
\begin{equation}
  p(\Delta v, r) \approx \delta\!\left(\Delta v - U_f
  \frac{r}{L_f}\right) + \frac{r}{L_f}\, \frac{1}{U_f}
  \,g\left(\frac{\Delta v}{U_{f}}\right)\,,
  \label{eq:approx_pdfincr}
\end{equation}
where $U_{f} = \mbox{\it Re}\,\nu/L_f$ is the typical velocity
associated to the scale $L_f$ and $g$ is a scaling function that is
determined explicitly in \cite{bmp95}. This approximation is in
agreement with the following qualitative picture. With a probability
almost equal to one, the two points $\vec{x}$ and $\vec{x}+r\,\vec{e}$
lie in the same cell; the velocity increment is then given by the
typical velocity gradient which, according to the approximation
(\ref{eq:approxLin}), is order $U_f/L_f$. With a probability $r/L_f$
the two points are sitting on different sides of a shock separating
two such cells and the associated velocity difference is of the order
of $U_f$.

The structure functions of the velocity field given by the various
moments of $\Delta v$ can be straightforwardly estimated from the
approximation (\ref{eq:approx_pdfincr}).  Their scaling behavior
$\langle \Delta v^p \rangle \sim r^{\zeta_p}$ at small separations $r$
display a bifractal behavior as sketched in
figure~\ref{fig:scalingexponents}. When $p<1$, the first term on the
right-hand side of (\ref{eq:approx_pdfincr}) dominates and $\langle
\Delta v^p \rangle \propto U_f^p\, (r/L_f)^p$. For $p>1$ the shock
contribution is dominating the small-$r$ behavior and thus $\langle
\Delta v^p \rangle \propto U_f^p\, (r/L_f)$.

This approach, which makes use of replica tricks, is as we have seen
able to catch the leading scaling behavior of velocity structure
functions in any dimension. It is based on approximations of the
velocity field by the superposition~(\ref{eq:approxLin}) of Gaussian
velocity potentials.  A first advantage of this method is that it
catches the generic aspect of the solution including the hierarchy of
high-order singularities appearing in the solution when $\mbox{\it
Re}\to\infty$ which was examined in
section~\ref{subs:singularities}. This method also gives predictions
regarding the dependence on $\mbox{\it Re}$ of the statistical
properties of the solution.  However, as stressed in~\cite{bmp95}, the
validity of this approximation is expected to hold only in the limit
of infinite space dimension $d$. In particular, it is known that for
$d\le2$ a full continuous replica-symmetry-breaking scheme is
needed~\cite{mp90}. Nevertheless, as we have seen, there is enough
evidence that this approach describes very well the qualitative
aspects of the solution.

\subsection{Dissipative anomaly and operator product expansion}
\label{subs:anomaly}

The replica-trick approach described in the previous subsection cannot
reproduce one of the main statistical features of the solution, namely
the tails of the velocity increments PDF $p(\Delta v, r)$.  Indeed the
prediction (\ref{eq:approx_pdfincr}) based on a variational
approximation of the velocity field implies that $p$ identically
vanishes when $\Delta v > U_f\,(r/L_f)$. In order to study the
quantitative behavior of the PDF $p(\Delta v, r)$ in the inviscid
limit $\nu\to0$ (or equivalently $\mbox{\it Re}\to\infty$),
Polyakov~\cite{p95} proposed to use an operator product
expansion. This approach leads to an explicit expression for $p(\Delta
v, r)$ and predicts a super-exponential tail at large positive values
and a power-law behavior for negative ones.  Such predictions have
immediate implications for the asymptotics of the PDF $p(\xi)$ of the
velocity gradient $\xi = \partial_x v$.  The work of Polyakov was the
starting point of a controversy on the value of the exponent of the
left tail of $p(\xi)$.  Before returning to this issue in the next
subsection, we give in the sequel a quick overview of the original
work by Polyakov.

We henceforth focus on the one-dimensional solutions to the Burgers
equation with Gaussian forcing whose autocorrelation is given by
(\ref{eq:correlforce}).  Following~\cite{p95} (see also
\cite{b98,blp04}) we introduce the characteristic function of the
$n$-point velocity distribution
\begin{equation}
  Z_n(\lambda_j, x_j;t) \equiv \left\langle
  {\rm e}\,^{\lambda_1\,v(x_1,t)+\cdots+\lambda_n\,v(x_n,t)}
  \right\rangle\,.
  \label{eq:defZn}
\end{equation}
For a finite value of the viscosity $\nu$, it is easily seen that this
quantity is a solution to a Fokker--Planck (master) equation obtained
by differentiating $Z_n$ with respect to $t$ and using the Burgers
equation and the fact that the forcing is Gaussian and
$\delta$-correlated in time. This leads to
\begin{eqnarray}
  \frac{\partial Z_n}{\partial t} &+& \sum_j
  \lambda_j\,\frac{\partial}{\partial\lambda_j} \left (
  \frac{1}{\lambda_j}\, \frac{\partial Z_n}{\partial x_j} \right ) =
  \nonumber \\ &=& \frac{1}{2} \sum_{i,j}
  b(x_i-x_j)\,\lambda_i\,\lambda_j\,Z_n + \mathcal{D}^{(n)}_\nu\,,
  \label{eq:mastereqZ}
\end{eqnarray}
where $b \equiv (\mathrm{d}^2 B) / (\mathrm{d}r^2)$ denotes the
spatial part of the correlation of the forcing applied to the velocity
field.  $\mathcal{D}^{(n)}_\nu$ denotes the contribution of the
dissipative term and reads
\begin{equation}
 \mathcal{D}^{(n)}_\nu \equiv \nu\, \left\langle\! \left[\sum_j
  \lambda_j\, \partial_{x_j}^2 v(x_j,t) \right ]\! {\rm
  e}^{\sum_j\lambda_j\,v(x_j,t)} \right\rangle\!.
  \label{eq:defD}
\end{equation}
This term does not vanish in the limit $\nu\to0$ since the solutions
to the Burgers equation develop singularities with a finite
dissipation. It has been proposed in~\cite{p95} to use an analogy with
the anomalies appearing in quantum field theory in order to tackle
this term in the inviscid limit. The important assumption is then made
that the singular term in the operator product expansion relates
linearly to the characteristic function $Z_n$. Since this expansion
should preserve the statistical symmetries of the Burgers equation, it
leads to the replacement in all averages of the singular limit $
\lim_{\nu\to0} \nu\,\lambda\,(\partial_{x}^2 v)\, {\rm e}^{\lambda\,
v}$ by the asymptotic expression
\begin{equation}
  \left(\frac{\mathsf{a}}{2} + \frac{\mathsf{b}-1}{\lambda}
  \,\frac{\partial}{\partial x} + \mathsf{c}\, \lambda
  \,\frac{\partial}{\partial\lambda} \right){\rm e}^{\lambda\, v}\,,
  \label{eq:anomaly}
\end{equation}
where the coefficients $\mathsf{a}$, $\mathsf{b}$ and $\mathsf{c}$ are
parameters that can be determined only indirectly.  However their
possible values can be restricted by requiring that $Z_n$ is the
characteristic function of a probability distribution which is
non-negative, finite, normalizable, and that the dissipative term
$\mathcal{D}^{(n)}_\nu$ acts as a positive operator. Finding these
coefficients is similar to an eigenvalue problem in quantum mechanics.

We now come to a crucial point in Polyakov's approach. Important
restrictions on the form of the different anomalous terms
in~(\ref{eq:anomaly}) result from the fact that the solutions to the
Burgers equation obey a certain form of Galilean invariance. A notion
of ``strong Galilean principle'' is introduced for invariance of the
$n$-point distribution of velocity under the transformation $v \mapsto
v + v_0$ with $v_0$ an arbitrary constant.  As a consequence, the
$n$-point characteristic function $Z_n$ has to be proportional to
$\delta(\lambda_1+\cdots+\lambda_n)$. The operators appearing in the
limit $\nu\to 0$ have to be consistent with such an invariance.
In~\cite{p95} it is argued that this symmetry is automatically broken
by the forcing that introduces a typical velocity $\langle v^2
\rangle^{1/2} \propto b^{1/3}(0) \, L^{1/3}$.  However Polyakov
assumes this ``strong Galilean principle'' to be asymptotically
recovered in the limit $L\to\infty$ of infinite-size systems.  In the
case of finite-size systems, when $L$ is of the order of the
correlation length $L_f$ of the forcing, the strong Galilean symmetry
is broken because of the conservation of the spatial average of $v$
which introduces a characteristic velocity $v_0 = (1/L) \int v(x,t)\,
\mathrm{d}x$. However, the Galilean symmetry should be recovered when
averaging the correlation functions with respect to the mean velocity
$v_0$. This symmetry restoration was introduced in~\cite{blp04} where
it is referred to as the ``weak Galilean principle''. The $n$-point
characteristic function associated to an average velocity $v_0$
relates to that associated to a vanishing mean velocity by
\begin{equation}
  Z_n(\lambda_j,x_j; t ;v_0) = {\rm e}^{v_0 \sum_j
  \lambda_j}\,Z_n(\lambda_j,x_j; t ;0)\,. \nonumber
\end{equation}
After averaging with respect to $v_0$, one obtains
\begin{equation}
  Z_n(\lambda_j,x_j; t) = 2\pi\, \delta\!\!\left(\sum_j \lambda_j
  \right)\,Z_n(\lambda_j,x_j; t ;0)\,.
  \label{eq:weakgalilean}
\end{equation}
One can easily check that (\ref{eq:mastereqZ}), together with the
dissipative term given by (\ref{eq:anomaly}), are compatible with this
expression for the $n$-point characteristic function $Z_n$. Moreover,
any higher-order term in the expansion (\ref{eq:anomaly}) of the
dissipative anomaly would violate Galilean invariance.

To obtain the statistical properties of the solution, one needs to
further restrict the values of the three free parameters $\mathsf{a}$,
$\mathsf{b}$, and $\mathsf{c}$ appearing in the expansion
(\ref{eq:anomaly}). Following \cite{p95} this can be done by
considering the case $n=2$ that corresponds to the equation for the
PDF of velocity differences. Performing the change of variables
$\lambda_{1,2} = \Lambda\pm\mu$ and $x_{1,2} = X\pm y/2$, and assuming
that $\lambda\ll \mu$ and $y\ll L_f$ (so that the spatial part of the
forcing correlation is to leading order $b(y) \simeq b_0 - b_1y^2$),
the stationary and space-homogeneous solutions to the master equation
(\ref{eq:mastereqZ})) satisfy
\begin{eqnarray}
  \frac{\partial^2 Z_2}{\partial\mu\partial y} & - &
  (2b_0\Lambda^2+b_1\mu^2y^2) Z_2 = \nonumber \\ & = & \mathsf{a}Z_2
  +\frac{2\mathsf{b}}{\mu} \frac{\partial Z_2}{\partial y} +
  \mathsf{c}\Lambda \frac{\partial Z_2}{\partial \Lambda} +
  \mathsf{c}\mu\frac{\partial Z_2}{\partial \mu}\,.
  \label{eq:mastereqZ2}
\end{eqnarray}
It is next assumed in \cite{p95} (see also \cite{blp04}) that the
velocity difference $v(x_1,t)-v(x_2,t)$ is statistically independent
of the mean velocity $(v(x_1,t)+v(x_2,t))/2$. This implies that the
two-point characteristic function factorizes as $Z_2 =
Z_2^+(\Lambda)Z_2^-(\mu,y)$, where the two functions $Z_2^+$ and
$Z_2^- $ satisfy the closed equations
\begin{eqnarray}
  -2b_0 \Lambda^2 Z_2^+ &=& \mathsf{c}\Lambda\frac{\partial
   Z_2^+}{\partial \Lambda}\,, \label{eq:mastereqZ2plus} \\
   \frac{\partial^2 Z_2^-}{\partial \mu \partial y} - b_1 \mu^2y^2
   Z_2^- &=& \mathsf{a} Z_2^- + \frac{2\mathsf{b}}{\mu} \frac{\partial
   Z_2^-}{\partial y} + \mathsf{c}\mu \frac{\partial Z_2^-}{\partial
   \mu}\,. \label{eq:mastereqZ2minus}
\end{eqnarray}
The solution to the first equation corresponds to a Gaussian
distribution which is normalizable only if $\mathsf{c}<0$. As shown
numerically in \cite{blp04} this distribution is representative of the
bulk of the one-point velocity PDF. Information on the solutions to
the second equation can be obtained assuming the scaling property
$Z_2^-(\mu,y) = \Phi(\mu y)$, which amounts to considering only those
contributions to the distribution of velocity differences stemming
from velocity gradients $\xi = \partial_x v$. This yields a prediction
the negative and positive tails of the PDF of velocity gradients:
\begin{eqnarray}
  && p(\xi) \propto |\xi|^{-\alpha} \ \mbox{ when }
  \xi\to-\infty\,, \label{eq:pdfximinus}\\ && p(\xi) \propto
  \xi^\beta\exp(-\mathsf{C}\,\xi^3) \ \mbox{ when } \xi\to+\infty\,,
  \label{eq:pdfxiplus}
\end{eqnarray}
where $\mathsf{C}$ is a constant, which depends only on the strength
of the forcing. The two exponents $\alpha$ and $\beta$ are related to
the coefficient $\mathsf{b}$ of the anomaly by
\begin{equation}
\alpha= 2\mathsf{b}+1 \mbox{ and } \beta = 2\mathsf{b}-1\,.
\end{equation}
The value of $\mathsf{b}$ remains undetermined but is prescribed to
belong to a certain range. This approach was first designed in
\cite{p95} for infinite-size systems where strong Galilean invariance
holds. In that case consistency with such an invariance leads to
dropping the third term in the operator product expansion (i.e.\
$\mathsf{c}=0$). Positivity and normalizability of the two-point
velocity PDF and non-positivity of the anomalous dissipation operator
imply that the two other coefficients form a one-parameter family with
$3/4\le \mathsf{b}\le 1$. In particular, this implies that the left
tail of the velocity gradient PDF with exponent $\alpha$ should be
shallower than $\xi^{-3}$. As we will see in the next section, strong
evidence has been obtained that $p(\xi)\propto \xi^{-7/2}$ for
$\xi\to-\infty$. This seems to contradict the approach based on
operator product expansion. However, as argued in \cite{blp04}, the
breaking of strong Galilean invariance occurring in finite-size
systems and resulting in the presence of the $\mathsf{c}$ anomaly
broadens the range of admissible values for $\mathsf{b}$. In
particular it allows for the value $\mathsf{b} = 5/4$ which
corresponds to the exponent $\alpha=7/2$.

\subsection{Tails of the velocity gradient PDF}
\label{subs:-7/2}

After the numerical work of Chekhlov and Yakhot~\cite{cy95b}, the
asymptotic behavior at large positive and negative values of the PDF
of velocity derivatives $\xi=\partial_x v$ for the one-dimensional
randomly forced Burgers equation attracted much attention. A broad
consensus emerged around the prediction of Polyakov~\cite{p95} that
$p(\xi)$ displays tails of the form (\ref{eq:pdfxiplus}) and
(\ref{eq:pdfximinus}), but the values of the exponents $\alpha$ and
$\beta$ were at the center of a controversy. Note that the presence of
a super-exponential tail $\propto\exp(-\mathsf{C}\,\xi^3)$ at large
positive arguments has been confirmed by the use of instanton
techniques~\cite{gm96} and that the only remaining uncertainty
concerns the exponent of the algebraic prefactor. A standard approach
to determine the exponents $\alpha$ and $\beta$ appearing in
(\ref{eq:pdfximinus}) and (\ref{eq:pdfxiplus}) makes use of the
stationary solutions to the inviscid limit of the Fokker--Planck
equation for the PDF, namely
\begin{equation}
-\partial_\xi\left(\xi^2p\right) -\xi p +\nu \partial_\xi
\left[\la\partial_{x}^3 v\,|\, \partial_x v\!=\!\xi\ra p \right] =
\tilde{b}\partial_{\xi}^2 p\,.
\label{eq:fokker-planck}
\end{equation}
Here the brackets $\la\cdot|\cdot\ra$ denote conditional averages and
the right-hand side expresses the diffusion of probability due to the
delta-correlation in time of the forcing. The main difficulty in
studying the solutions of (\ref{eq:fokker-planck}) stems from the
treatment of the dissipative term $D^\nu(\xi) = \nu \partial_\xi
\left[\la\partial_{x}^3 v|\partial_x v \!=\! \xi\ra p \right]$ in the
limit $\nu\to0$. The value $\alpha = 3$ is obtained if a piecewise
linear approximation is made for the solutions of the Burgers
equation~\cite{bm96}. Gotoh and Kraichnan~\cite{gk98} argued that the
dissipative term is to leading order negligible and presented
analytical and numerical arguments in favor of $\alpha = 3$ and $\beta
= 1$. However, the inviscid limit of (\ref{eq:fokker-planck}) contains
anomalies due to the singular behavior of $D^\nu(\xi)$ in the limit
$\nu\to0$.  As we have seen in previous section, the approach based on
the use of an operator product expansion~\cite{p95} leads to a
relation involving unknown coefficients which must be determined,
e.g., from numerical simulations~\cite{yc96,b98,blp04}, and restricts
the possible values to $5/2 \leq \alpha \leq 3$~\cite{b01}.  Anomalies
cannot be studied without a complete description of the singularities
of the solutions, such as shocks, and a thorough understanding of
their statistical properties.

E, Khanin, Mazel and Sinai made a crucial observation in~\cite{ekms97}
that large negative gradients stem mainly from {\em preshocks}, that
is the cubic-root singularities in the velocity preceding the
formation of shocks (see section~\ref{subs:singularities}). They then
used a simple argument for determining the fraction of space-time
where the velocity gradient is less than some large negative value.
This leads to $\alpha=7/2$, provided preshocks do not cluster. Later
on, this approach has been refined by E and Vanden Eijnden who
proposed to determine the dissipative anomaly of
(\ref{eq:fokker-planck}) using formal matched
asymptotics~\cite{eve99a} or bounded variation
calculus~\cite{eve00}. As we shall see below, with the assumption that
shocks are born with a zero amplitude, that their strengths add up
during collisions, and that there ar no accumulations of preshocks,
the value $\alpha=7/2$ was confirmed~\cite{eve00}. Other attempts to
derive this value using also isolated preshocks have been
made~\cite{kl00,b01}. Note that there are simpler instances, including
time-periodic forcing~\cite{bfk00} (see section~\ref{s:timeperiodic})
and decaying Burgers turbulence with smooth random initial
conditions~\cite{bf00,eve00} (see section~\ref{subs:density}), which
fall in the universality class $\alpha=7/2$, as can be shown by
systematic asymptotic expansions using a Lagrangian approach.

We give here the flavor of the approach used in~\cite{eve99a} in order
to estimate the dissipative anomaly $D^0(\xi) = \lim_{\nu\to0}
D^\nu(\xi)$.  One first notices that for $|\xi|\gg \tilde{b}^{1/3}$,
the forcing term in the right-hand side of (\ref{eq:fokker-planck})
becomes negligible, so that stationary solutions to the Fokker--Planck
equation satisfy
\begin{equation}
 p(\xi) \approx |\xi|^{-3}\int_{-\infty}^\xi
 \!\!\mathrm{d}\xi^\prime\, \xi^\prime D^\nu(\xi^\prime).
 \label{eq:defpfnanomaly}
\end{equation}
A straightforward consequence of this asymptotic expression is that,
if the integral in the right-hand side decreases as $\xi\to-\infty$
(i.e.\ if $\xi D^\nu(\xi)$ is integrable), then $p(\xi)$ decreases
faster than $|\xi|^{-3}$, and thus $\alpha>3$.

To get some insight into the behavior of $D^\nu$ as $\nu\to0$, one
next observes that the solutions to the one-dimensional Burgers
equation contain smooth regions where viscosity is negligible, which
are separated by thin shock layers where dissipation takes place. The
basic idea consists in splitting the velocity field $v$ into the sum
of an outer solution away from shocks and of an inner solution near
them for which boundary layer theory applies. Matched asymptotics are
then used to construct a uniform approximation of $v$. To construct
the inner solution near a shock centered at $x=x_\star$, one performs
the change of variable $x\mapsto \tilde{x}=(x\!-\!x_\star)/\nu$ and
looks for an expression of $\tilde{v}(\tilde{x},t) =
v(x_\star\!+\!\nu\tilde{x},t)$ in the form of a Taylor expansion in
powers of $\nu$: $\tilde{v} = \tilde{v}_0 + \nu \tilde{v}_1 +
\mathrm{o}(\nu)$. At leading order, the inner solution satisfies
\begin{equation}
  \left[\tilde{v}_0-v_\star \right] \partial_{\tilde{x}}
  \tilde{v}_0 = \partial^2_{\tilde{x}}\tilde{v}_0,
\end{equation}
where $v_\star = (\mathrm{d}x_\star)/(\mathrm{d}t)$. This expression
leads to the well-known hyperbolic tangent velocity profile
\begin{equation}
  \tilde{v}_0 = v_\star - \frac{s}{2} \tanh
  \!\left(\frac{s\tilde{x}}{4}\right).
\end{equation}
Here, $s = v(x_\star+,t)-v(x_\star-,t)$ denotes here the velocity jump
across the shock and is given by matching conditions to the outer
solution. The term of order $\nu$ is then a solution of
\begin{equation}
  \partial_t \tilde{v}_0 + \left[\tilde{v}_0-v_\star\right]
  \partial_{\tilde{x}} \tilde{v}_1 = \partial^2_{\tilde{x}}\tilde{v}_1
  + f(x,t).
\end{equation}
In order to evaluate the dissipative anomaly, it is convenient to
assume spatial ergodicity so that the viscous term in
(\ref{eq:fokker-planck}) can be written as
\begin{equation}
  D^\nu(\xi)  =  \nu \partial_\xi\!  \lim_{L\to\infty} \frac{1}{2L}
  \int_{-L}^L \!\!  \mathrm{d}x\,\, \partial_x^3 v\,\,
  \delta(\partial_x v\!-\!\xi).
\end{equation}
In the limit $\nu\to0$ the only remaining contribution stems from
shocks and is thus given by the inner solution. Using the expansion of
the solution up to the first order in $\nu$, this leads to writing the
dissipative term in the limit of vanishing viscosity as (see Appendix
of~\cite{eve00b} for details)
\begin{equation}
   D^0(\xi) = \frac{\rho}{2} \int_{-\infty}^0 \!\!\!
  \mathrm{d}s\,s\,[p^+(s,\xi) + p^-(s,\xi)]\,,
  \label{eq:dissipppm}
\end{equation}
where $\rho$ is the density of shocks and $p^+$ (respectively $p^-$)
is the joint probability of the shock jump and of the value of the
velocity gradient at the right (respectively left) of the shock. This
expression guarantees the finiteness of the dissipative anomaly, and in
particular the fact that the integral in the right-hand side of
(\ref{eq:defpfnanomaly}) is finite in the limit $\nu\to0$ and
converges to 0. As a consequence, this gives a proof that the exponent
$\alpha$ of the left tail of the velocity gradient PDF is larger than
$3$.

To proceed further, E and Vanden Eijnden proposed to estimate the
probability densities $p^+$ and $p^-$ by deriving master equations for
the joint probability of the shock strength $s$, its velocity
$v_\star$ and the values $\xi^{\pm}$ of the velocity gradient at its
left and at its right. This is done in \cite{eve00} using a
formulation of Burgers dynamics stemming from bounded variation
calculus.  More precisely, it is shown in~\cite{v67} that the Burgers
equation is equivalent to considering the solutions to the partial
differential equation
\begin{equation}
  \partial_t v + \bar{v} \partial_x v = f\,,
\end{equation}
where $\bar{v}(x,t) = (v(x+,t) + v(x-,t))/2$. Basically this means
that Burgers dynamics can be formulated in terms of the transport of
the velocity field by its average $\bar{v}$. This formulation
straightforwardly yields a master equation for $v(x\pm,t)$ and
$\partial_x v(x\pm,t)$ which is then used to estimate $p^\pm$ and the
dissipative anomaly (\ref{eq:dissipppm}). Although the treatment of
the master equation does not involve any closure hypothesis, it is not
fully rigorous: in particular it requires the assumption that shocks
are created with zero amplitude and that shock amplitudes add up
during collision. However such an approaches strongly suggests that
$\alpha=7/2$ and $\beta=1$.

Obtaining numerically clean scaling for the PDF of gradients is not
easy with standard schemes.  Let us observe that any method involving
a small viscosity, either introduced explicitly (e.g.\ in a spectral
calculation) or stemming from discretization (e.g.\ in a finite
difference calculation), may lead to the presence of a power-law range
with exponent $-1$ at very large negative gradients~\cite{gk98}. This
behavior makes the inviscid $|\xi|^{-\alpha}$ range appear shallower
than it actually is, unless extremely high spatial resolution is
used. In contrast, methods that directly capture the inviscid limit
with the appropriate shock conditions, such as the fast Legendre
transform method~\cite{nv94}, lead to delicate interpolation
problems. They have been overcome in the case of time-periodic
forcing~\cite{bfk00} but with white-noise-in-time forcing, it is
difficult to prevent spurious accumulations of preshocks leading to
$\alpha=3$.

To avoid such pitfalls, a Lagrangian particle and shock tracking
method was developed in~\cite{b01}. This method is able to separate
shocks and smooth parts of the solution and is particularly effective
for identifying preshocks. The main idea is to consider the evolution
of a set of $N$ massless point particles accelerated by a
discrete-in-time approximation of the forcing with a uniform time
step. When two of these particles intersect, they merge and create a
new type of particle, a shock, characterized by its velocity (half sum
of the right and left velocities of merging particles) and its
amplitude.  The particle-like shocks then evolve as ordinary
particles, capture further intersecting particles and may merge with
other shocks. In order not to run out of particles too quickly, the
initial small region where particles have the least chance of being
subsequently captured is determined by localization of the global
minimizer of the Lagrangian action (see
section~\ref{subs:stationary}).  The calculation is then restarted
from $t=0$ for the same realization of forcing but with a vastly
increased number of particles in that region. This particle and shock
tracking method gives complete control over shocks and preshocks.

\begin{figure}[ht]
\centerline{\includegraphics[width=0.4\textwidth]{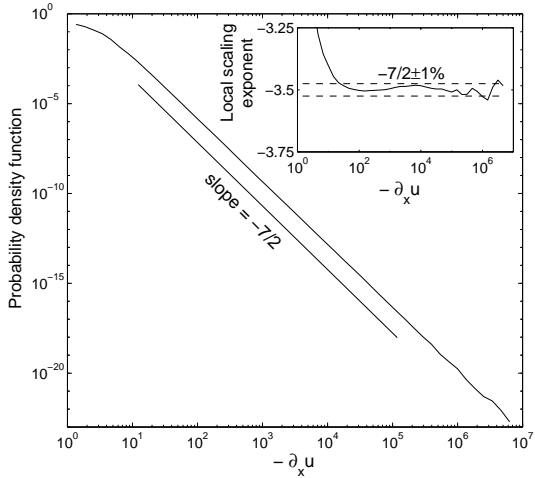}}
\caption{PDF of the velocity gradient at negative values in log-log
  coordinates obtained by averaging over 20 realizations and a time
  interval of 5 units of time (after relaxation of transients). The
  simulation involves up to $N=10^5$ particles and the forcing is
  applied at discrete times separated by $\delta t=10^{-4}$.  Upper
  inset: local scaling exponent (from ~\cite{b01}).}
\label{figpdf}
\end{figure}
Figure \ref{figpdf} shows the PDF of the velocity gradients in log-log
coordinates at negative values, for a Gaussian forcing restricted to
the first three Fourier modes with equal variances such that the
large-scale turnover time is order unity. Quantitative information
about the value of the exponent is obtained by measuring the ``local
scaling exponent'', i.e.\ the logarithmic derivative of the PDF
calculated in this case using least-square fits on half-decades.  It
is seen that over about five decades, the local exponent is within
less than 1\% of the value $\alpha = 7/2$ predicted by E {\it et
al.}~\cite{ekms97}.

\subsection{Self-similar forcing and multiscaling}
\label{subs:selfsimilar}

As we have seen in section~\ref{subs:intermittency}, the solutions to
the Burgers equation in a finite domain and with a large-scale forcing
have structure functions (moments of the velocity increment)
displaying a bifractal scaling behavior. Such a property can be easily
interpreted by the presence of a finite number of shocks with a size
order unity in the finite system.  Somehow this double scaling and its
relationship with singularities gives some insight on the multiscaling
properties that are expected in the case of turbulent incompressible
hydrodynamics flows. There is a general consensus that the turbulent
solutions to the Navier--Stokes equations display a full multifractal
spectrum of singularities which are responsible for a nonlinear
$p$-dependence of the scaling exponents $\zeta_p$ associated to the
scaling behavior of the $p$-th order structure
function~\cite{f95}. The construction of simple tractable models which
are able to reproduce such a behavior has motivated much work during
the last decades. Significant progress, both analytical and numerical,
has been made in confirming multiscaling in passive-scalar and
passive-vector problems~(see, e.g., \cite{fgv01} for a
review). However, the linearity of the passive-scalar and
passive-vector equations is a crucial ingredient of these studies, so
it is not clear how they can be generalized to fluid turbulence and
the Navier--Stokes equation.

After the work of Chekhlov and Yakhot~\cite{cy95}, it appeared that
the Burgers equation with self-similar forcing could be the simplest
nonlinear partial differential equation which has the potential to
display multiscaling of velocity structure functions. We report in
this section various works that tried to confirm or to weaken this
statement.  Let us consider the solutions to the one-dimensional
Burgers equation with a forcing term $f(x,t)$ which is random,
space-periodic, Gaussian and whose spatial Fourier transform has
correlation
\begin{eqnarray}
\langle{\hat f}(k,t){\hat f}(k^\prime,t^\prime)\rangle =
        2D_0\,|k|^{\beta}\,\delta(t-t^\prime)\,\delta(k+k^\prime).
\label{eq:force}
\end{eqnarray}
The exponent $\beta$ determines the scaling properties of the
forcing. When $\beta>0$ the force acts at small scales; for instance
$\beta=2$ corresponds to thermal noise for the velocity potential, and
thus to the KPZ model for interface growth~\cite{kpz86}. It is well
known in this case (see, e.g., \cite{bs95}) that the solution displays
simple scaling (usually known as KPZ scaling), such that $\zeta_q=q$
for all $q$. More generally, the case $\beta>0$ can be exactly solved
using a one-loop renormalization group approach~\cite{mhkz89}.

As stressed in \cite{hj96}, renormalization group techniques fail when
$\beta<0$ and the forcing acts mostly at large scales and non-linear
terms play a crucial role. When $\beta<-3$, the forcing is
differentiable in the space variable, the solution is piecewise smooth
and contains a finite number of shocks with sizes order unity. The
scaling exponents are then $\zeta_p = \min\,(1,\,p)$. In the case of
non-differentiable forcing ($-3<\beta<0$), the presence of order-unity
shocks and dimensional arguments suggest that the scaling exponents
are $\zeta_p = \min\,(1,\, -p\beta/3)$. However, very little is known
regarding the distribution of shocks with intermediate sizes. In
particular, there is no clear evidence whether or not they form a
self-similar structure at small scales. We summarize here some studies
which were done on Burgers turbulence with self-similar forcing to
show how difficult it might be to measure scaling laws of structure
functions and in particular how logarithmic corrections can masquerade
anomalous scaling.

\begin{figure}[ht]
  \centerline{\includegraphics[width=0.35\textwidth]{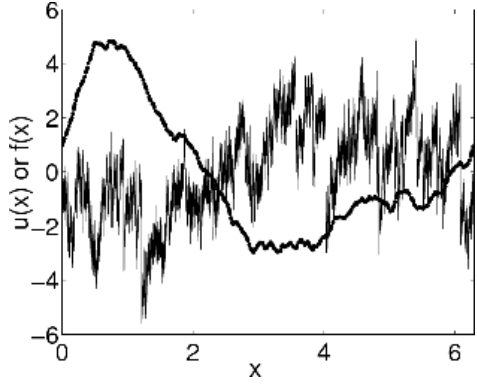}}
  \caption{Representative snapshots of the velocity $v$ (jagged line)
    in the statistically stationary r\'egime, and of the integral of
    the force $f$ over a time step (rescaled for plotting purposes).}
  \label{fig:ufpic}
\end{figure}
For this we focus on the case $\beta=-1$ which has attracted much
attention; indeed, dimensional analysis suggests that $\zeta_p = p/3$
when $p\le 3$, leading to a K41-type $-5/3$ energy spectrum. Early
studies~\cite{cy95b,cy95} seemed to confirm this prediction using
pseudo-spectral viscous numerical simulations at rather low
resolutions (around ten thousands gridpoints). It was moreover argued
in~\cite{hj96,hj97} that a self-similar forcing with $-1 < \beta < 0$,
could lead to genuine multifractality.  The lack of accuracy in the
determination of the scaling exponents left open the question of a
weak anomalous deviation from the dimensional prediction.  This
question was recently revisited in~\cite{mbpf05} with high-resolution
inviscid numerical simulations using the fast Legendre transform
algorithm (see section~\ref{sssec:flt}). A typical snapshot of the
forcing and of the solution in the stationary r\'{e}gime are
represented in figure~\ref{fig:ufpic}. It is clear that because of
shocks the velocity develops small-scale fluctuations much stronger
than those present in the force. However one notices that shock
dynamics and spatial finiteness of the system lead, as predicted, to
the presence of few shocks with order-unity sizes.

\begin{figure}[ht]
  \centerline{\includegraphics[width=0.35\textwidth]{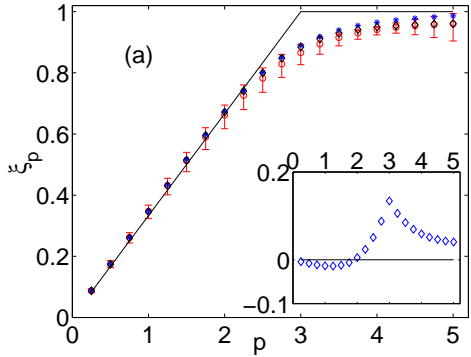}}
  \caption{Scaling exponents $\zeta_p$ versus order $p$ for $N =
    2^{16} (\diamond), \,2^{18} (\ast)$, and $2^{20} (\circ)$ grid
    points. Error bars (see text) are shown for the case $N =
    2^{20}$. The deviation of $\zeta_p$ from the exponents for
    bifractal scaling (full lines), shown as an inset, naively
    suggests multiscaling (from~\cite{mbpf05})}
  \label{fig:xip_inset}
\end{figure}
Structure functions were measured with high accuracy. They typically
exhibit a power-law behavior over nearly three decades in length
scale; this is more than two decades better than in~\cite{cy95}.  In
principle one expects to be able to measure the scaling exponents with
enough accuracy to decide between bifractality and
multiscaling. Surprisingly the naive analysis summarized in
figure~\ref{fig:xip_inset} does suggest multiscaling: the exponents
$\zeta_p$ deviate significantly from the bifractal-scaling prediction
(full lines).  Since the goal here is to have a precise handle on the
scaling properties of velocity increments, it is important to
carefully define how the scaling exponents are measured.  They are
estimated from the average logarithmic derivative of $S^{\rm abs}_p(r)
= \langle |v(x+r)-v(x)|^p \rangle$ over almost two decades in the
separation $r$. The error bars shown are given by the maximum and
minimum deviations from this mean value in the fitting range.  Note
also that the observed multiscaling is supported by the fact that
there is no substantial change in the value of the exponents when
changing the number $N$ of grid points in the simulation from $2^{16}$
to $2^{20}$: any dependence of $\zeta_p$ upon $N$ is much less than
the error bars determined through the procedure described above.

\begin{figure}[ht]
  \centerline{\includegraphics[width=0.35\textwidth]{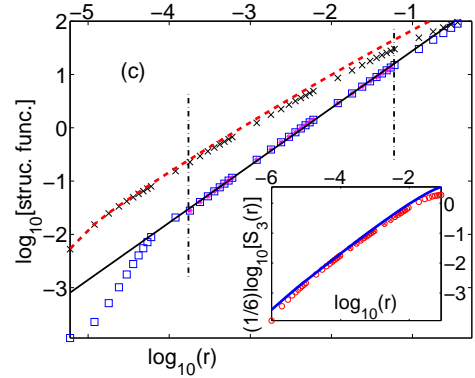}}
  \caption{Log-log plots of $S_3^{\rm abs}(r)$ (dashed line), $S_3(r)$
    (crosses), and $\langle (\delta^{+}v)^3\rangle $ (squares) versus
    $r$.  The continuous line is a least-square fit to the range of
    points limited by two vertical dashed lines in the plot.  Inset:
    An explicit check of the von K\'arm\'an--Howarth relation
    (\ref{eq:exact}) from the simulations with $N = 2^{20}$ reported
    in~\cite{mbpf05}. The dashed curve is the integral of the spatial
    part of the forcing correlation and the circles represent the
    numerical computation of the left-hand side.}
  \label{fig:s3all_inset}
\end{figure}
As found in~\cite{mbpf05}, the observed deviations of the scaling
exponents from bifractality are actually due to the contamination by
subleading terms in $S^{\rm abs}_p(r)$. To quantify this effect, let
us focus on the third-order structure function ($p=3$) for which one
measures $\zeta_3 \approx 0.85 \pm 0.02$ over nearly four decades (see
figure~\ref{fig:s3all_inset}). To estimate subleading terms we first
notice that the third-order structure function $S_3(r) \equiv \langle
(v(x+r) - v(x))^3 \rangle$, which is defined, this time, without the
absolute value, obeys an analog of the von K\'arm\'an--Howarth
relation in fluid turbulence, namely
\begin{equation}
\frac{1}{6} S_3(r) = \int_0^{r} b(r^\prime) \mathrm{d}r^\prime,
\label{eq:exact}
\end{equation}
where $b(\cdot)$ denotes the spatial part of the force correlation
function, i.e.\ $\langle f(x+r,t^{\prime})f(x,t)\rangle =
b(r)\delta(t-t^{\prime})$.  This relation, together with the
correlation (\ref{eq:force}) and $\beta = -1$, implies the behavior
$S_3(r)\sim r \ln r$ at small separations $r$. As seen in
figure~\ref{fig:s3all_inset}, the graph of $S_3(r)$ in log-log
coordinates indeed displays a significant curvature which is a
signature of logarithmic corrections. The next step consists in
decomposing the velocity increments $\delta_r v = v(x+r,t)-v(x,t)$
into their positive $\delta_r^+v$ and negative $\delta_r^-v$ parts. It
is clear that
\begin{eqnarray}
  && S^{\rm abs}_3(r) = - \la (\delta_r^-v)^3\ra + \la (\delta_r^+v)^3
  \ra, \nonumber \\ && S_3(r) = \la (\delta_r^-v)^3 \ra + \la
  (\delta_r^+v)^3 \ra ,
  \label{eq:posnegpartsS3}
\end{eqnarray}
so that $\la (\delta_r^+v)^3 \ra = (S^{\rm abs}_3(r) +S_3(r))/2$. As
seen in figure~\ref{fig:s3all_inset} the log-log plot of $\la
(\delta_r^+v)^3 \ra$ as a function of $r$ is nearly a straight line
with slope $\approx 1.07$ very close to unity. This observation is
confirmed in~\cite{mbpf05} by independently measuring the PDFs of
positive and negative velocity increments. Assuming that $\la
(\delta_r^+v)^3 \ra \sim B\,r$, one obtains the following prediction
for the small-$r$ behaviors of the third-order structure functions
\begin{eqnarray}
  &&S^{\rm abs}_3(r) \sim - A r \ln r + B\,r, \nonumber \\ && S_3(r)
  \sim A r \ln r + B\,r.
  \label{eq:behaveS3}
\end{eqnarray}
This suggests that the only difference in the small-separation
behaviors of $S^{\rm abs}_3(r)$ and $S_3(r)$ is the sign in the
balance between the leading term $\propto r \ln r$ and the subleading
term $\propto r$.  In a log-log plot this difference amounts to
shifting the graph away from where it is most curved and thus makes it
straighter, albeit with a (local) slope which is not unity. This
explains why significant deviations from $1$ are observed for
$\zeta_3$. Note that a similar approach can be used for higher-order
structure functions. It leads for instance to $S_4(r) \approx C r - D
r^{4/3}$, where $C$ and $D$ are two positive constants.  The negative
sign before the sub-leading term~$(r^{4/3})$ is crucial. It implies
that, for any finite $r$, a naive power-law fit to $S_4$ can yield a
scaling exponent less than unity. The presence of sub-leading,
power-law terms with opposite signs also explains the small apparent
``anomalous'' scaling behavior observed for other values of $p$ in the
simulations.  Note that similar artifacts involving two competing
power-laws have been described in~\cite{bclst04,bch07}.

The work reported in this section indicates that a naive
interpretation of numerical measurements might result in predicting
artificial anomalous scaling laws. In the case of Burgers turbulence
for which high-resolution numerics are available and statistical
convergence of the averages can be guaranteed, we have seen that it is
not too difficult to identify the numerical artifacts which are
responsible for such a masquerading. However this is not always the
case. For instance, it seems reasonable enough to claim that attacking
the problem of multiscaling in spatially extended nonlinear systems,
such as Navier--Stokes turbulence, requires considerable theoretical
insight that must supplement sophisticated and heavy numerical
simulations and experiments. Note finally that, up to now, the
question of the presence or not of anomalous scaling laws in the
Burgers equation with a self-similar forcing with exponent
$-1<\beta<0$ remains largely open.

%%%%%%%%%%%%%%%%%%%%%%%%%%%%%%%%%%%%%%%%%%%%%%
\section{Concluding remarks and open questions}
\label{s:conclusion}

This review summarizes recent work connected with the Burgers
equation. Originally this model was introduced as a simplification of
the Navier--Stokes equation with the hope of shedding some light on
issues such as turbulence. This hope did not materialize. Nevertheless
many of the interesting questions that have been addressed for Burgers
turbulence are eventually transpositions of similar questions for
Navier--Stokes turbulence. One particularly important instance is the
issue of universality with respect to the form of the forcing and of
the initial condition. For Burgers turbulence most of the universal
features, such as scaling exponents or functional forms of PDF tails
are dominated by the presence of shocks and other singularities in the
solution. This applies both to the case of decaying turbulence driven
by random initial conditions and randomly forced turbulence. In the
latter case one is mostly interested in analysis of stationary
properties of solutions, for example stationary distribution for
velocity increments or gradients. Another set of questions is
motivated by more mathematical considerations.  It mainly concerns the
construction of a stationary invariant measure when Burgers dynamics
in a finite-size domain is supplemented by an external random source
of energy. Again it has been shown that the presence of shocks, and in
particular of global shocks, plays a crucial role in the construction
of the statistically stationary solution. Both physical and
mathematical questions lead to a similar answer: one first needs to
describe and control shocks. The main message to retain for
hydrodynamical turbulence is hence a strong confirmation of the common
wisdom that it cannot be fully understood without a detailed
description of singularities. Moreover, the behavior depends not only
on the local structure of singularities, but also on their
distribution at larger scales.  Here a word of caution: for
incompressible fully developped Navier--Stokes turbulence, we have no
evidence that the universal scaling properties observed in experiments
and simulations stem from real singularities. Indeed the issue of a
finite-time blow-up of the three-dimensional Euler equation is still
open (see, e.g.~\cite{fmb03}). Another important observation that can
be drawn from the study of Burgers turbulence is that both the tools
used and the answers obtained strongly depend on the kind of setting
one considers: decay versus forced turbulence, finite-size versus
infinite-size systems, smooth versus self-similar forcing, etc.

Besides turbulence, the random Burgers equation has various
applications in cosmology, in non-equilibrium statistical physics and
in disordered media. Among them, the connection to the problem of
directed polymers has attracted much attention. As already noted in
the Introduction, there is a mathematical equivalence between the
zero-viscosity limit of the forced Burgers equation and the
zero-temperature limit for directed polymers. We have seen in
section~\ref{subs:extended} that the so-called KPZ scaling, which
usually is derived for a finite temperature, can be established can be
established also in the zero-temperatur limit, using the action
minimizer representation. Such an observation leads to two related
questions: to what extent can the limit of zero temperature give an
insight into finite-temperature polymer dynamics and how can the
global minimizer formalism be extended to tackle the
finite-temperature setting?  It looks plausible that in polymer
dynamics, or more generally in the study of random walks in a random
potential, the trajectories carrying most of the Gibbs probability
weight are defining corridors in space time. These objects can
concentrate near the trajectories of global minimizers but, at the
moment, there is no formalism to describe them, nor attempts to
quantify their contribution to the Gibbs statistics.

Another important open question concerns the multi-dimensional
extensions of the Burgers equation. As we have seen, when the forcing
is potential, the potential character of the velocity field is
conserved by the dynamics. This leads to the construction of
stationary solutions which carry many similarities with the
one-dimensional case. Up to now there is only limited understanding of
what happens when the potentiality assumption of the flow is
dropped. This problem has of course concrete applications in gas
dynamics and for disperse inelastic granular media (see, e.g.,
\cite{bcdr99}).  An interesting question concerns the construction of
the limit of vanishing viscosity, given that the Hopf--Cole
transformation is inapplicable in the non-potential case.
Understanding extensions of the viscous limiting procedure to the
non-potential case might give new insight into the problem of the
large Reynolds number limit in incompressible turbulence.  Another
question related to non-potential flow concerns the interactions
between vorticity and shocks. For instance, in two dimensions the
vorticity is transported by the flow. This results in its growth in
the highly compressible regions of the flow. The various singularities
of the velocity field should hence be strongly affected by the flow
rotation and, in particular, the shocks are expected to have a
spiraling structure.

We finish with few remarks on open mathematical problems. As we have
seen in the one-dimensional case, one can rigorously prove
hyperbolicity of the global minimizer. In the multi-dimensional case
it is also possible to establish the existence and, in many cases,
uniqueness of the global minimizer. However, the very important
question of its hyperbolicity is still an open problem. If proven,
hyperbolicity would allow for rigorous analysis of the regularity
properties of the stationary solutions and of the topological shocks.
There are many interesting problems -- even basic issues of existence
and uniqueness -- in the non-compact case where at present a
mathematical theory is basically absent.  Finally, a very challenging
open problem concerns the extension of the results on the evolution of
matter inside shocks to the case of general Hamilton-Jacobi equations.

\begin{ack}
Over the years of our work on Burgers turbulence, we profited a lot
from numerous discussions with Uriel Frisch whose influence on our
work is warmly acknowledged. We also want to express our sincere
gratitude to all of our collaborators: W.\ E, U.\ Frisch, D.\ Gomes,
V.H.~Hoang, R.\ Iturriaga, D.\ Khmelev, A.\ Mazel, D.~Mitra, P.\
Padilla, R.\ Pandit, Ya.\ Sinai, A.\ Sobolevski{\u\i}, and B.\
Villone.\\ While writing this article, we benefited from discussions
with M.\ Blank, I.\ Bogaevsky, K.\ Domelevo, V.\ Epstein, and
A.~Sobolevski{\u\i}. Finally, our thanks go to Itamar Procaccia whose
encouragements and patience are greatly appreciated.
\end{ack}

\bibliographystyle{plain}
\bibliography{burgulence}

\begin{thebibliography}{100}

\bibitem{an01}
S.~Anisov.
\newblock Lowerbounds for the complexity of some 3-dimensional manifolds.
\newblock preprint, 2001.

\bibitem{a63}
V.~Arnol'd.
\newblock Small denominators and problems of stability of motion in classical
  and celestial mechanics.
\newblock {\em Uspekhi Mat.\ Nauk}, 18:91--192, 1963.

\bibitem{a83}
S.~Aubry.
\newblock The twist map, the extended {F}renkel--{K}ontorova model and the
  devil's staircase.
\newblock {\em Physica D}, 7:240--258, 1983.

\bibitem{afnb97}
E.~Aurell, U.~Frisch, A.~Noullez, and M.~Blank.
\newblock Bifractality of the {D}evil's staircase appearing in {B}urgers
  equation with {B}rownian initial velocity.
\newblock {\em J.\ Stat.\ Phys.}, 88:1151--1164, 1997.

\bibitem{bs95}
A.-L. Barab{\'{a}}si and H.E. Stanley.
\newblock {\em Fractal Concepts in Surface Growth}.
\newblock Cambridge University Press, 1995.

\bibitem{b01}
J.~Bec.
\newblock Universality of velocity gradients in forced {B}urgers turbulence.
\newblock {\em Phys.\ Rev.\ Lett.}, 87:104501 (1--4), 2001.

\bibitem{bch07}
J.~Bec, M.~Cencini, and R.~Hillebrand.
\newblock Heavy particles in incompressible flows: the large {S}tokes number
  asymptotics.
\newblock {\em Physica D}, 226:11--22, 2007.

\bibitem{bf00}
J.~Bec and U.~Frisch.
\newblock Pdf's of derivatives and increments for decaying {B}urgers
  turbulence.
\newblock {\em Phys.\ Rev.\ E}, 61:1395--1402, 2000.

\bibitem{bfk00}
J.~Bec, U.~Frisch, and K.~Khanin.
\newblock Kicked {B}urgers turbulence.
\newblock {\em J.\ Fluid Mech.}, 416:239--267, 2000.

\bibitem{bik02}
J.~Bec, R.~Iturriaga, and K.~Khanin.
\newblock Topological shocks in {B}urgers turbulence.
\newblock {\em Phys.\ Rev.\ Lett.}, 89:024501 (1--4), 2002.

\bibitem{bk03}
J.~Bec and K.~Khanin.
\newblock Forced {B}urgers equation in an unbounded domain.
\newblock {\em J.\ Stat.\ Phys.}, 113:741--759, 2003.

\bibitem{bcdr99}
E.~Ben-Naim, S.~Y. Chen, G.~D. Doolen, and S.~Redner.
\newblock Shocklike dynamics of inelastic gases.
\newblock {\em Phys.\ Rev.\ Lett.}, 83:4069--4072, 1999.

\bibitem{bb00}
J.-D. Benamou and Y.~Brenier.
\newblock A computational fluid dynamics solution to the {M}onge--{K}antorovich
  mass transfer problem.
\newblock {\em Numerische Mathematik}, 84:375--393, 2000.

\bibitem{bgk96}
D.~Bernard, K.~Gaw{\c{e}}dzki, and A.~Kupiainen.
\newblock Anomalous scaling in the {N}-point functions of a passive scalar.
\newblock {\em Phys.\ Rev.\ E}, 54:2564--2572, 1996.

\bibitem{b81}
D.P. Bertsekas.
\newblock A new algorithm for the assignment problem.
\newblock {\em Math.\ Programming}, 21:152--171, 1981.

\bibitem{bclst04}
L.~Biferale, M.~Cencini, A.~Lanotte, and M.~Sbragaglia.
\newblock Anomalous scaling and universality in hydrodynamic systems with
  power-law forcing.
\newblock {\em New J.\ Phys.}, 4:37, 2004.

\bibitem{boga-physicad}
I.A. Bogaevsky.
\newblock Perestroikas of shocks and singularities of minimum functions.
\newblock {\em Physica D}, 173:1--28, 2002.

\bibitem{bog04}
I.A. Bogaevsky.
\newblock Matter evolution in burgulence.
\newblock preprint, 2004.

\bibitem{b98}
S.~Boldyrev.
\newblock {B}urgers turbulence, intermittency, and non-universality.
\newblock {\em Phys.\ Plasmas}, 5:1681--1687, 1998.

\bibitem{blp04}
S.~Boldyrev, T.~Linde, and A.~Polyakov.
\newblock Velocity and velocity-difference distributions in {B}urgers
  turbulence.
\newblock {\em Phys.\ Rev.\ Lett.}, 93:184503, 2004.

\bibitem{bm96}
J.-P. Bouchaud and M.~M{\'{e}}zard.
\newblock Velocity fluctuations in forced {B}urgers turbulence.
\newblock {\em Phys.\ Rev.~E}, 54:5116--5121, 1996.

\bibitem{bmp95}
J.-P. Bouchaud, M.~M{\'{e}}zard, and G.~Parisi.
\newblock Scaling and intermittency in {B}urgers turbulence.
\newblock {\em Phys.\ Rev.\ E}, 52:3656--3674, 1995.

\bibitem{b89}
Y.~Brenier.
\newblock Un algorithme rapide pour le calcul de transform\'ees de
  {L}egendre-{F}enchel discr\`etes.
\newblock {\em C.\ R.\ Acad.\ Sci.\ Paris S\'er.\ I Math.}, 308:587--589, 1989.

\bibitem{b91}
Y.~Brenier.
\newblock Polar factorization and monotone rearrangement of vector-valued
  functions.
\newblock {\em Comm.\ Pure Appl.\ Math.}, 44:375--417, 1991.

\bibitem{bfhlmms03}
Y.~Brenier, U.~Frisch, M.~H{\'{e}}non, G.~Loeper, S.~{Matarrese},
  R.~{Mohayaee}, and A.~Sobolevski{\u\i}.
\newblock {Reconstruction of the early Universe as a convex optimization
  problem}.
\newblock {\em Mon.\ Not.\ Roy.\ Astron.\ Soc.}, 346:501--524, 2003.

\bibitem{b39}
J.M. Burgers.
\newblock Mathematical examples illustrating relations occuring in the theory
  of turbulent fluid motion.
\newblock {\em Verhand.\ Kon.\ Neder.\ Akad.\ Wetenschappen, Afd.\ Natuurkunde,
  Eerste Sectie}, 17:1--53, 1939.

\bibitem{b74}
J.M. {B}urgers.
\newblock {\em The Nonlinear Diffusion Equation}.
\newblock D. Reidel, Dordrecht, 1974.

\bibitem{cefv00}
M.~Chaves, G.~Eyink, U.~Frisch, and M.~Vergassola.
\newblock Universal decay of scalar turbulence.
\newblock {\em Phys.\ Rev.\ Lett.}, 86:2305--2308, 2000.

\bibitem{cy95b}
A.~Chekhlov and V.~Yakhot.
\newblock Kolmogorov turbulence in a random-force-driven {B}urgers equation.
\newblock {\em Phys.\ Rev.\ E}, 51:R2739--R2742, 1995.

\bibitem{cy95}
A.~Chekhlov and V.~Yakhot.
\newblock Kolmogorov turbulence in a random-force-driven {B}urgers equation:
  anomalous scaling and probability density functions.
\newblock {\em Phys.\ Rev.\ E}, 52:5681--5684, 1995.

\bibitem{cfkl95}
M.~Chertkov, G.~Falkovich, I.~Kolokolov, and V.~Lebedev.
\newblock Normal and anomalous scaling of the fourth-order correlation function
  of a randomly advected passive scalar.
\newblock {\em Phys.\ Rev.\ E}, 52:4924--4941, 1995.

\bibitem{css00}
D.~Chowdhury, L.~Santen, and A.~Schadschneider.
\newblock Statistical physics of vehicular traffic and some related systems.
\newblock {\em Phys.\ Rep.}, 329:199--329, 2000.

\bibitem{c51}
J.D. Cole.
\newblock On a quasi-linear parabolic equation occurring in aerodynamics.
\newblock {\em Quart.\ Appl.\ Math.}, 9:225--236, 1951.

\bibitem{cl95}
P.~Coles and F.~Lucchin.
\newblock {\em Cosmology : the origin and evolution of cosmic structures}.
\newblock J.~Wiley and sons, 1995.

\bibitem{dl85}
R.~Dautray and J.-L. Lions.
\newblock {\em Analyse math\'ematique et calcul num\'erique pour les sciences
  et les techniques.\ {T}ome 3}.
\newblock Masson, Paris, 1985.

\bibitem{e99}
W.~E.
\newblock {A}ubry--{M}ather theory and periodic solutions for the forced
  {B}urgers equation.
\newblock {\em Comm.\ Pure Appl.\ Math.}, 52:0811--0828, 1999.

\bibitem{ekms97}
W.~E, K.~Khanin, A.~Mazel, and Ya. Sinai.
\newblock Probability distribution functions for the random forced {B}urgers
  equation.
\newblock {\em Phys.\ Rev.\ Lett.}, 78:1904--1907, 1997.

\bibitem{ekms00}
W.~E, K.~Khanin, A.~Mazel, and Ya. Sinai.
\newblock Invariant measures for {B}urgers equation with stochastic forcing.
\newblock {\em Ann.\ Math.}, 151:877--960, 2000.

\bibitem{eve99a}
W.~E and E.~Vanden~Eijnden.
\newblock Asymptotic theory for the probability density functions in {B}urgers
  turbulence.
\newblock {\em Phys.\ Rev.\ Lett.}, 83:2572--2575, 1999.

\bibitem{eve99b}
W.~E and E.~Vanden~Eijnden.
\newblock On the statistical solution of the riemann equation and its
  implications for {B}urgers turbulence.
\newblock {\em Phys.\ Fluids}, 11:2149--2153, 1999.

\bibitem{eve00b}
W.~E and E.~Vanden~Eijnden.
\newblock Another note on {B}urgers turbulence.
\newblock {\em Phys.\ Fluids}, 12:149--154, 2000.

\bibitem{eve00}
W.~E and E.~Vanden~Eijnden.
\newblock Statistical theory for the stochastic {B}urgers equation in the
  inviscid limit.
\newblock {\em Comm.\ Pure Appl.\ Math.}, 53:852--901, 2000.

\bibitem{e89}
L.~Evans.
\newblock The perturbed test function method for viscosity solutions of
  nonlinear {P}{D}{E}.
\newblock {\em Proc.\ Roy.\ Soc.\ Edinburgh Sect.\ A}, 111:359--375, 1989.

\bibitem{et00}
G.L. Eyink and D.J. Thomson.
\newblock Free decay of turbulence and breakdown of self-similarity.
\newblock {\em Phys.\ Fluids}, 12:477--479, 2000.

\bibitem{ex00}
G.L. Eyink and J.~Xin.
\newblock Self-similar decay in the {K}raichnan model of a passive scalar.
\newblock {\em J.\ Stat.\ Phys.}, 100:679--741, 2000.

\bibitem{fgv01}
G.~Falkovich, K.~Gaw{\c{e}}dzki, and M.~Vergassola.
\newblock Particles and fields in fluid turbulence.
\newblock {\em Rev.\ Mod.\ Phys.}, 73:913--976, 2001.

\bibitem{f97}
A.~Fathi.
\newblock Th\'eor\`eme {K}{A}{M} faible et th\'eorie de {M}ather sur les
  syst\`emes lagrangiens.
\newblock {\em C.\ R.\ Acad.\ Sci.\ Paris S\'er.\ I Math.}, 324:1043--1046,
  1997.

\bibitem{fbook}
A.~Fathi.
\newblock {\em Weak {K}{A}{M} theorem in {L}agrangian dynamics}.
\newblock Cambridge University Press, Cambridge, 2003.

\bibitem{fellervol2}
W.~Feller.
\newblock {\em An introduction to probability theory and its applications},
  volume~2.
\newblock J.~Wiley and sons, 1995.

\bibitem{fs93}
W.~Fleming and H.~Soner.
\newblock {\em Controlled {M}arkov processes and viscosity solutions}.
\newblock Springer--Verlag, New York, 1993.

\bibitem{ff83}
J.-D. Fournier and U.~Frisch.
\newblock L'{\'{e}}quation de {B}urgers d{\'{e}}terministe et statistique.
\newblock {\em J.\ M{\'{e}}c.\ Th{\'{e}}or.\ Appl.}, 2:699--750, 1983.

\bibitem{fk39}
J.~Frenkel and T.~Kontorova.
\newblock On the theory of plastic deformation and twinning.
\newblock {\em Acad.\ Sci.\ U.S.S.R.\ J.\ Phys.}, 1:137--149, 1939.

\bibitem{f95}
U.~Frisch.
\newblock {\em Turbulence\,: the Legacy of {A}.{N}.~{K}olmogorov}.
\newblock Cambridge University Press, 1995.

\bibitem{fbv01}
U.~Frisch, J.~Bec, and B.~Villone.
\newblock Universal law for the distribution of density in the
  {B}urgers/adhesion model.
\newblock {\em Physica D}, 152--153:620--635, 2001.

\bibitem{fmms02}
U.~Frisch, S.~{Matarrese}, R.~{Mohayaee}, and A.~Sobolevski{\u\i}.
\newblock {A reconstruction of the initial conditions of the Universe by
  optimal mass transportation}.
\newblock {\em Nature}, 417:260--262, 2002.

\bibitem{fmb03}
U.~Frisch, T.~Matsumoto, and J.~Bec.
\newblock Singularities of {E}uler flow? {N}ot out of the blue!
\newblock {\em J.\ Stat.\ Phys.}, 113:761--781, 2003.

\bibitem{gk95}
K.~Gaw{\c{e}}dzki and A.~Kupiainen.
\newblock Anomalous scaling of the passive scalar.
\newblock {\em Phys.\ Rev.\ Lett.}, 75:3834--3837, 1995.

\bibitem{gikp05}
D.~Gomes, R.~Iturriaga, K.~Khanin, and P.~Padilla.
\newblock Viscosity limit of stationary distributions for the random forced
  {B}urgers equation.
\newblock {\em Moscow Math. Journal}, 5:613--631, 2005.

\bibitem{gk98}
T.~Gotoh and R.H. Kraichnan.
\newblock Steady-state {B}urgers turbulence with large-scale forcing.
\newblock {\em Phys.\ Fluids}, 10:2859--2866, 1998.

\bibitem{gm96}
V.~Gurarie and A.~Migdal.
\newblock Instantons in the {B}urgers equation.
\newblock {\em Phys.\ Rev.\ E}, 54:4908--4914, 1996.

\bibitem{gms91}
S.~Gurbatov, A.~Malakhov, and A.~Saichev.
\newblock {\em Nonlinear Random Waves and Turbulence in Nondispersive Media:
  Waves, Rays, Particles}.
\newblock Manchester University Press, 1991.

\bibitem{gs84}
S.~Gurbatov and A.~Saichev.
\newblock Probability distribution and spectra of potential hydrodynamic
  turbulence.
\newblock {\em Radiophys.\ Quant.\ Electr.}, 27(4):303--313, 1984.

\bibitem{gsaft97}
S.~Gurbatov, S.~Simdyankin, E.~Aurell, U.~Frisch, and G.~Toth.
\newblock On the decay of {B}urgers turbulence.
\newblock {\em J.\ Fluid Mech.}, 344:339--374, 1997.

\bibitem{hj96}
F.~Hayot and C.~Jayaprakash.
\newblock Multifractality in the stochastic {B}urgers equation.
\newblock {\em Phys.\ Rev.\ E}, 54:4681--4684, 1996.

\bibitem{hj97}
F.~Hayot and C.~Jayaprakash.
\newblock From scaling to multiscaling in the stochastic {B}urgers equation.
\newblock {\em Phys.\ Rev.\ E}, 56:4259--4262, 1997.

\bibitem{hk03}
V.H. Hoang and K.~Khanin.
\newblock Random {B}urgers equation and {L}agrangian systems in non-compact
  domains.
\newblock {\em Nonlinearity}, 16:819--842, 2003.

\bibitem{h50}
E.~Hopf.
\newblock The partial differential equation $u_t+uu_x=u_{xx}$.
\newblock {\em Comm.\ Pure Appl.\ Math.}, 3:201--230, 1950.

\bibitem{ik01}
R.~Iturriaga and K.~Khanin.
\newblock Two results on invariant measures for random {L}agrangian systems and
  random {B}urgers equation.
\newblock In {\em Proceedings of the Third European Congress of Mathematics,
  Barcelona July 2000}. Birkhauser, 2001.
\newblock Progress in Mathematics.

\bibitem{ik03}
R.~Iturriaga and K.~Khanin.
\newblock {B}urgers turbulence and random {L}agrangian systems.
\newblock {\em Comm.\ Math.\ Phys.}, 232:377--428, 2003.

\bibitem{jkm99}
H.R. Jauslin, H.O. Kreiss, and J.~Moser.
\newblock On the forced {B}urgers equation with periodic boundary conditions.
\newblock In {\em Differential equations: La Pietra 1996 (Florence)}, pages
  133--153. Amer.\ Math.\ Soc., Providence, RI, 1999.

\bibitem{jetal98}
A.~Jenkins, C.S. Frenk, F.R. Pearce, P.A. Thomas, J.M. Colberg, S.D. White,
  H.M. Couchman, J.A. Peacock, G.~Efstathiou, and A.H. Nelson.
\newblock Evolution of structure in cold dark matter {U}niverses.
\newblock {\em Astrophys.\ J.}, 499:20--40, 1998.

\bibitem{js98}
J.V. Jos{\'{e}} and E.J. Saletan.
\newblock {\em Classical dynamics: a contemporary approach}.
\newblock Cambridge University Press, 1998.

\bibitem{k42}
L.V. Kantorovich.
\newblock On the translocation of masses.
\newblock {\em C.\ R.\ (Doklady) Acad.\ Sci.\ USSR}, 37:199--201, 1942.

\bibitem{kpz86}
M.~Kardar, G.~Parisi, and Y.-C. Zhang.
\newblock Dynamical scaling of growing interfaces.
\newblock {\em Phys.\ Rev.\ Lett.}, 56:889--892, 1986.

\bibitem{kz87}
M.~Kardar and Y.-C. Zhang.
\newblock Scaling of directed polymers in random media.
\newblock {\em Phys.\ Rev.\ Lett.}, 58:2087--2090, 1987.

\bibitem{kks03}
K.~Khanin, D.~Khmelev, and A.~Sobolevski{\u\i}.
\newblock A blow-up phenomenon in the {H}amilton--{J}acobi equation in an
  unbounded domain.
\newblock In {\em Proceedings of ``Idempotent Mathematics and Mathematical
  Physics'' workshop at the Erwin Schr{\"o}dinger Institute in Vienna}, 2003.

\bibitem{k79}
S.~Kida.
\newblock Asymptotic properties of {B}urgers turbulence.
\newblock {\em J.\ Fluid Mech}, 93:337--377, 1979.

\bibitem{kpsm92}
L.~Kofman, D.~Pogosyan, S.~Shandarin, and A.L. Melott.
\newblock Coherent structures in the universe and the adhesion model.
\newblock {\em Astrophys.\ J.}, 393:437--449, 1992.

\bibitem{k41decay}
A.N. Kolmogorov.
\newblock On degeneration (decay) of isotropic turbulence in an incompressible
  viscous liquid.
\newblock {\em Dokl.\ Akad.\ Nauk SSSR}, 31:538--540, 1941.

\bibitem{k57}
A.N. Kolmogorov.
\newblock Th\'eorie g\'en\'erale des syst\`emes dynamiques et m\'ecanique
  classique.
\newblock In {\em Proceedings of the International Congress of Mathematicians,
  Amsterdam, 1954, Vol.\ 1}, pages 315--333. Erven P. Noordhoff N.V.,
  Groningen, 1957.

\bibitem{kl00}
I.~Kolokolov and V.~Lebedev.
\newblock About probability tails in forced {B}urgers turbulence.
\newblock preprint, 2000.

\bibitem{k75}
S.N. Kru{\v{z}}kov.
\newblock Generalized solutions to {H}amilton--{J}acobi equations of the
  eikonal type.
\newblock {\em Mat.\ Sb.\ (N.S.)}, 98:450--493, 1975.

\bibitem{l57}
P.D. Lax.
\newblock Hyperbolic systems of conservation laws {II}.
\newblock {\em Comm.\ Pure Appl.\ Math.}, 10:537--566, 1957.

\bibitem{l97}
M.~Lesieur.
\newblock {\em Turbulence in Fluids}.
\newblock Fluid Mechanics and Its Applications~40. Kluwer, 1997.

\bibitem{l82}
P.-L. Lions.
\newblock {\em Generalized solutions of {H}amilton-{J}acobi equations},
  volume~69.
\newblock Pitman Research Notes in Math., 1982.

\bibitem{m82}
J.N. Mather.
\newblock Existence of quasi-periodic orbits for twist homeomorphisma of the
  annulus.
\newblock {\em Topology}, 21:457--467, 1982.

\bibitem{ma90}
S.~Matveev.
\newblock Complexity theory of three-dimensional manifolds.
\newblock {\em Acta Appl.\ Math.}, 19(2):101--130, 1990.

\bibitem{mhkz89}
E.~Medina, T.~Hwa, M.~Kardar, and Y.C. Zhang.
\newblock Burgers equation with correlated noise: renormalization-group
  analysis and applications to directed polymers and interface growth.
\newblock {\em Phys.\ Rev.\ A}, 39:3053--3075, 1989.

\bibitem{mp90}
M.~M{\'{e}}zard and G.~Parisi.
\newblock Interfaces in a random medium and replica symmetry breaking.
\newblock {\em J.\ Phys.\ A}, 23:L1229--L1234, 1990.

\bibitem{mpv87}
M.~M{\'{e}}zard, G.~Parisi, and M.A. Virasoro.
\newblock {\em Spin glass theory and beyond}.
\newblock World Scientific, Singapore, 1987.

\bibitem{mbpf05}
D.~Mitra, J.~Bec, R.~Pandit, and U.~Frisch.
\newblock Is multiscaling an artifact in the stochastically forced {B}urgers
  equation.
\newblock {\em Phys.\ Rev.\ Lett.}, 84:194501, 2005.

\bibitem{msw95}
S.A. Molchanov, D.~Surgailis, and W.A. Woyczynski.
\newblock Hyperbolic asymptotics in {B}urgers' turbulence and extremal
  processes.
\newblock {\em Comm.\ Math.\ Phys.}, 168:209--226, 1995.

\bibitem{m62}
J.~Moser.
\newblock On invariant curves of area-preserving mappings of an annulus.
\newblock {\em Nachr.\ Akad.\ Wiss.\ G\"ottingen Math.-Phys.\ K1.II 1}, pages
  1--20, 1962.

\bibitem{nv94}
A.~Noullez and M.~Vergassola.
\newblock A fast {L}egendre transform algorithm and applications to the
  adhesion model.
\newblock {\em J.\ Sci.\ Comput.}, 9:259--281, 1994.

\bibitem{o57}
O.A. Oleinik.
\newblock Discontinuous solutions of non-linear differential equations.
\newblock {\em Uspekhi Mat.\ Nauk (N.S.)}, 12(3):3--73, 1957.

\bibitem{o77}
S.A. Orszag.
\newblock Statistical theory of turbulence.
\newblock In R.~Balian and J.L. Peube, editors, {\em Fluid Dynamics, Les
  Houches 1973}, pages 237--374. Gordon and Breach, 1977.

\bibitem{ol00}
S.~Ossia and M.~Lesieur.
\newblock Energy backscatter in {L}{E}{S} of 3d incompressible isotropic
  turbulence.
\newblock {\em J.\ Turbulence}, 1:007, 2000.

\bibitem{p93}
P.J.E. Peebles.
\newblock {\em Principles of Physical Cosmology}.
\newblock Princeton University Press, 1993.

\bibitem{p95}
A.~Polyakov.
\newblock Turbulence without pressure.
\newblock {\em Phys.\ Rev.\ E}, 52:6183--6188, 1995.

\bibitem{pr54}
I.~Proudman and W.H. Reid.
\newblock On the decay of a normally distributed and homogeneous turbulent
  velocity field.
\newblock {\em Phil.\ Trans.\ R.\ Soc.\ Lond.~A}, 247:163--189, 1954.

\bibitem{pagep95}
N.~Provatas, T.~Ala-Nissila, M.~Grant, K.~Elder, and L.~Piche.
\newblock Flame propagation in random media.
\newblock {\em Phys.\ Rev.\ E}, 51:4232--4236, 1995.

\bibitem{rs78}
H.A. Rose and P.-L. Sulem.
\newblock Fully developed turbulence and statistical mechanics.
\newblock {\em J.\ Phys.\ France}, 39:441--484, 1978.

\bibitem{ss95}
B.~Schraiman and E.~Siggia.
\newblock Anomalous scaling of a passive scalar in turbulent flow.
\newblock {\em C.R.\ Acad.\ Sci.}, 321:279--284, 1995.

\bibitem{saf92}
Z.S. She, E.~Aurell, and U.~Frisch.
\newblock The inviscid {B}urgers equation with initial data of {B}rownian type.
\newblock {\em Commun.\ Math.\ Phys.}, 148:623--641, 1992.

\bibitem{s92}
Ya. Sinai.
\newblock Statistics of shocks in solutions of inviscid {B}urgers equation.
\newblock {\em Commun.\ Math.\ Phys.}, 148:601--622, 1992.

\bibitem{s99}
A.~Sobolevski{\u\i}.
\newblock Periodic solutions of the {H}amilton--{J}acobi equation with a
  periodic non-homogeneous term and {A}ubry--{M}ather theory.
\newblock {\em Mat.\ Sbornik}, 190:1487--1504, 1999.

\bibitem{vdfn94}
M.~Vergassola, B.~Dubrulle, U.~Frisch, and A.~Noullez.
\newblock {B}urgers' equation, {D}evil's staircases and the mass distribution
  for large-scale structures.
\newblock {\em Astron.\ Astrophys.}, 289:325--356, 1994.

\bibitem{v67}
A.I. Vol'pert.
\newblock Spaces {B}{V} and quasilinear equations.
\newblock {\em Mat.\ Sb.}, 73:255--302, 1967.

\bibitem{kh38}
T.~von K\'arm\'an and L.~Howarth.
\newblock On the statistical theory of isotropic turbulence.
\newblock {\em Proc.\ R.\ Soc.\ Lond.~A}, 164:192--215, 1938.

\bibitem{w98}
W.~Woyczy{\'n}ski.
\newblock {\em Burgers-{K}{P}{Z} Turbulence: Gottingen lectures}.
\newblock Springer--Verlag, New York, 1998.

\bibitem{yc96}
V.~Yakhot and A.~Chekhlov.
\newblock Algebraic tails of probability density functions in the
  random-force-driven {B}urgers turbulence.
\newblock {\em Phys.\ Rev.\ Lett.}, 77:3118--3121, 1996.

\bibitem{z70}
Ya. Zel'dovich.
\newblock Gravitational instability: an approximate theory for large density
  perturbations.
\newblock {\em Astron.\ Astrophys.}, 5:84--89, 1970.

\end{thebibliography}

\end{document}